\documentclass[]{spie}  
\usepackage{amsmath,amsfonts,amssymb}
\usepackage{authblk}
\usepackage{multicol}
\usepackage{blindtext,graphicx}
\usepackage[absolute]{textpos}
\usepackage{bm}
\usepackage[colorlinks=true, allcolors=blue]{hyperref}
\usepackage{multicol}
\usepackage{floatrow}
\usepackage{subcaption}
\graphicspath{{./Images/}}

\begin{document}
	
	\begin{textblock}{20}(4,1)
		\normalsize Distribution Statement A: Approved for public release: distribution is unlimited 
		\vspace{-22mm}
	\end{textblock}
	
	\title{\vspace{9mm}Comparison of Possibilistic Fuzzy Local Information C-Means and Possibilistic K-Nearest Neighbors for Synthetic Aperture Sonar Image Segmentation}
	
	\author[a]{Joshua Peeples}
	\author[a]{Matthew Cook}
	\author[a]{Daniel Suen}
	\author[a]{Alina Zare}
	\author[b]{James Keller}
	\affil[a]{Electrical and Computer Engineering, University of Florida,Gainesville, FL, 32611}
	\affil[b]{Computer Science and Electrical Engineering, University of Missouri,Columbia, MO, 65211}
	
	
	\pagestyle{empty} 
	\setcounter{page}{301} 

	\maketitle
	
	\begin{abstract}
		Synthetic aperture sonar (SAS) imagery can generate high resolution images of the seafloor. Thus, segmentation algorithms can be used to partition the images into different seafloor environments. In this paper, we compare two possibilistic segmentation approaches. Possibilistic approaches allow for the ability to detect novel or outlier environments as well as well known classes. The Possibilistic Fuzzy Local Information C-Means (PFLICM) algorithm has been previously applied to segment SAS imagery.  Additionally, the Possibilistic $K$-Nearest Neighbors (P$K$NN) algorithm has been used in other domains such as landmine detection and hyperspectral imagery.  In this paper, we compare the segmentation performance of a semi-supervised approach using PFLICM and a supervised method using Possibilistic $K$-NN.  We include final segmentation results on multiple SAS images and a quantitative assessment of each algorithm.	
	\end{abstract}
	
	
	\section{Introduction}
	
	Automatic target recognition (ATR) systems have a long history of study\cite{bhanu1986automatic,
		roth1990survey}, and finds continued importance in synthetic aperture sonar (SAS) images
	\cite{del2009automatic,williams2014exploiting,groen2010model,lyons2018comparison}. 
	SAS produces high resolution images of the sea environment \cite{hayes2009synthetic}. High 
	quality data allows environmental context information to be extracted which has proven effective 
	in boosting overall performance of machine learning systems in several applications 
	\cite{zare2009context,williams2014exploiting,du2015possibilistic,Liu2018Fully}. To create this 
	environmental context, the SAS imagery needs to be segmented into the various textures that 
	compose the seafloor. Numerous segmentation algorithms \cite{williams2009bayesian,fandos2011high}
	have been previously applied to partition the sea environment from SAS imagery, recently, 
	possibilistic clustering algorithms have been investigated \cite{zare2017possibilistic,
		peeples2018possibilistic}.
	
\begin{figure}[htb]
		\centering
		\includegraphics[width=.5\textwidth]{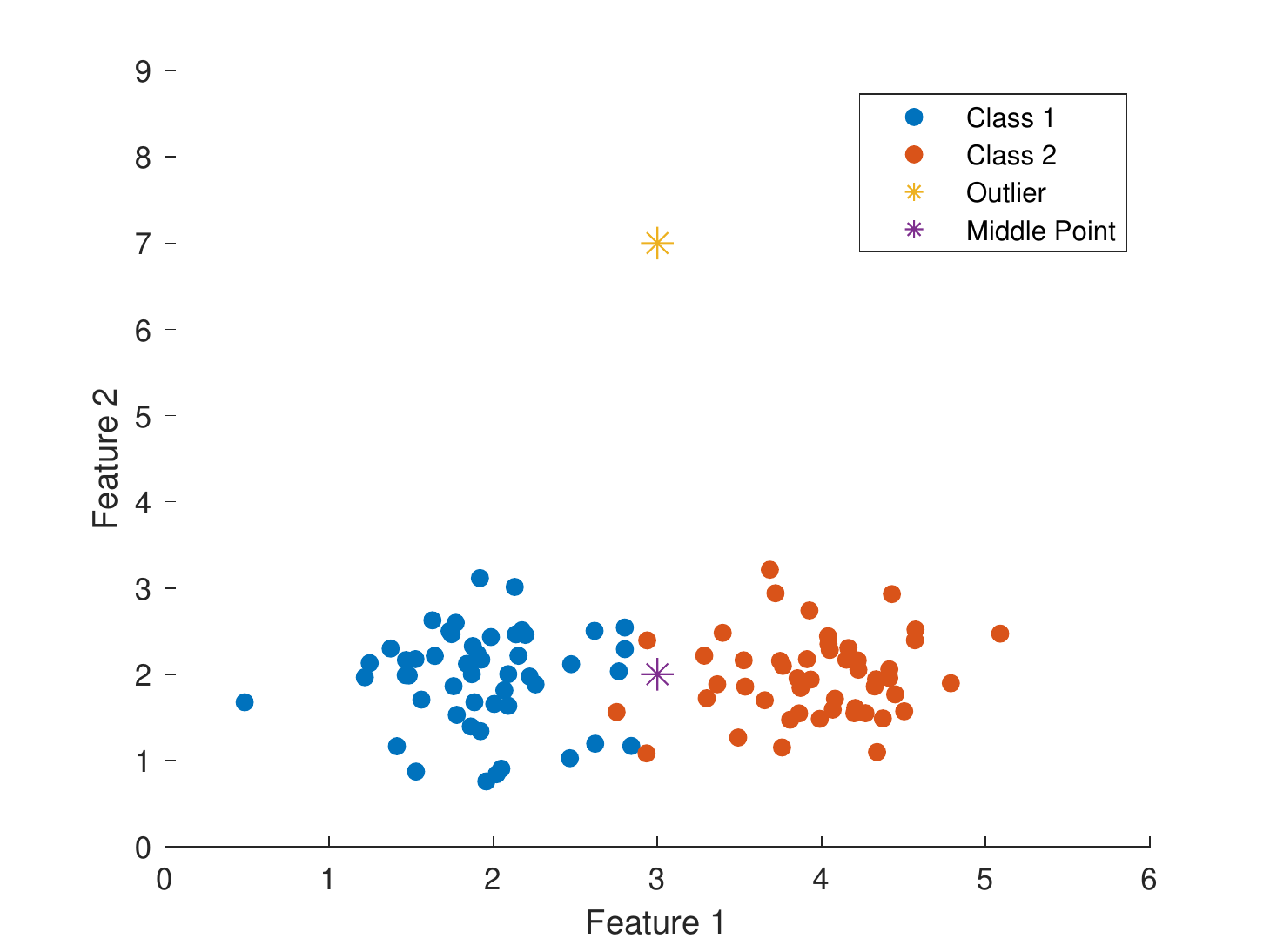}
		\caption{Example to illustrate differences between possibilistic and probabilistic
			weight assignments.}
		\label{PSSPRBex}
	\end{figure}

	Possibility theory was first proposed in 1978 by Zadeh\cite{zadeh1978fuzzy} and various 
	algorithms have been developed to apply these principles since then
	\cite{krishnapuram1993possibilistic,zhang2004improved,jenhani2008decision}. Possibility 
	theory is similar to probablistic theory in that possibility assignments, called 
	typicalities, must fall in the closed range $[0, 1]$. However, the main advantage to 
	possibilities over common probabilistic models such as mixture models is that the sum to one 
	constraint is relaxed. This allows outliers to receive low possibilities whereas in the 
	probabilistic model these points might be assigned uniform membership across all classes. On 
	the other hand, samples that are very similar to multiple classes can be assigned high 
	possibilities in multiple classes, while in sum-to-one constrained models the assigned value 
	for each class would be much lower. An example of this is shown in Figure \ref{PSSPRBex}. 
	In this figure, if a Gaussian mixture model were used to assign memberships to the outlier and 
	the middle point, the value would be 0.5 in both cases since the points are equidistant from 
	each cluster center, whereas if the membership was possibilistic small weights would be 
	assigned to the outlier, while values larger than 0.5 would be assigned to the middle point 
	since the data is close to both clusters.
	
	The Possibilistic Fuzzy Local Information C-Means (PFLICM) \cite{zare2017possibilistic} and 
	Possibilistic $K$-Nearest Neighbors (P$K$NN) \cite{frigui2009detection} algorithms both use 
	possibility theory and have shown success in various applications. PFLICM has shown 
	success in distinguishing seafloor textures 
	\cite{zare2017possibilistic,peeples2018possibilistic} while P$K$NN has been used in 
	landmine detection \cite{frigui2009detection}. Both of these possibilistic approaches could 
	be extended to improve ATR for SAS imagery by providing soft labels that characterize the 
	contents of the seafloor texture. The soft labels could then be used by other algorithms 
	that use seafloor texture information to improve their performance.
	
	In the two previous works that use PFLICM on SAS data, Zare et al 2017 paper
	\cite{zare2017possibilistic} and Peeples et al 2018 paper \cite{peeples2018possibilistic}, PFLICM
	was applied in an unsupervised manner, and worked well in that situation. However, in the 
	environmental context problem, it is sometimes good to know what exactly each cluster 
	represents. In this way, a separate ATR pipeline could be implemented for each of the known 
	environments detected. For this reason, PFLICM has been modified in this paper to allow 
	semi-supervised learning to occur for the purpose of explicitly labelling each of the learned 
	cluster centers. In this paper, P$K$NN will be used to segment SAS imagery and be compared with PFLICM for that task. The PFLICM and P$K$NN algorithms are detailed in Sections \ref{sec:PFLICM} and 
	\ref{sec:PKNN} respectively. 
	\section{METHODS}
	\label{sec:Methods}
	\subsection{PFLICM}
	\label{sec:PFLICM}
	
	The Possibilistic Fuzzy Local Information C-Means (PFLICM) algorithm integrates two previous clustering algorithms, Fuzzy Local Information C-Means (FLICM)\cite{krinidis2010robust} and Possibilisitic Fuzzy C-Means (PFCM)\cite{pal2005possibilistic}.  The objective function for PFLICM is shown in (\ref{eq:PFLICM}): \begin{equation}
	J = \sum_{c = 1}^{C}\sum_{n=1}^{N}a{u_{cn}^m\big({||\textbf{x}_n-\textbf{c}_c||}^2_2} + G_{cn}\big) + bt_{cn}^q{||\textbf{x}_n-\textbf{c}_c||^2_2} + \sum_{c = 1}^{C}\gamma_c\sum_{n=1}^{N}\big{(1-t_{cn})}^q, \label{eq:PFLICM}
	\end{equation} with the following constraints
	\begin{equation}
	u_{cn}\geq0 \hspace{10mm}\forall n = 1,...,N \hspace{10mm}\sum_{c=1}^{C}u_{cn} = 1,
	\end{equation} where $u_{cn}$ is the membership of the $n^{th}$ pixel in the $c^{th}$ cluster, $\textbf{x}_n$ is a $d \times 1$ vector for the nth pixel, $\textbf{c}_c$ is a $d \times 1$ vector of the $c^{th}$ cluster, $t_{cn}$ is the typicality value of the $n^{th}$ pixel in the $c^{th}$ cluster, $a$, $b$ and $\gamma_c$ are weights on the membership and typicality terms respectively, and $m$ and $q$ control the degree of the membership values for each cluster and identification of outliers in the data respectively. The $G_{cn}$ term follows from Krinidis\cite{krinidis2010robust} and incorporates local spatial information:
	\begin{equation}
	G_{cn} = \underset{k\neq n}{\sum_{{k\in\mathcal{N}_n}}}\frac{1}{d_{nk}+1}\big(1-u_{ck})^m||\textbf{x}_k-\textbf{c}_c||^2_2,
	\end{equation}
	where $\textbf{x}_n$ is the center pixel of a local window, $\mathcal{N}_n$ is the neighborhood around the center pixel, and $d_{nk}$ is the Euclidean distance between the center pixel and one of the neighboring pixels ($\textbf{x}_k$). The objective function is comprised of fuzzy membership and typicality terms. For the fuzzy membership terms, $\sum_{c = 1}^{C}\sum_{n=1}^{N}a{u_{cn}^m\big({||\textbf{x}_n-\textbf{c}_c||}^2_2} + G_{cn}\big)$, the membership of a data point will be higher to cluster centers that are closer (i.e. ${||\textbf{x}_n-\textbf{c}_c||}^2_2$ is small). The $G_{cn}$ term encourages neighboring pixels to have similar membership values by incorporating local spatial information (i.e. $d_{nk}$) and also serves as a penalty in the objective function. If the pixels are close (i.e. $d_{nk}$ is small) and membership is low, the $G_{cn}$ term will be large. If a neighbor is far away (i.e. $d_{nk}$ is large), the $G_{cn}$ value of that pixel should be small resulting in a small contribution to the overall objective function. Also, if the neighbors have high membership values, the $G_{cn}$ term will also be small and have little influence on the objective function. Similarly, the typicality terms, $bt_{cn}^q{||\textbf{x}_n-\textbf{c}_c||^2_2} + \sum_{c = 1}^{C}\gamma_c\sum_{n=1}^{N}\big{(1-t_{cn})}^q$, follow the same principle. The typicality of a data point will be higher to cluster centers that are close. The only difference is that the typicality values do not have the sum to one constraint. This is useful in the identification of new texture types because the typicality of a given data point will be low in all clusters.
	
\begin{figure}[h]
		\centering
		\includegraphics[scale=.33]{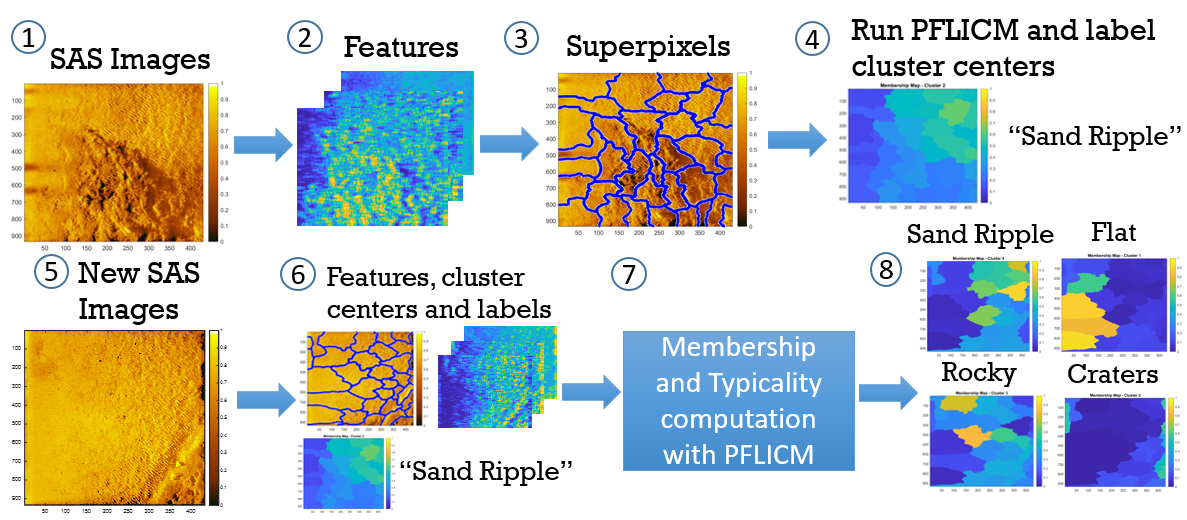}
		\caption{PFLICM: Creating Environmental Context}
		\label{fig:PFLICM_ext}
	\end{figure}

	The cluster centers, membership and typicality values are updated by alternating optimization. After random initialization, the partial derivative with respect to the $\textbf{c}_c$, $u_{cn}$, and $t_{cn}$ is calculated. After setting each expression equal to 0, the following update equations are obtained. For the update equation for the membership values, a Lagrange multiplier term was added to enforce the sum-to-one constraint:
	\begin{equation}
	\textbf{c}_c = \sum_{n}\cfrac{\big(au_{cn}^m + bt_{cn}^q\big)\textbf{x}_n}{\big(au_{cn}^m + bt_{cn}^q\big)},
	\end{equation}
	\begin{equation}
	u_{cn} = \cfrac{1}{\sum_{k=1}^{C}\bigg(\cfrac{(\textbf{x}_n-\textbf{c}_c)(\textbf{x}_n-\textbf{c}_c)^T+G_{cn}}{(\textbf{x}_n-\textbf{c}_k)(\textbf{x}_n-\textbf{c}_k)^T+G_{kn}}\bigg)},
	\end{equation}
	\begin{equation}
	t_{cn} = \cfrac{1}{1+\bigg(\frac{b}{\gamma_c}||\textbf{x}_k-\textbf{c}_c||^2_2\bigg)^{\frac{1}{q-1}}}, 
	\end{equation}
	$\gamma_c$ is the mean of the separation of all the data points in the corresponding cluster ($||\textbf{x}_k-\textbf{c}_c||^2_2$). The fuzzy factor term, $G$, is treated as a constant in the membership update equation\cite{celik2013comments}. PFLICM produces membership, typicality and cluster center values for the SAS imagery. The clusters centers can be assigned meaningful labels after clustering the ``training" images and then used to compute membership and typicality values for new or ``test" SAS images. The semi-supervised PFLICM extension is shown in Figure \ref{fig:PFLICM_ext}.
	\subsection{PKNN}
	\label{sec:PKNN}
	
	Possibilistic $K$ Nearest Neighbors\cite{frigui2009detection} extends the traditional
	$K$-Nearest Neighbors\cite{cover1967nearest} approach to return typicalities for each 
	class instead of crisp labels. The biggest difference comes during the initialization
	of P$K$NN. Here, a fuzzy membership is assigned to each sample in the training 
	data, this weight takes the form
	\begin{equation}
	\tilde{\mu}^i(y) = \left\{
	\begin{matrix}
	0.51 + \left(\frac{n_i}{K}\right) \times 0.49, &i = j\\
	\left(\frac{n_i}{K}\right) \times 0.49, &i \neq j
	\end{matrix}
	\right. .
	\label{muassign}
	\end{equation}
	Where $\tilde{\mu}(y)$ is the fuzzy membership of sample $y$, $j$ represents the actual
	class label of sample $y$, $i$ is the class for which the current fuzzy weight is being
	calculated for, and $n_i$ represents the number of $K$ nearest samples belonging to class
	$i$.
	
	In particular, instead of choosing the most often occurring
	class label in the $K$ nearest training samples, P$K$NN assigns a typicality computed as      
	\begin{equation}
	\text{Conf}^i(x^*) = \frac{1}{K}\sum\limits_{k=1}^K\tilde{\mu}^i(y_k)w_p(x^*,y_k),
	\label{Cassign}
	\end{equation}
	where Conf\textsuperscript{$i$}($x^*$) is the confidence that the current test point, $x^*$, 
	belongs to the i\textsuperscript{th} class in the training data and $y_k$ is the 
	$k^\text{th}$ nearest training sample. 
	
\begin{figure}[h]
		\centering
		\includegraphics[scale=.33]{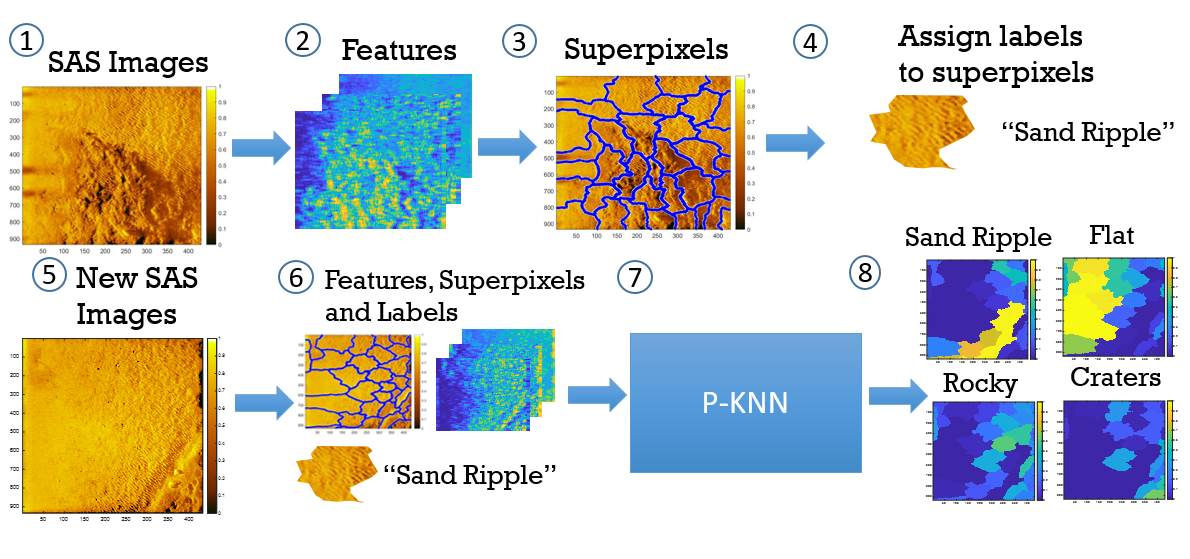}
		\caption{P$K$NN: Creating Environmental Context}
		\label{fig:PKNN_ext}
	\end{figure}

	The function $w_p(x,y_k)$ defines the possibilistic weight of each training sample for 
	the given test point and is related to the distance between the test point and each
	of the $k$ nearest neighbors. In particular the weight is calculated as 
	\begin{equation}
	w_p(x,y_k) = \left(1 + \max(0, \|x-y_k\|-\eta)^{2/(m-1)}\right)^{-1},
	\label{Wassign}
	\end{equation}
	where $w_p$ is the possibilistic weight for each class, $m\in(1,\infty)$ controls the 
	fuzziness of the weight function, and $\eta$ controls how much data is considered to
	be very close to the test sample. Note that this weight function does not depend on 
	the class label of the training samples and is used instead to determine how important 
	each training sample is in determining the correct label for x. This algorithm returns a confidence value for each class in the training data. These values do 
	not sum to one, which allows points that belong to no class to be assigned low confidence 
	values in every class. In contrast to PFLICM where the cluster centers are assigned labels, 
	each superpixel of the training data is given a label. The P$K$NN segmentation process is 
	shown in Figure \ref{fig:PKNN_ext}.
	\section{EXPERIMENTAL RESULTS}
	
	The experimental design for our evaluation of each algorithm is as follows: a dataset of 98 SAS images was used and three fold cross validation was performed. Folds two and three contained 33 images while fold one had 32 images. The classes represented in the dataset were sand ripples, flat, rocky, and craters. Before applying PFLICM and P$K$NN to each SAS image, feature extraction and superpixel segmentation were implemented. For each image, a 34 dimensional feature vector was computed using Sobel features\cite{frigui2009detection} (eight edge orientations and square mask sizes five, nine, eleven and fifteen) and lacunarity\cite{williams2015fast} (window sizes of [31,21] and [21,11]) to aid in capturing texture information. After the features are extracted, superpixels are computed on the grayscale image. Superpixels are pixels that are grouped together based on shared characteristics such as proximity, similarity, good continuation and other metrics\cite{ren2003learning}. Superpixel segmentation reduces computational costs because an algorithm can be applied at the superpixel level as opposed to each individual pixel. The fast normalized cuts with linear constraints algorithm was used for superpixel segmentation\cite{xu2009fast}. Once the superpixels are computed, the feature vectors of each pixel that correspond to a given superpixel are averaged. The resulting mean feature vector is assigned as the feature vector of the superpixel. Once the features and superpixels are computed, PFLICM and P$K$NN are applied to each feature vector at the superpixel level.
	
\begin{figure}[ht!]
		\begin{subfigure}{.32\textwidth}{
				\includegraphics[width=\textwidth]{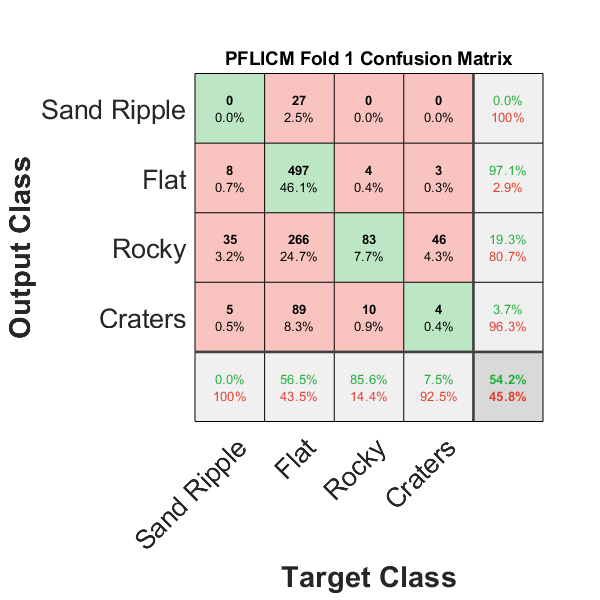}
				\caption{PFLICM Fold One}
				\label{fig:PFLICM_F1}
			}
		\end{subfigure}
		\begin{subfigure}{.32\textwidth}{
				\includegraphics[width=\textwidth]{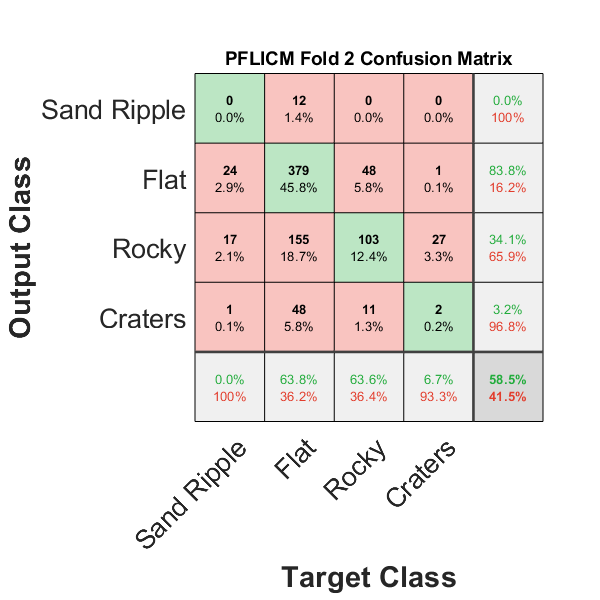}
				\caption{PFLICM Fold Two}
				\label{fig:PFLICM_F2}
			}
		\end{subfigure}
		\begin{subfigure}{.32\textwidth}{
				\includegraphics[width=\textwidth]{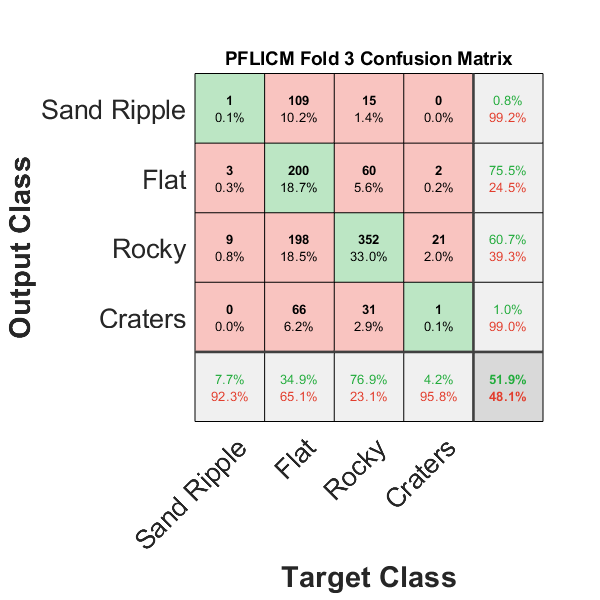}
				\caption{PFLICM Fold Three}
				\label{fig:PFLICM_F3}
			}
		\end{subfigure} 
		
		\begin{subfigure}{.32\textwidth}{
				\includegraphics[width=\textwidth]{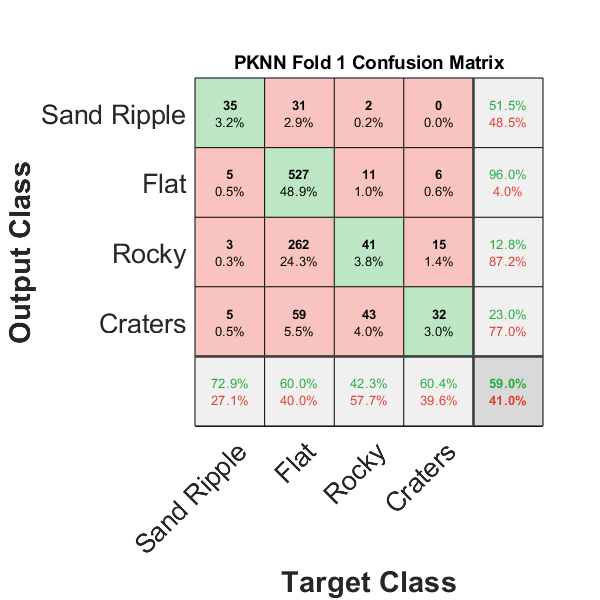}
				\caption{$P$KNN Fold One}
				\label{fig:PKNN_F1}
			}
		\end{subfigure}
		\begin{subfigure}{.32\textwidth}{
				\includegraphics[width=\textwidth]{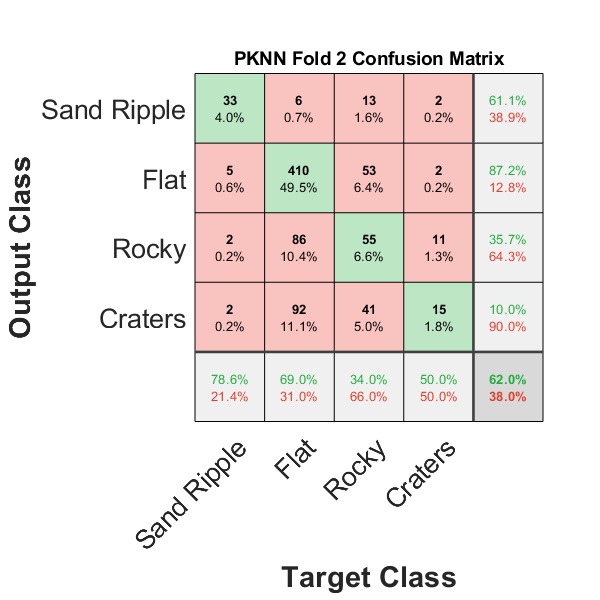}
				\caption{$P$KNN Fold Two}
				\label{fig:PKNN_F2}
			}
		\end{subfigure}
		\begin{subfigure}{.32\textwidth}{
				\includegraphics[width=\textwidth]{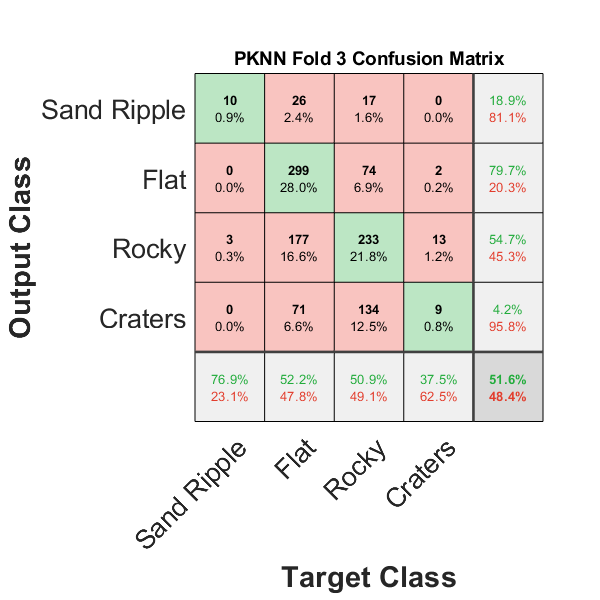}
				\caption{$P$KNN Fold Three}
				\label{fig:PKNN_F3}
			}
		\end{subfigure}
		
		\caption{Confusion matrices of PFLICM (\ref{fig:PFLICM_F1}-\ref{fig:PFLICM_F3}) and P$K$NN (\ref{fig:PKNN_F1}-\ref{fig:PKNN_F3}) on folds one, two and three respectively.}
		\centering
		\label{fig:ConfMats}
	\end{figure}

\begin{figure}[htb]
		\centering
		\begin{subfigure}{.19\textwidth}
			\centering
			\includegraphics[width=\textwidth]{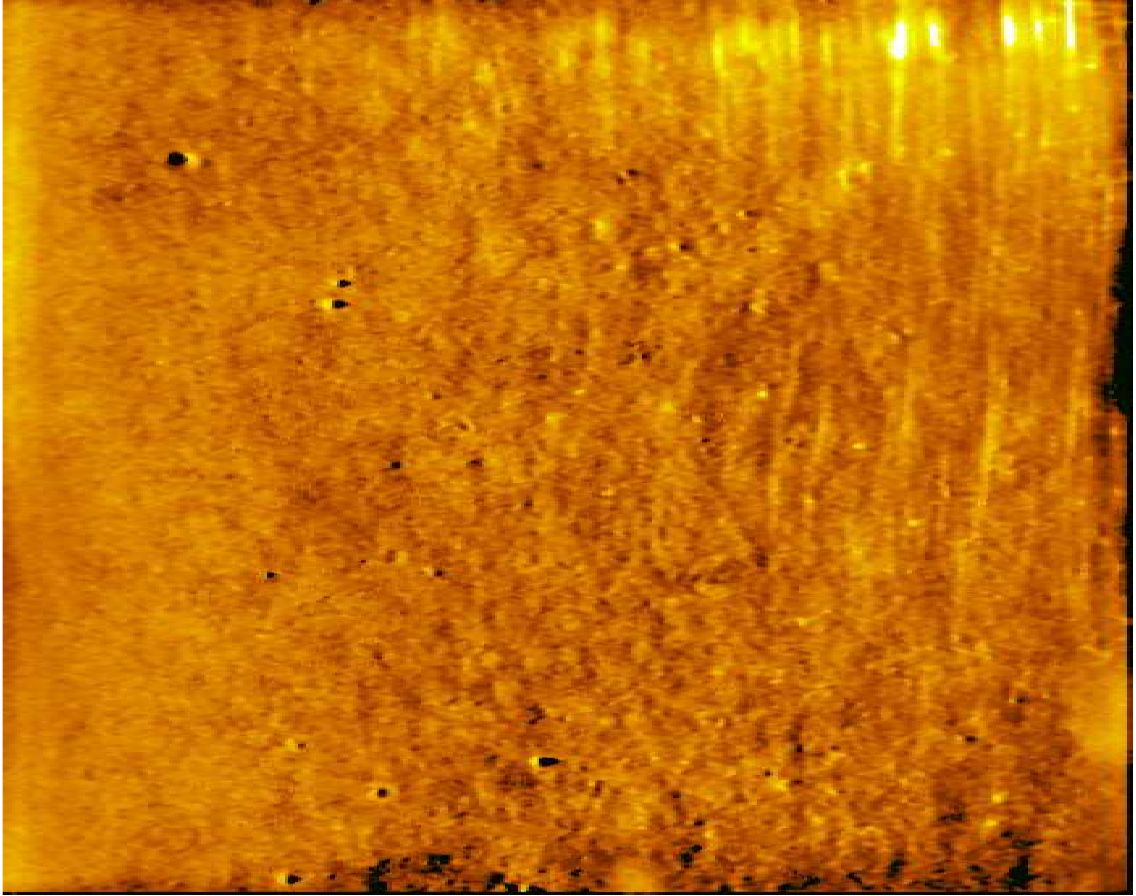}
			\caption{Original image}
		\end{subfigure}  
		\begin{subfigure}{.19\textwidth}
			\centering
			\includegraphics[width=\textwidth]{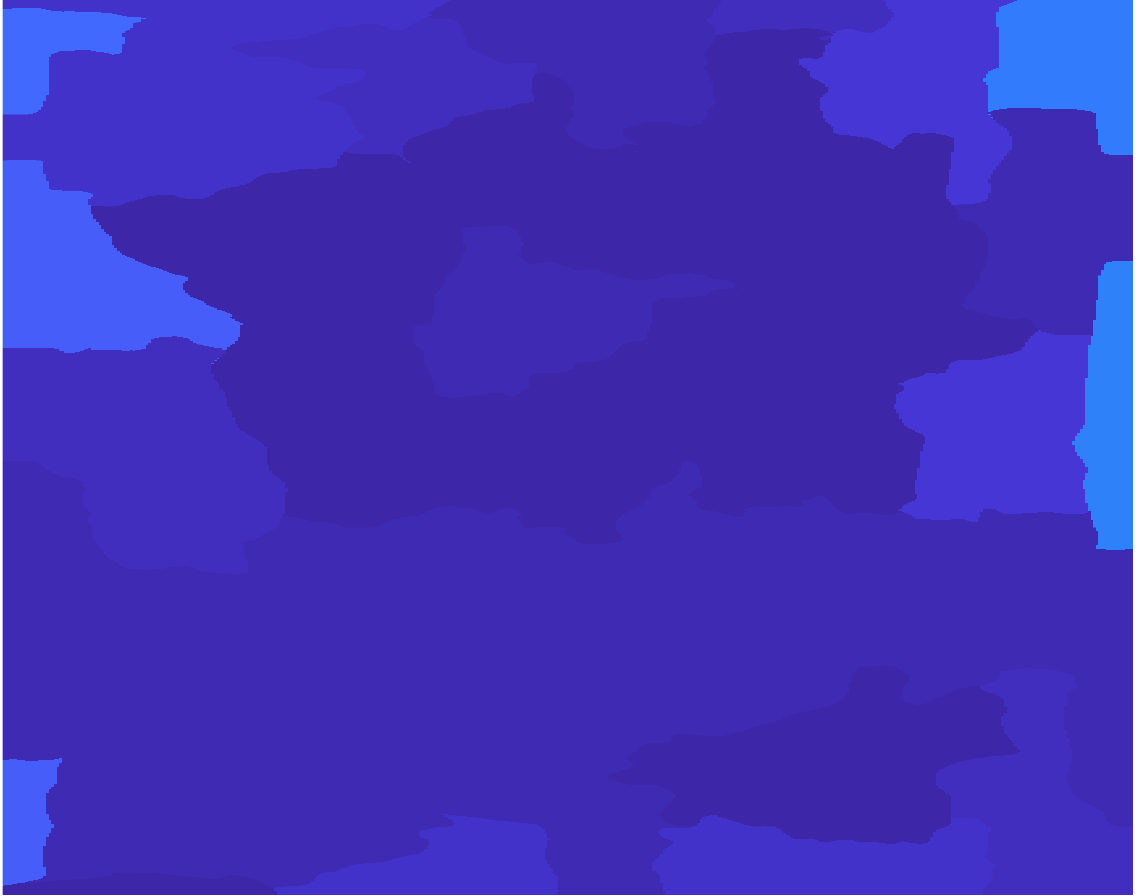}
			\caption{PFLICM - Crater}
		\end{subfigure}
		\begin{subfigure}{.19\textwidth}
			\centering
			\includegraphics[width=\textwidth]{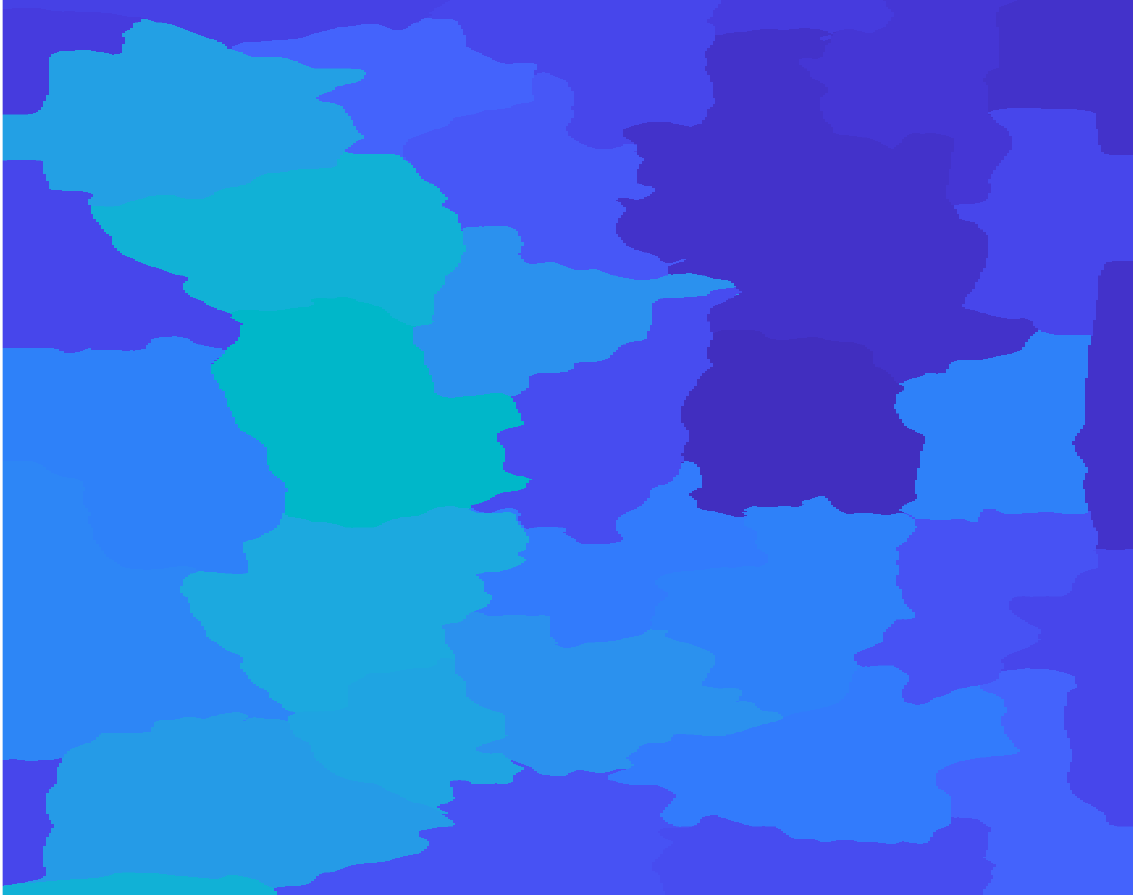}
			\caption{PFLICM - Flat}
		\end{subfigure}
		\begin{subfigure}{.19\textwidth}
			\centering
			\includegraphics[width=\textwidth]{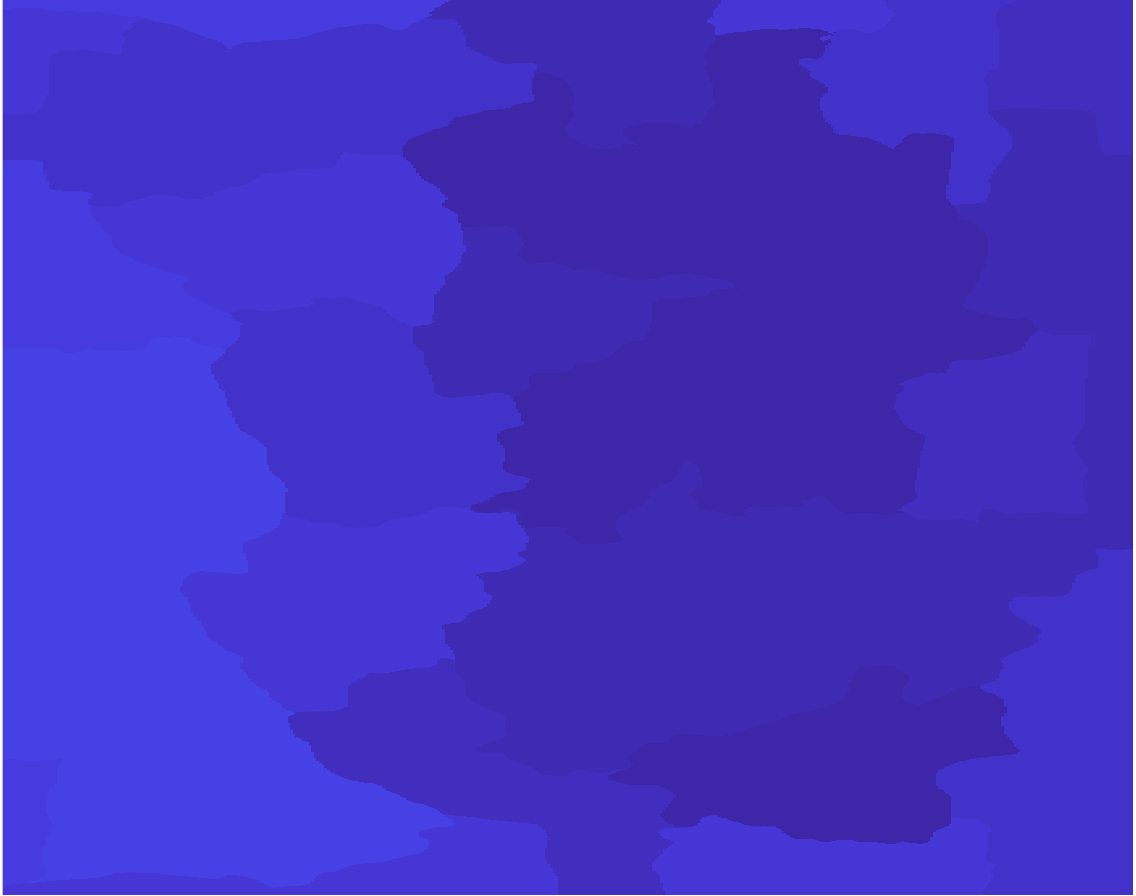}
			\caption{PFLICM - Ripples}
		\end{subfigure}
		\begin{subfigure}{.19\textwidth}
			\centering
			\includegraphics[width=\textwidth]{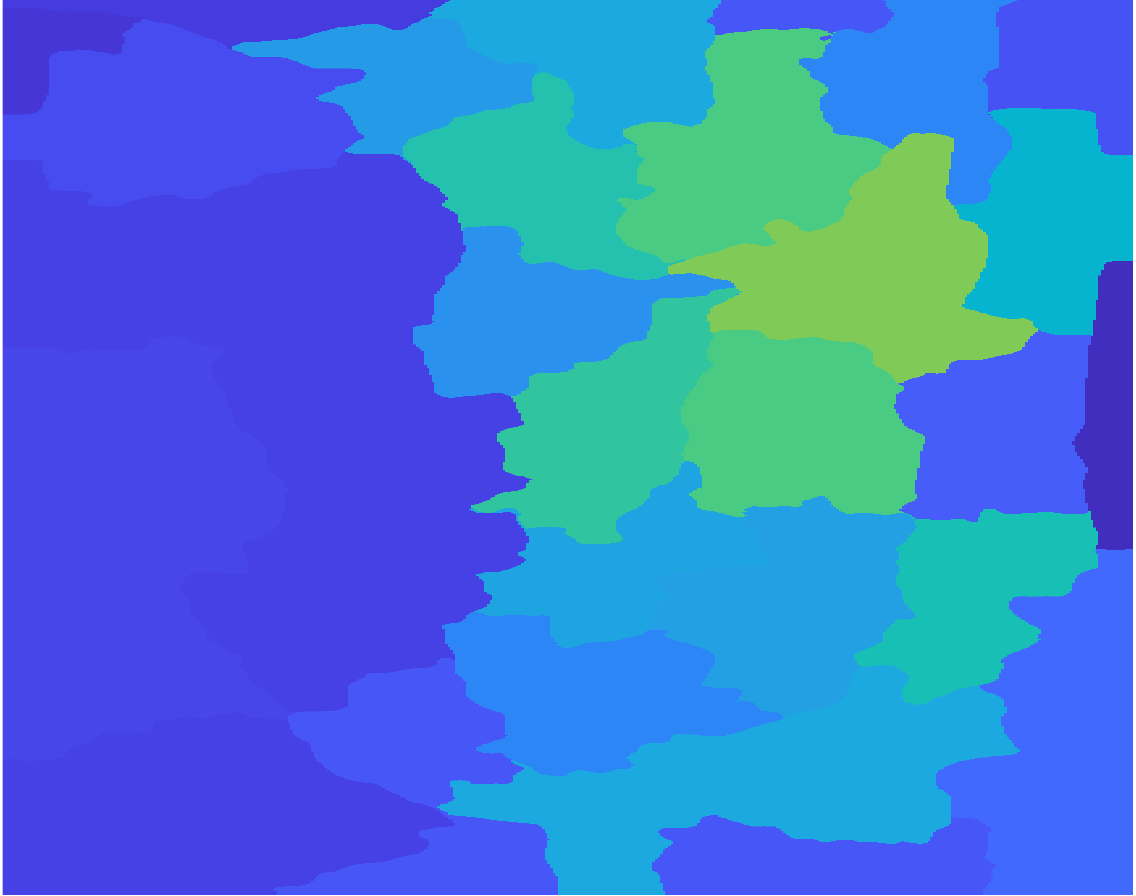}
			\caption{PFLICM - Rocky}
		\end{subfigure}
		
		\begin{subfigure}{.19\textwidth}
			\centering
			\includegraphics[width=\textwidth]{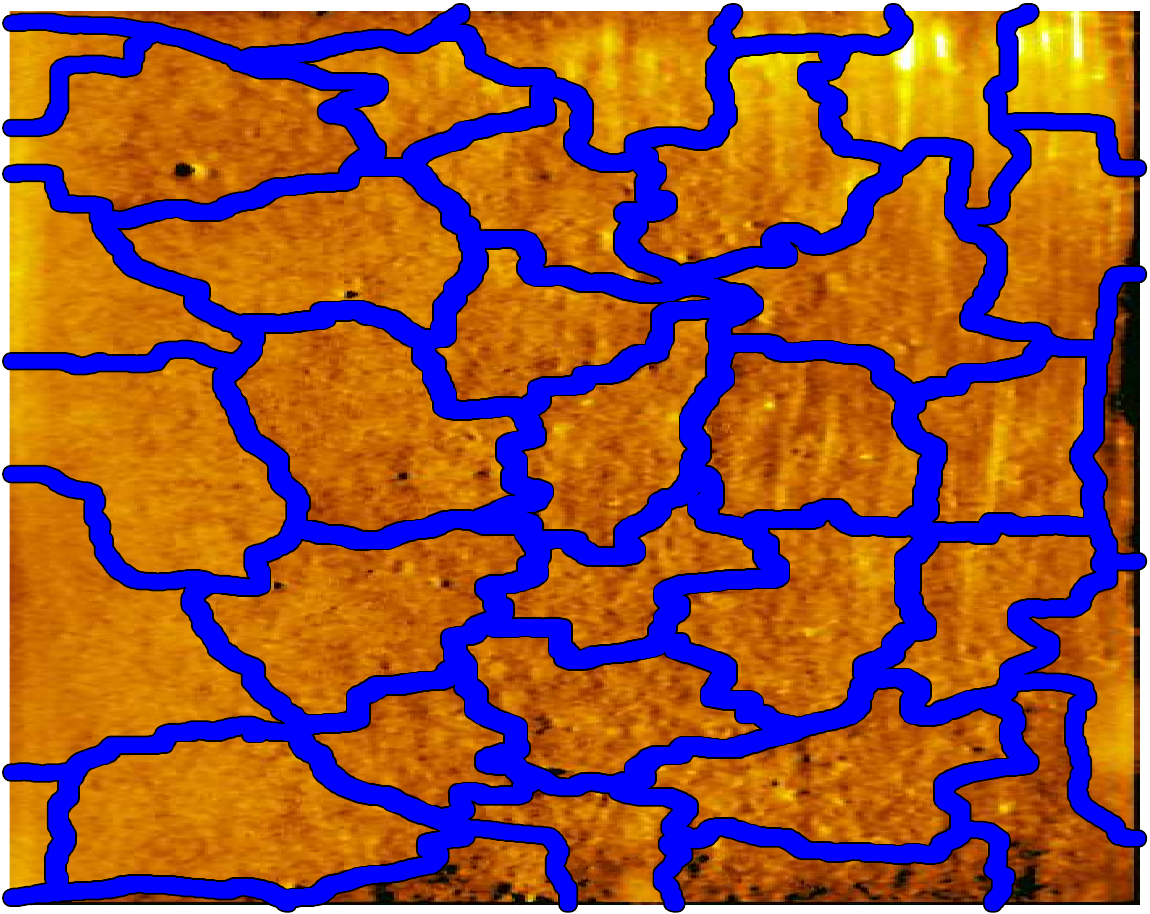}
			\caption{Superpixels}
		\end{subfigure}
		\begin{subfigure}{.19\textwidth}
			\centering
			\includegraphics[width=\textwidth]{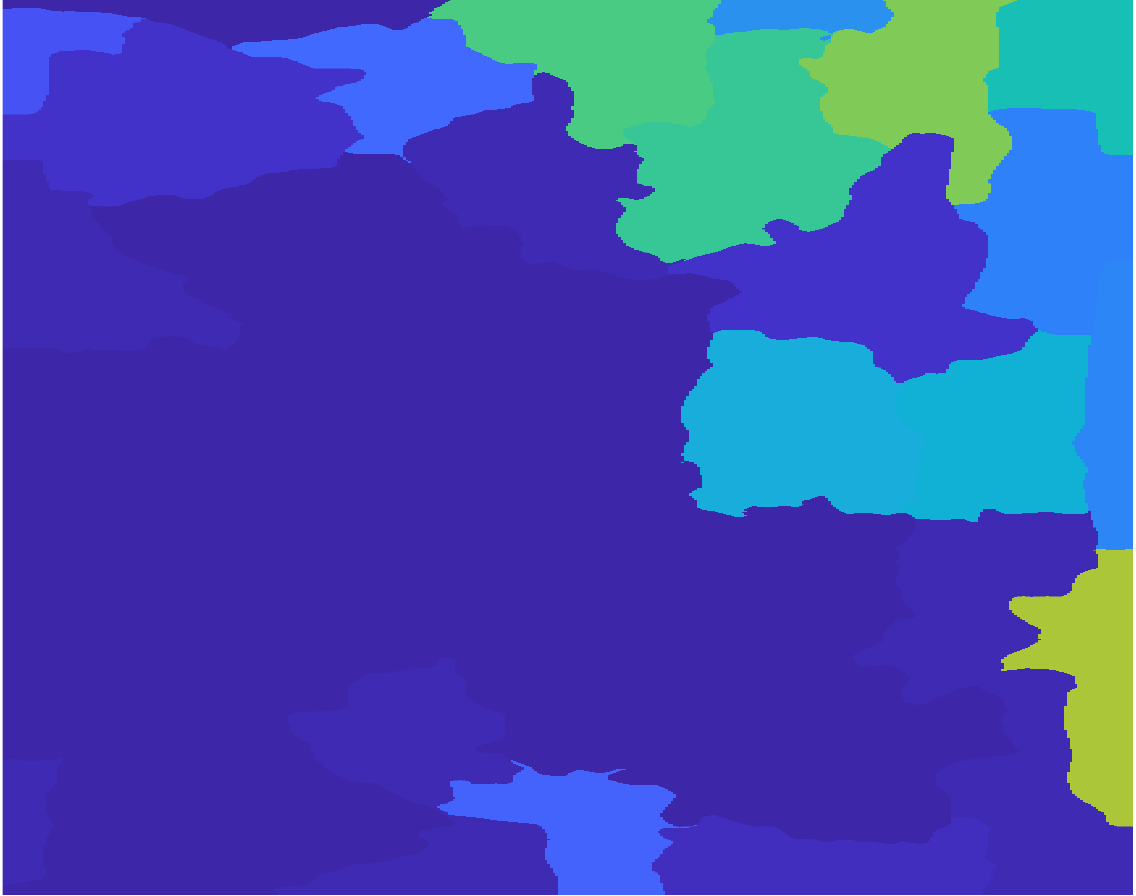}
			\caption{P$K$NN - Crater}
		\end{subfigure}
		\begin{subfigure}{.19\textwidth}
			\centering
			\includegraphics[width=\textwidth]{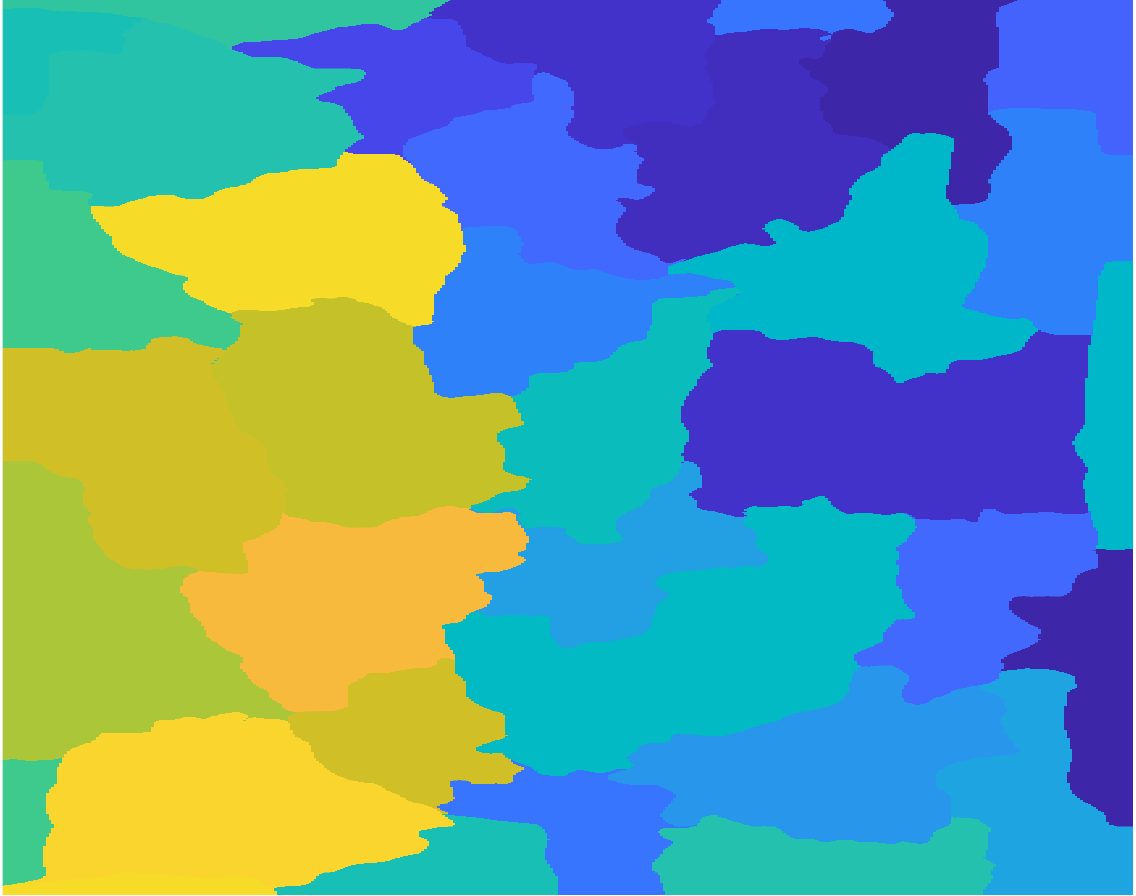}
			\caption{P$K$NN - Flat}
		\end{subfigure}
		\begin{subfigure}{.19\textwidth}
			\centering
			\includegraphics[width=\textwidth]{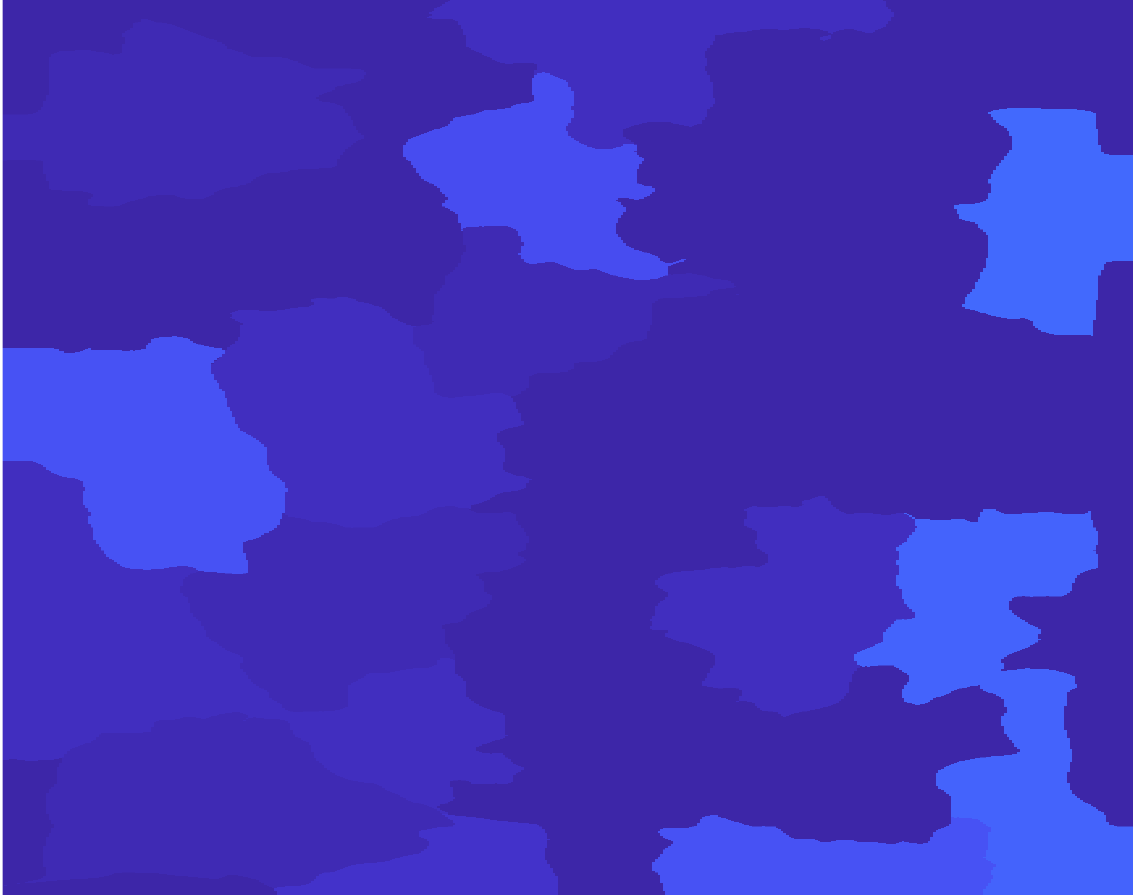}
			\caption{P$K$NN - Ripples}
		\end{subfigure}
		\begin{subfigure}{.19\textwidth}
			\centering
			\includegraphics[width=\textwidth]{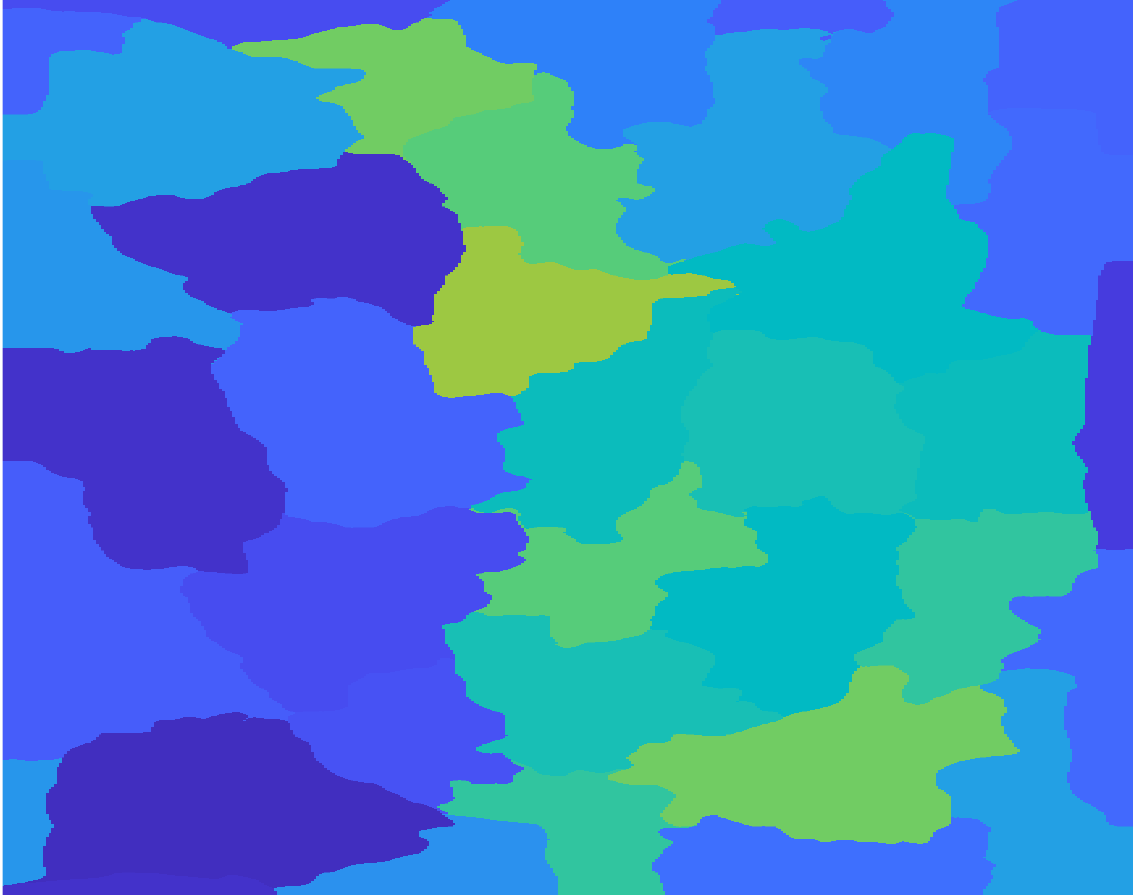}
			\caption{P$K$NN - Rocky}
		\end{subfigure}
		\caption{Example image one from fold one. The first column is actual image and superpixel segmentation. Of the remaining columns, the top row is PFLICM product maps and the bottom row is P$K$NN typicality maps.}
		\centering
		\label{fig:Img1}
	\end{figure}

\begin{figure}[htb]
		\centering
		\begin{subfigure}{.19\textwidth}
			\centering
			\includegraphics[width=\textwidth]{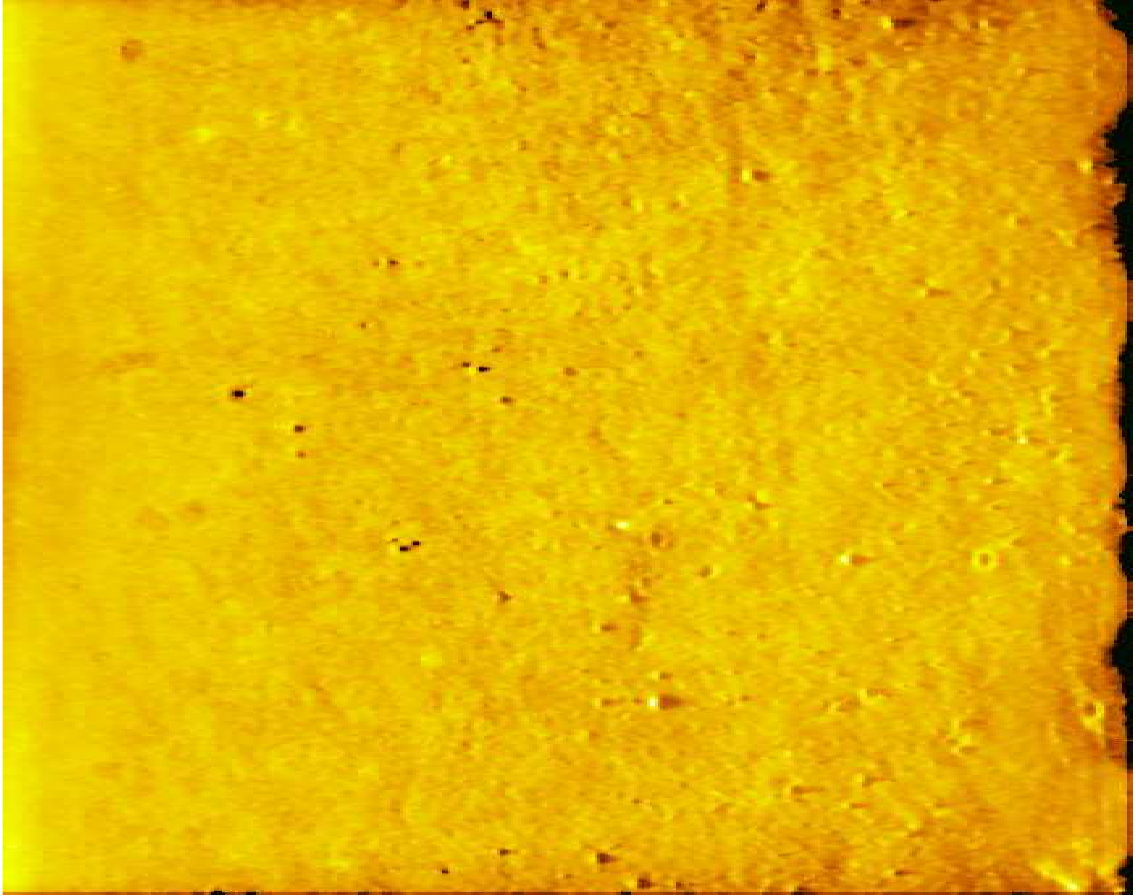}
			\caption{Original image}
		\end{subfigure}  
		\begin{subfigure}{.19\textwidth}
			\centering
			\includegraphics[width=\textwidth]{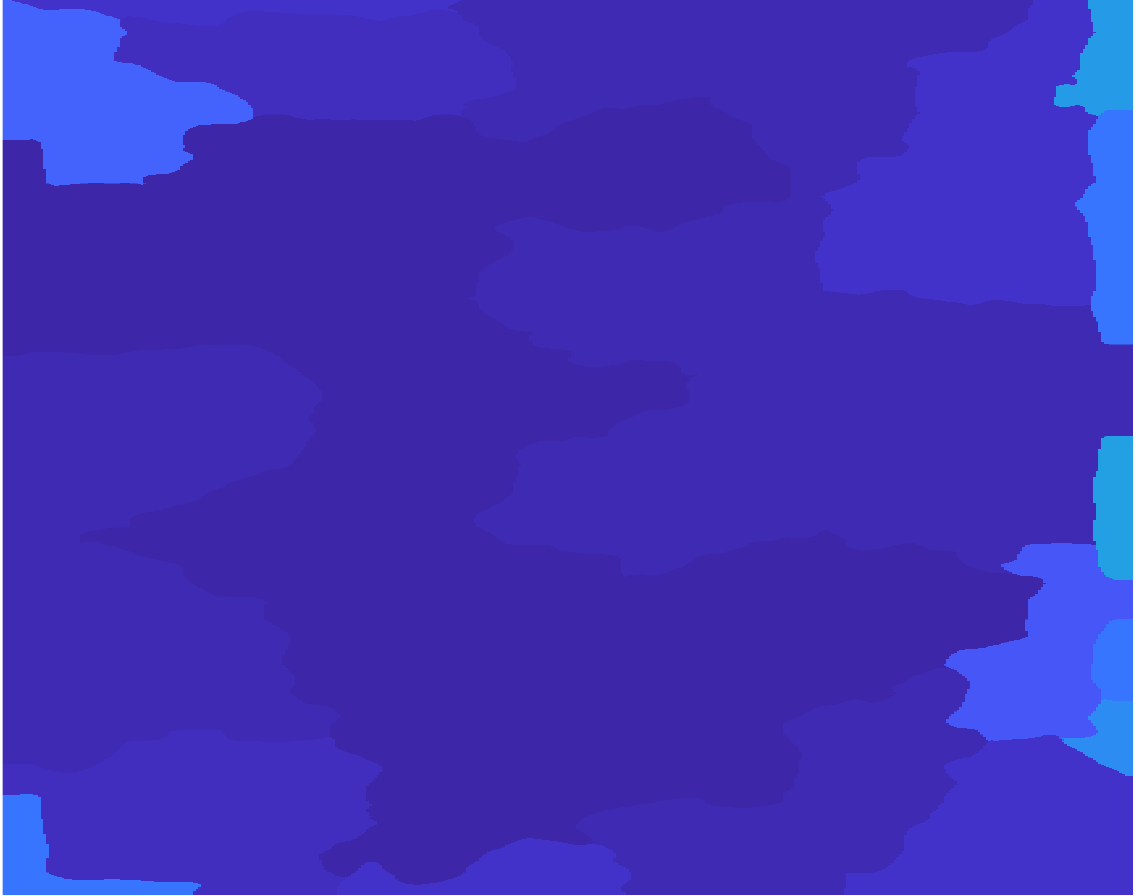}
			\caption{PFLICM - Crater}
		\end{subfigure}
		\begin{subfigure}{.19\textwidth}
			\centering
			\includegraphics[width=\textwidth]{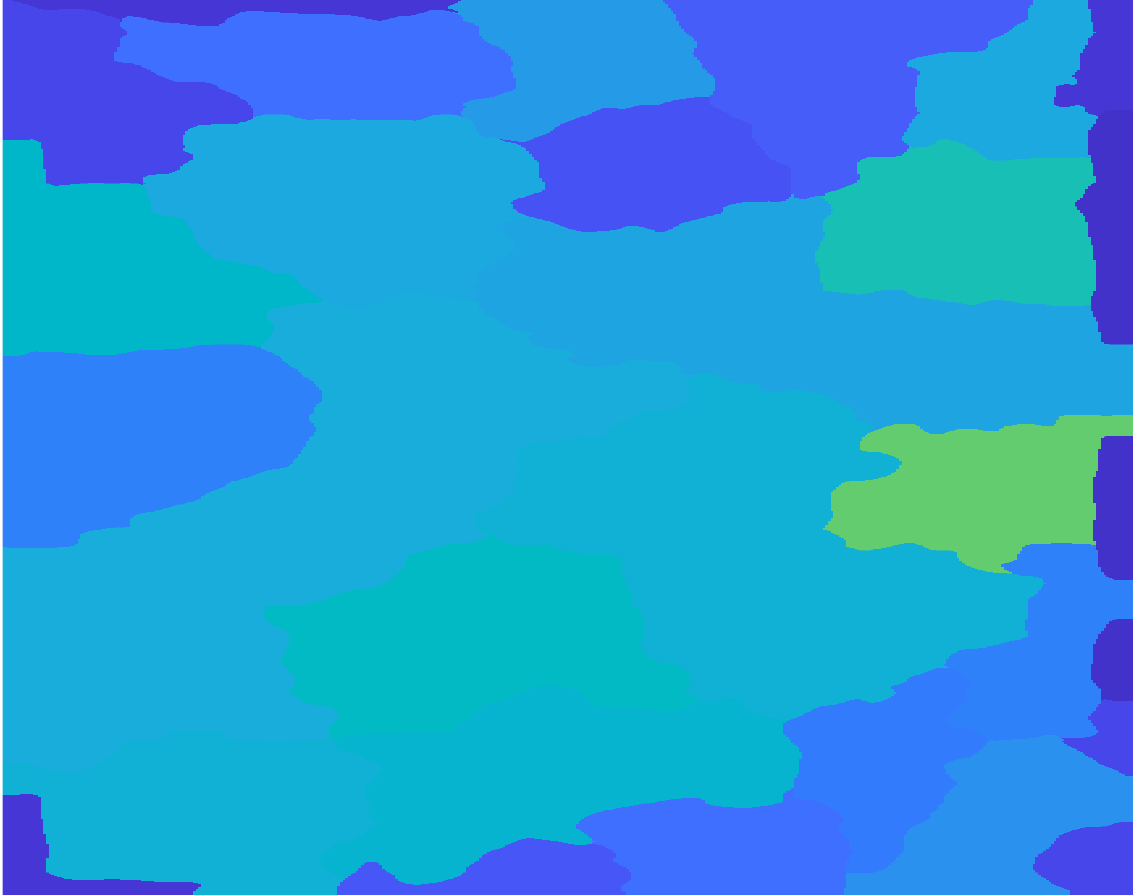}
			\caption{PFLICM - Flat}
		\end{subfigure}
		\begin{subfigure}{.19\textwidth}
			\centering
			\includegraphics[width=\textwidth]{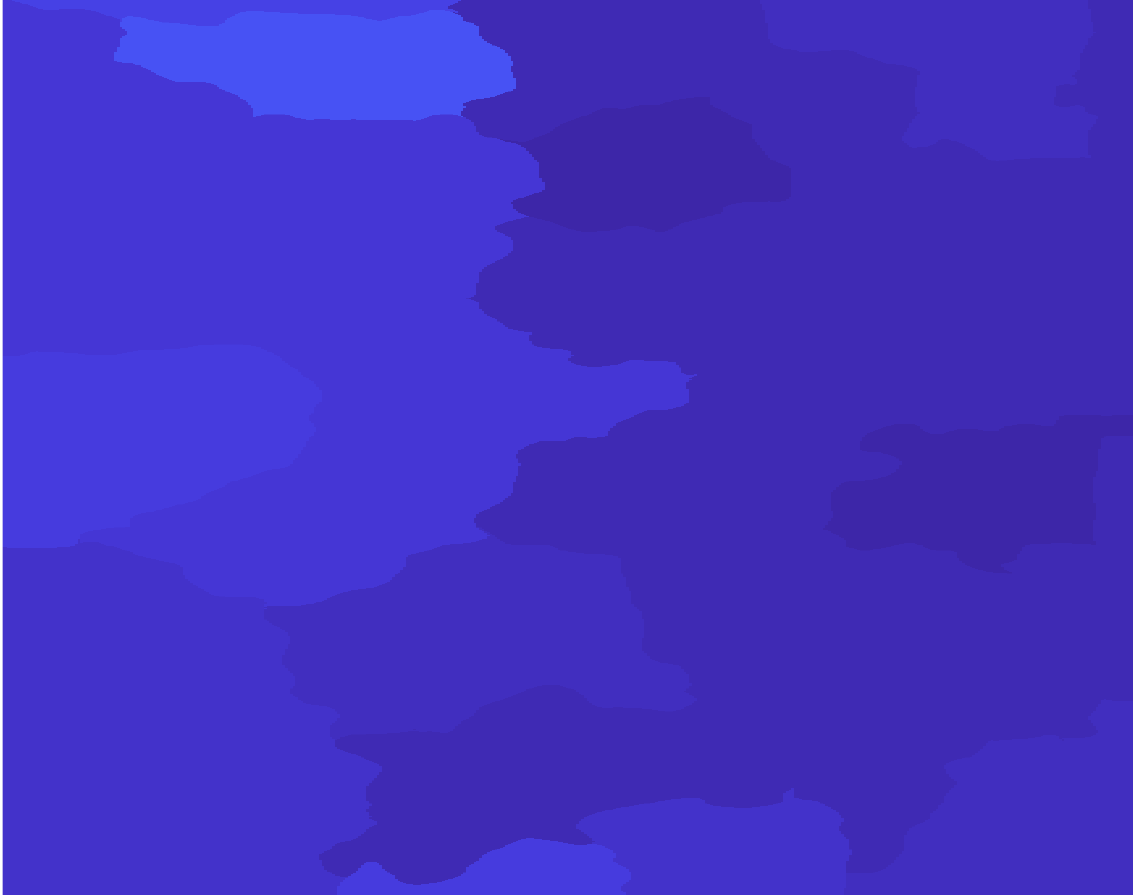}
			\caption{PFLICM - Ripples}
		\end{subfigure}
		\begin{subfigure}{.19\textwidth}
			\centering
			\includegraphics[width=\textwidth]{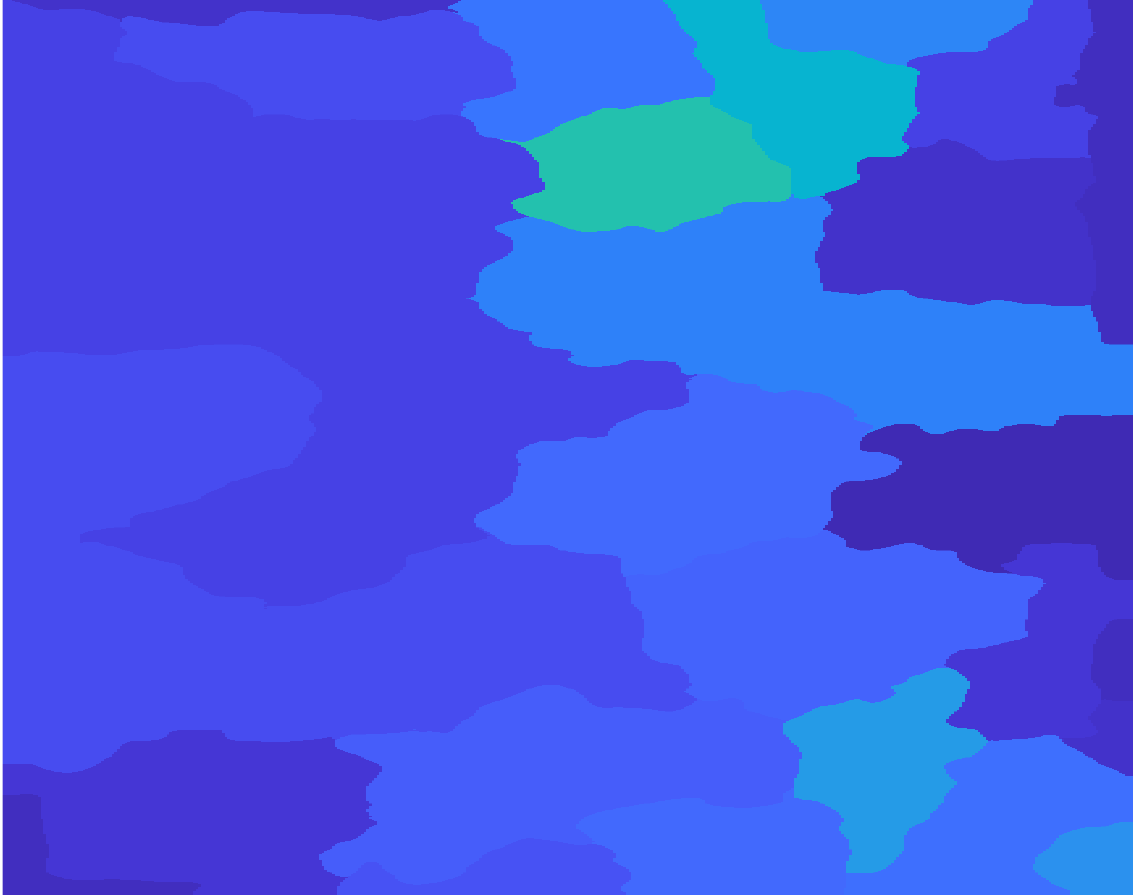}
			\caption{PFLICM - Rocky}
		\end{subfigure}
		
		\begin{subfigure}{.19\textwidth}
			\centering
			\includegraphics[width=\textwidth]{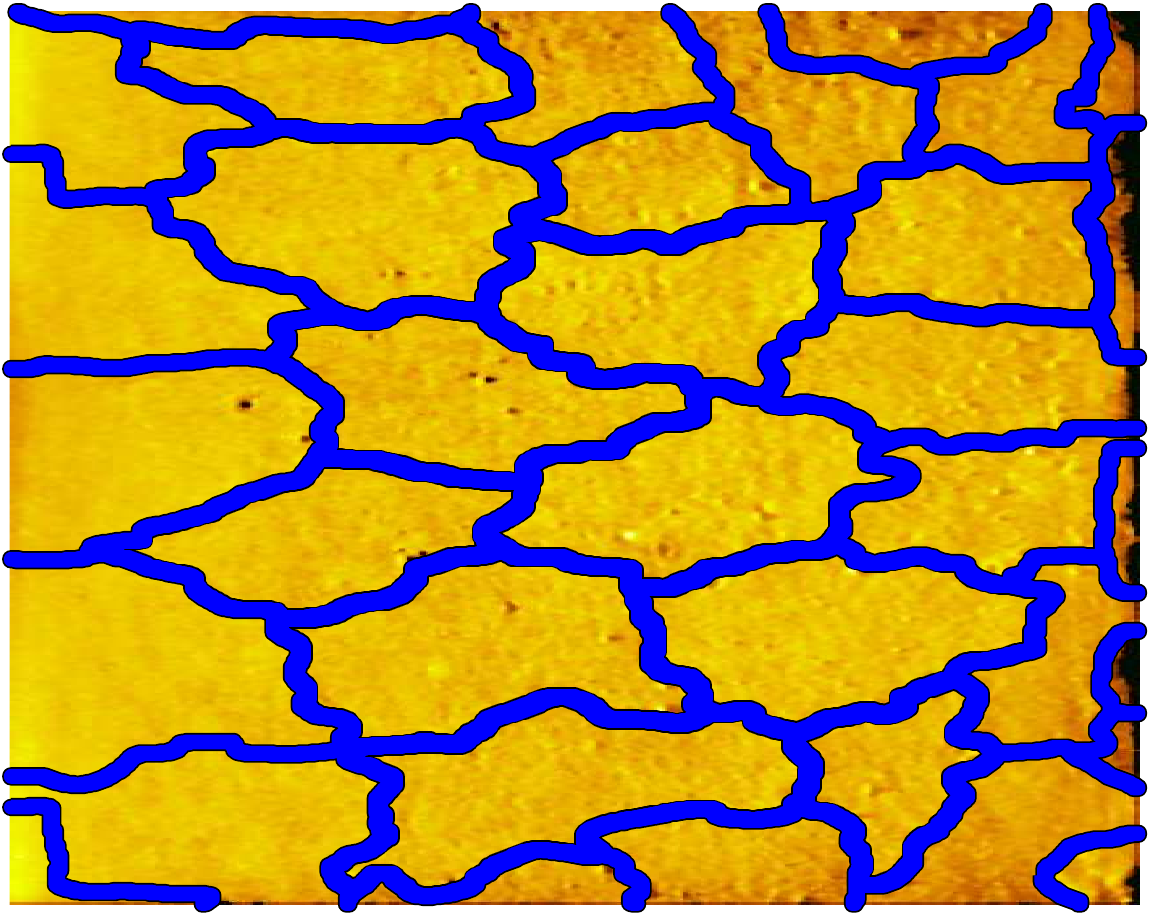}
			\caption{Superpixels}
		\end{subfigure}
		\begin{subfigure}{.19\textwidth}
			\centering
			\includegraphics[width=\textwidth]{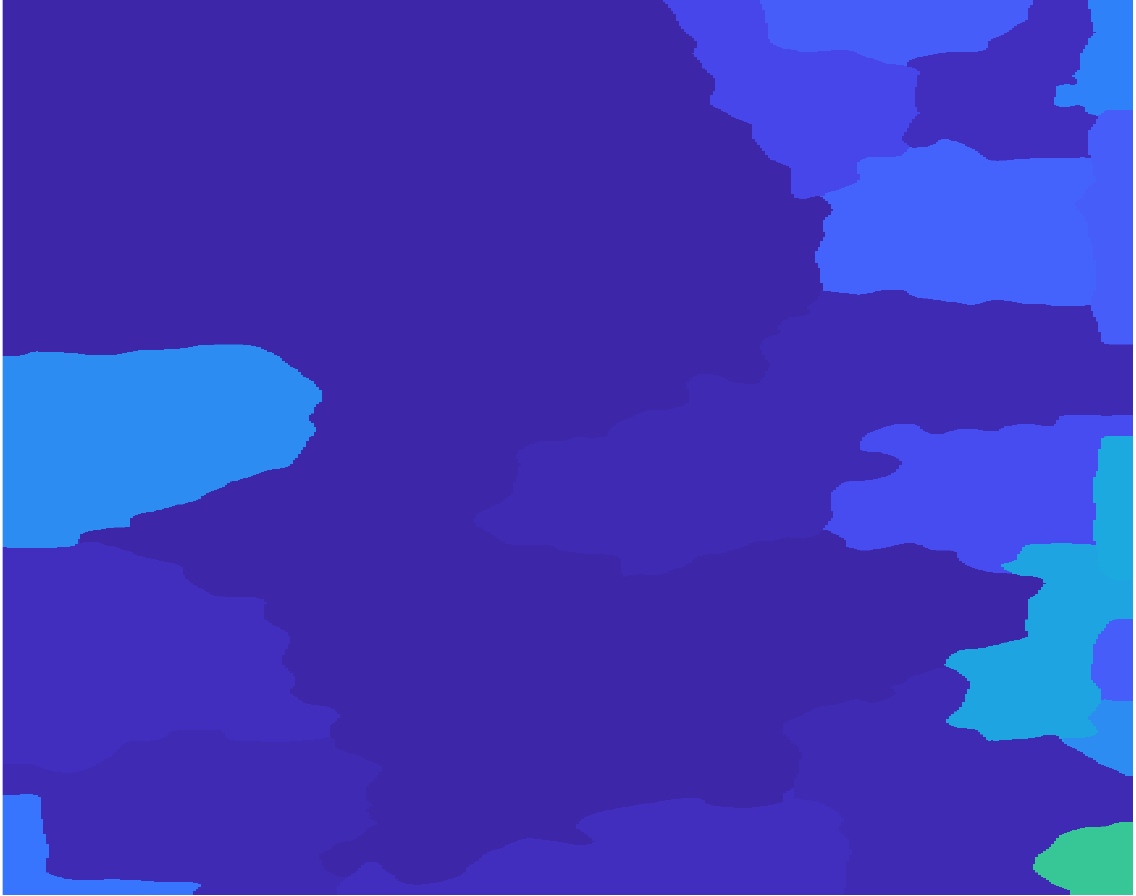}
			\caption{P$K$NN - Crater}
		\end{subfigure}
		\begin{subfigure}{.19\textwidth}
			\centering
			\includegraphics[width=\textwidth]{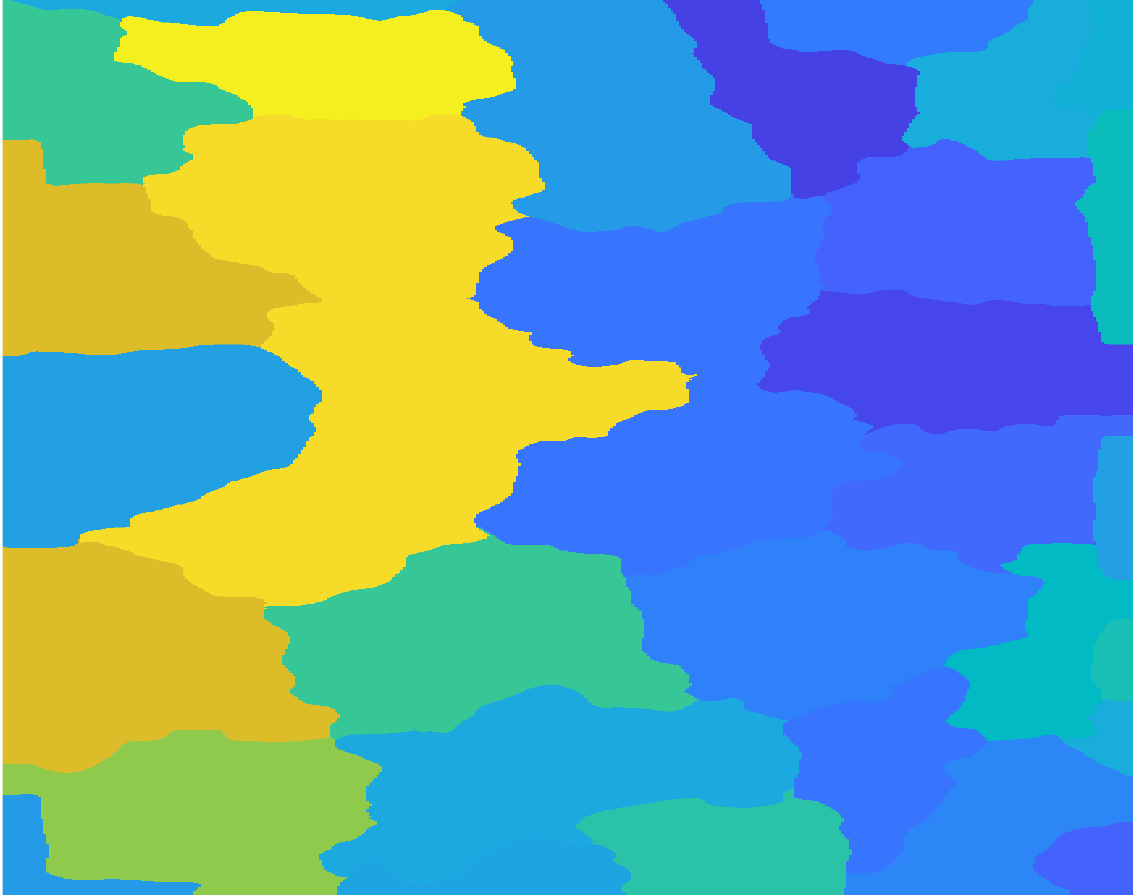}
			\caption{P$K$NN - Flat}
		\end{subfigure}
		\begin{subfigure}{.19\textwidth}
			\centering
			\includegraphics[width=\textwidth]{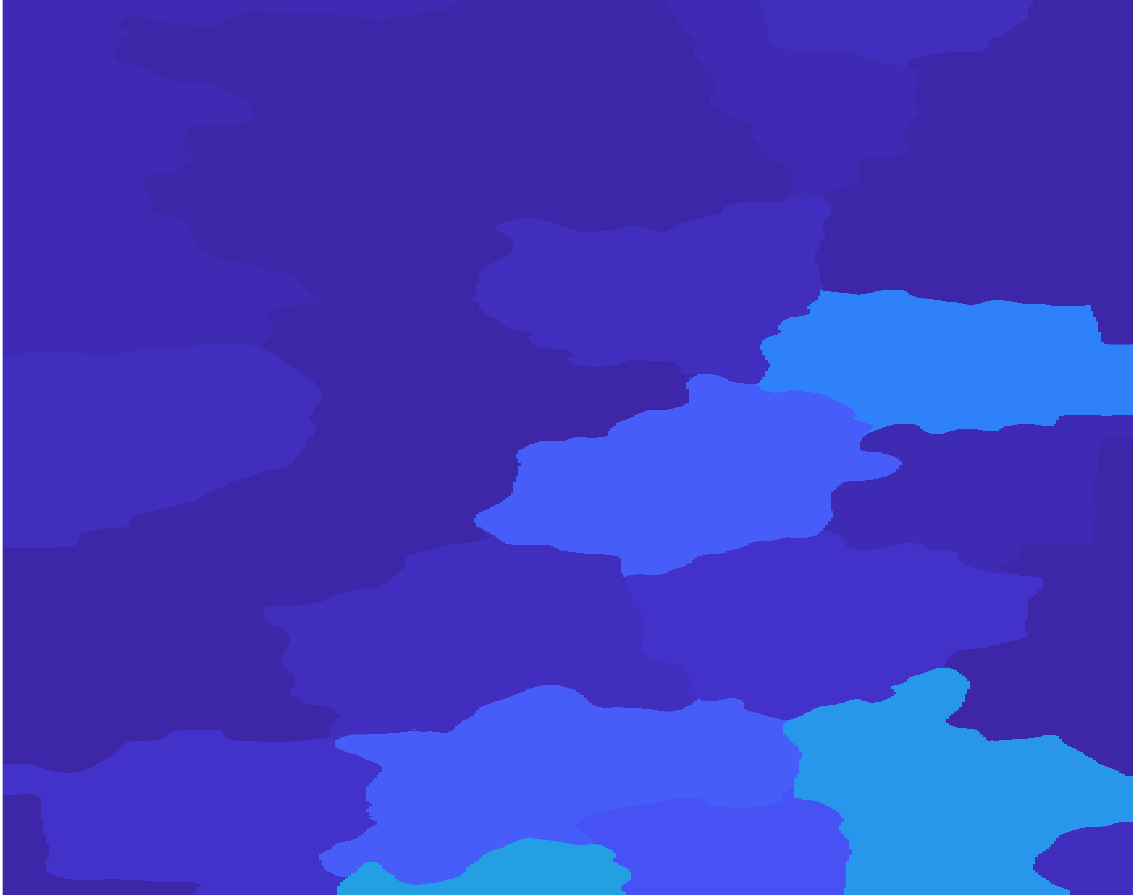}
			\caption{P$K$NN - Ripples}
		\end{subfigure}
		\begin{subfigure}{.19\textwidth}
			\centering
			\includegraphics[width=\textwidth]{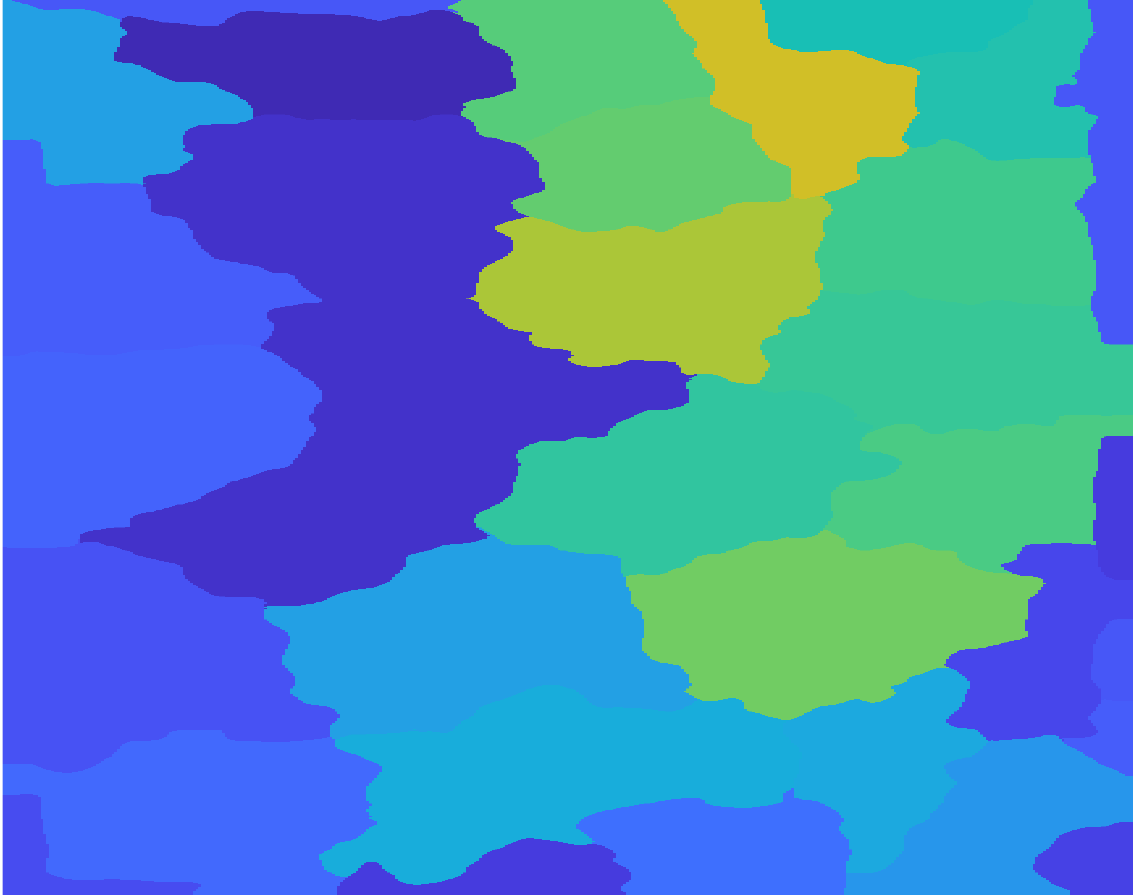}
			\caption{P$K$NN - Rocky}
		\end{subfigure}
		\caption{Example image two from fold one. The first column is actual image and superpixel segmentation. Of the remaining columns, the top row is PFLICM product maps and the bottom row is P$K$NN typicality maps.}
		\centering
		\label{fig:Img2}
	\end{figure}

\begin{figure}[htb]
		\centering
		\begin{subfigure}{.19\textwidth}
			\centering
			\includegraphics[width=\textwidth]{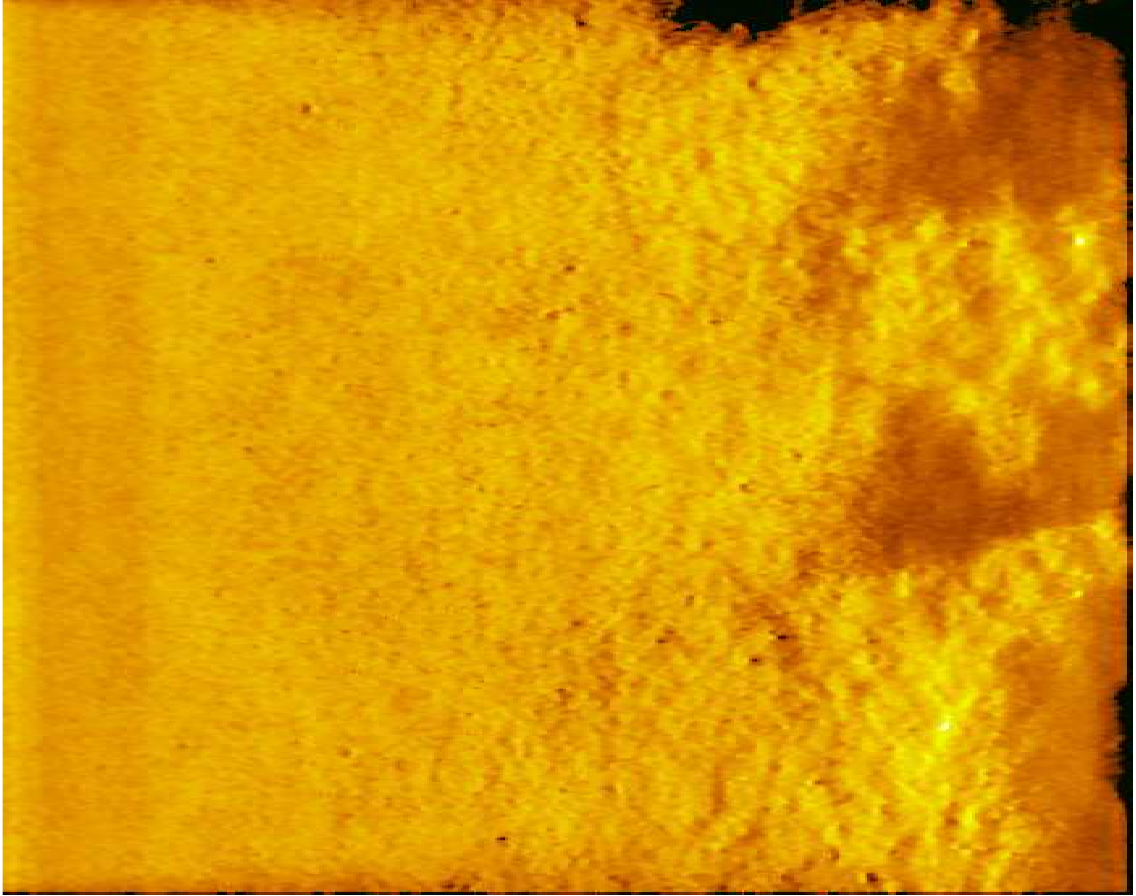}
			\caption{Original image}
		\end{subfigure}  
		\begin{subfigure}{.19\textwidth}
			\centering
			\includegraphics[width=\textwidth]{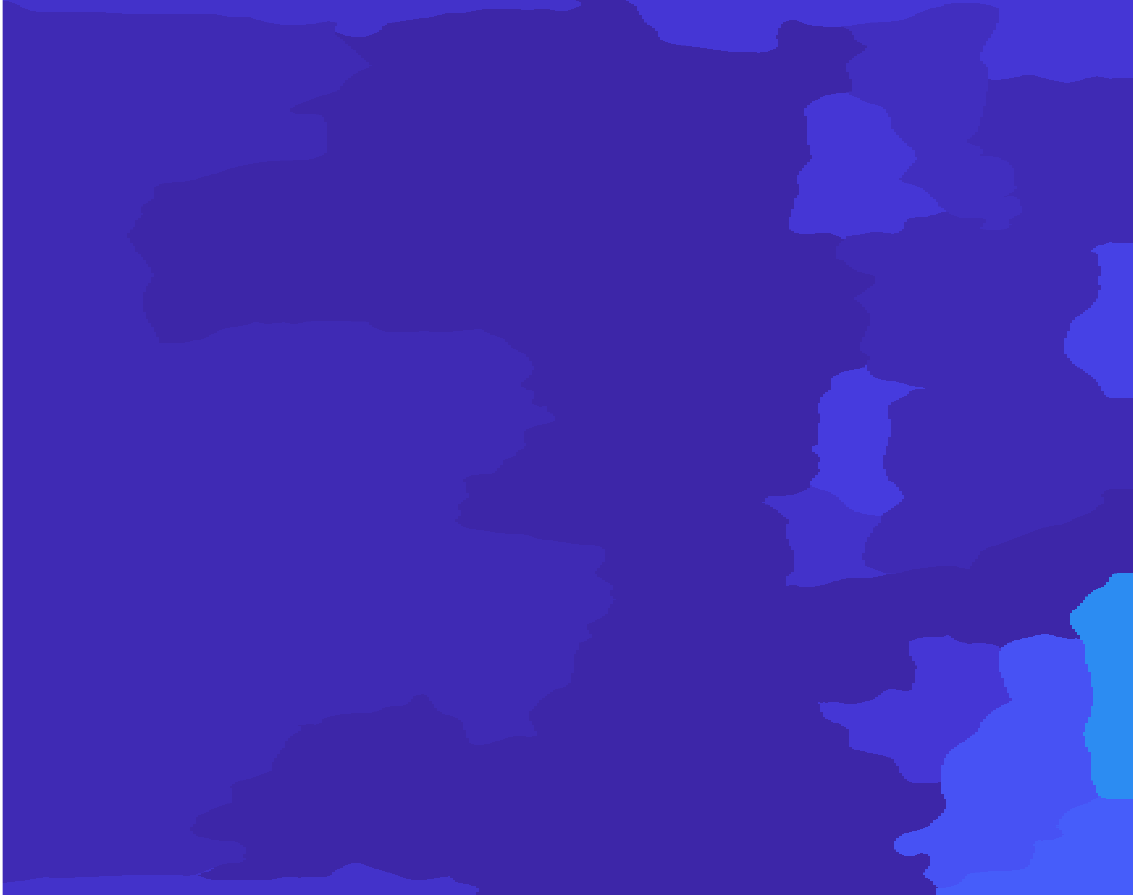}
			\caption{PFLICM - Crater}
		\end{subfigure}
		\begin{subfigure}{.19\textwidth}
			\centering
			\includegraphics[width=\textwidth]{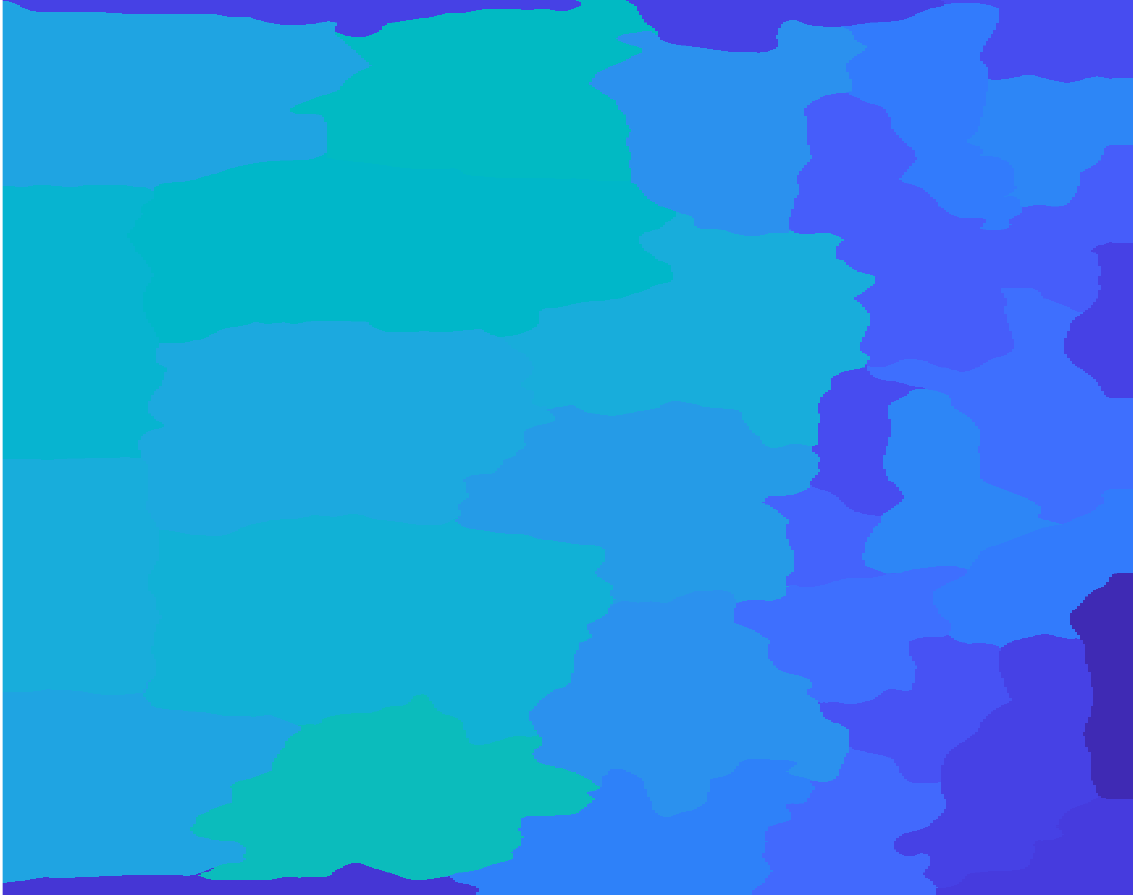}
			\caption{PFLICM - Flat}
		\end{subfigure}
		\begin{subfigure}{.19\textwidth}
			\centering
			\includegraphics[width=\textwidth]{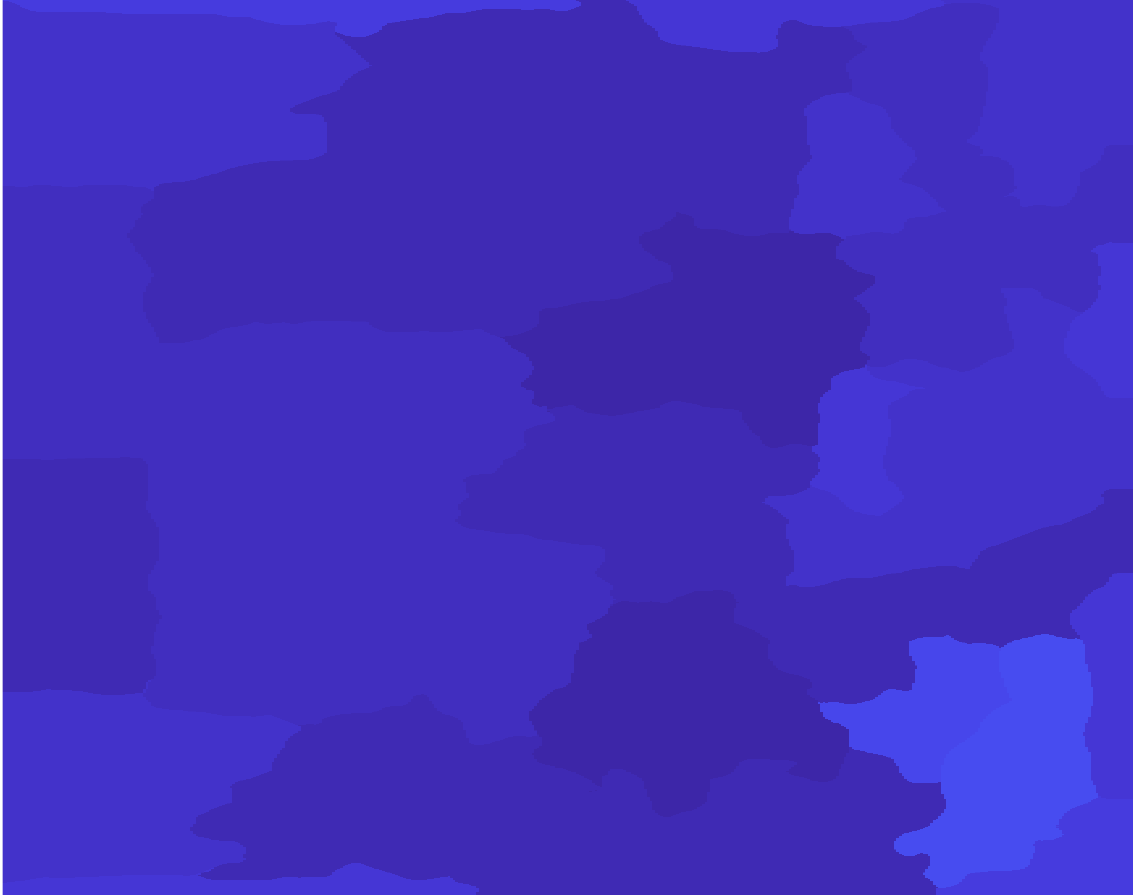}
			\caption{PFLICM - Ripples}
		\end{subfigure}
		\begin{subfigure}{.19\textwidth}
			\centering
			\includegraphics[width=\textwidth]{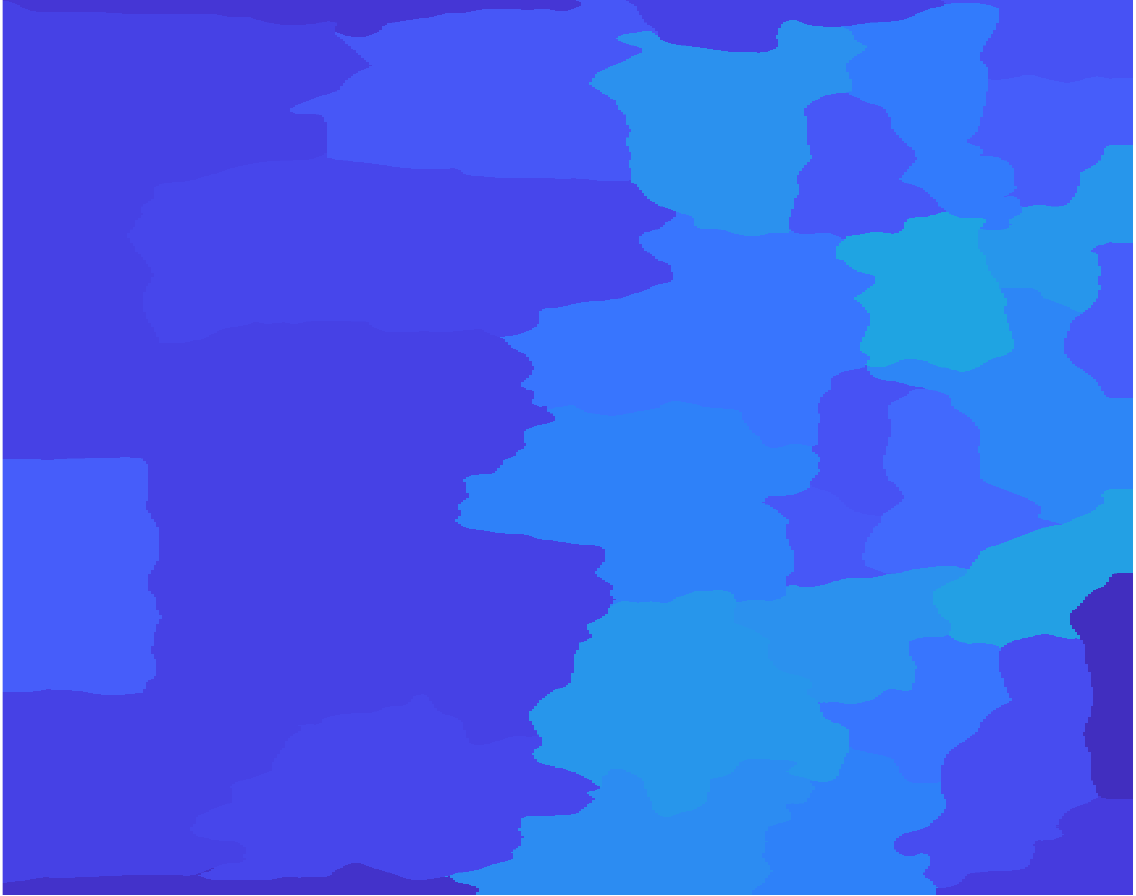}
			\caption{PFLICM - Rocky}
		\end{subfigure}
		
		\begin{subfigure}{.19\textwidth}
			\centering
			\includegraphics[width=\textwidth]{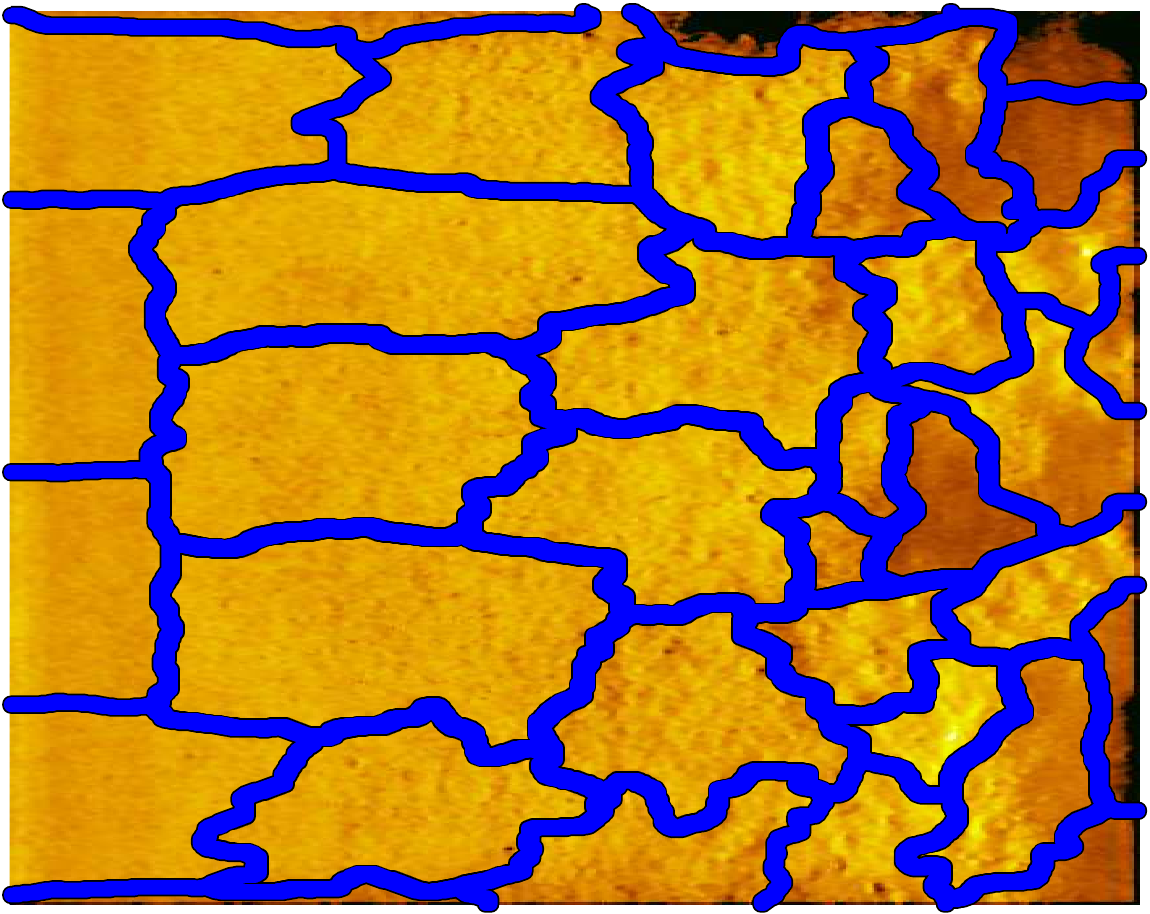}
			\caption{Superpixels}
		\end{subfigure}
		\begin{subfigure}{.19\textwidth}
			\centering
			\includegraphics[width=\textwidth]{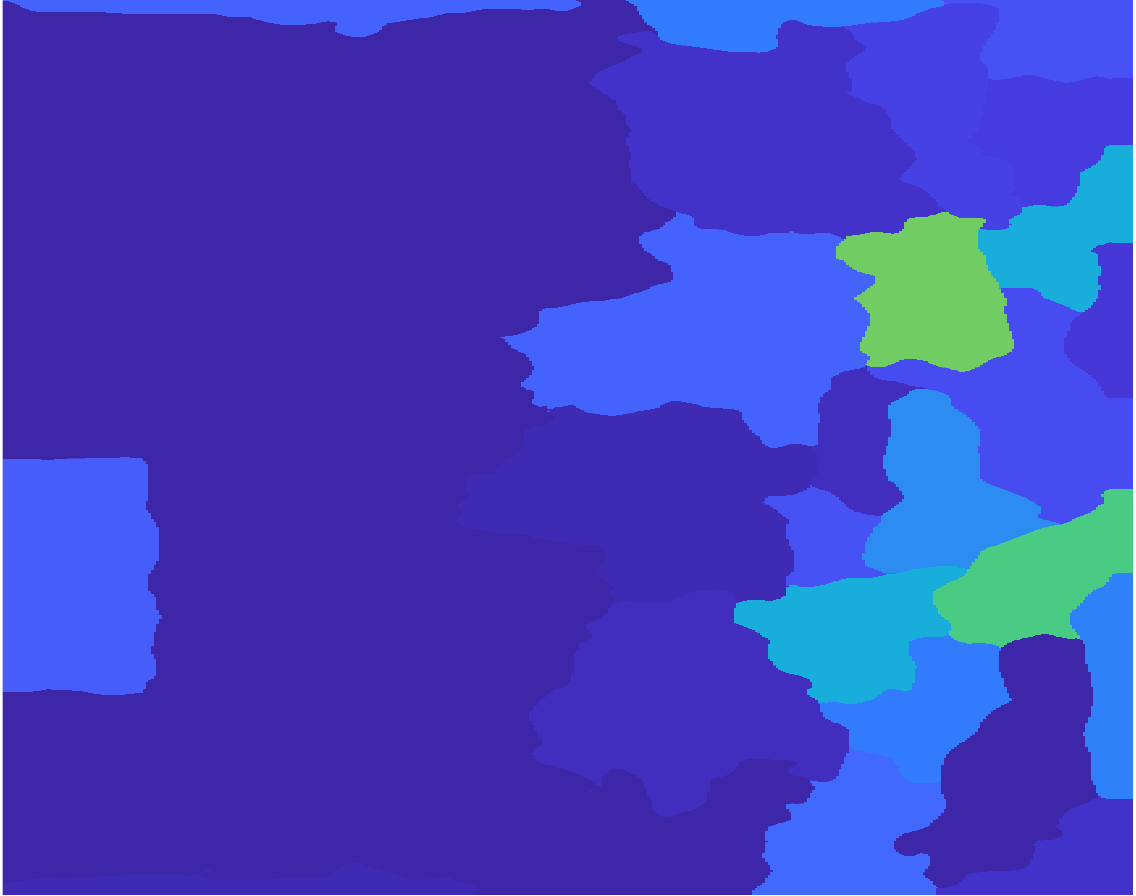}
			\caption{P$K$NN - Crater}
		\end{subfigure}
		\begin{subfigure}{.19\textwidth}
			\centering
			\includegraphics[width=\textwidth]{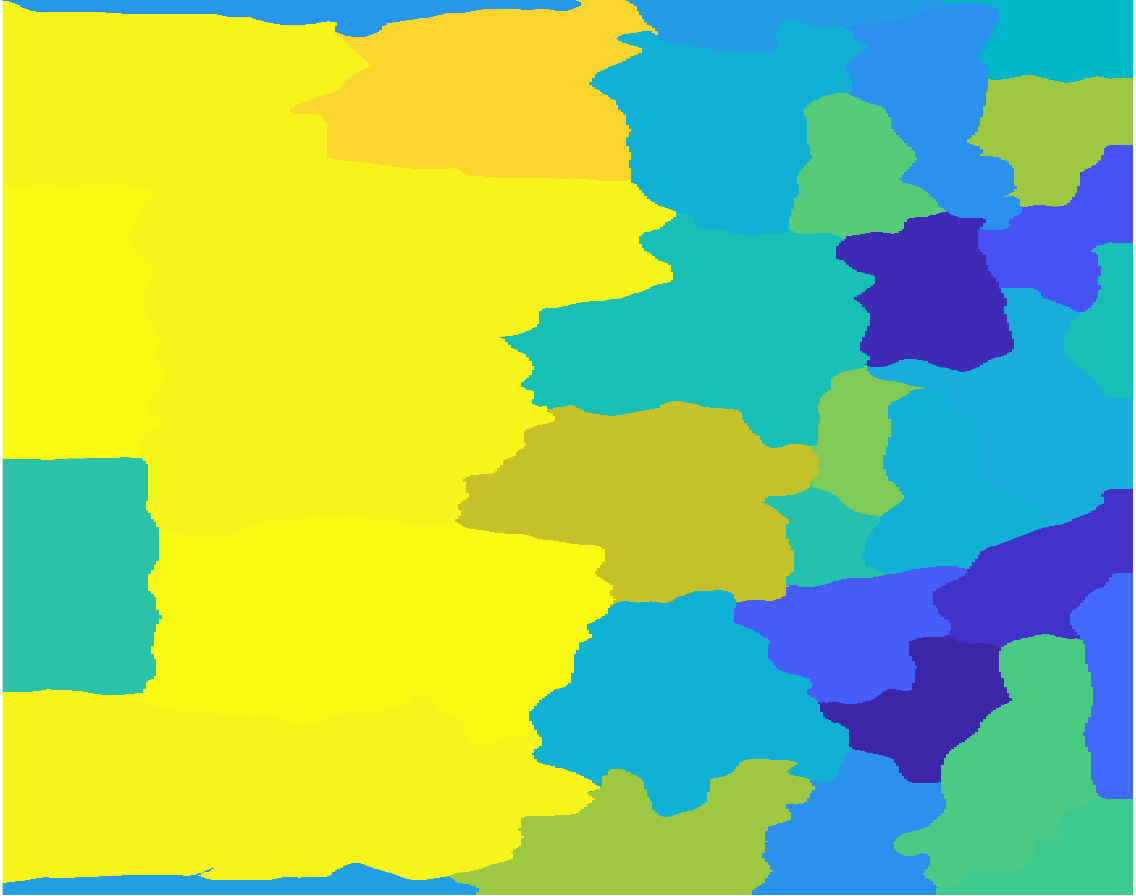}
			\caption{P$K$NN - Flat}
		\end{subfigure}
		\begin{subfigure}{.19\textwidth}
			\centering
			\includegraphics[width=\textwidth]{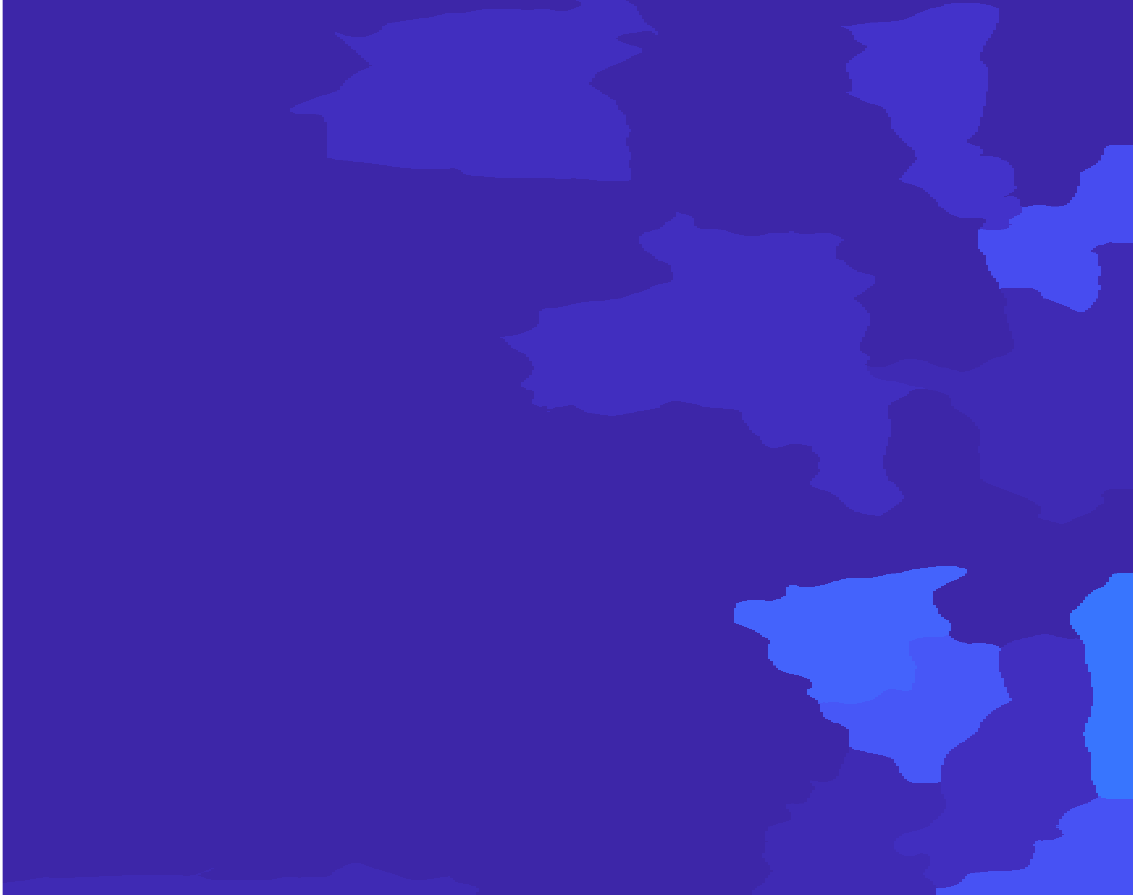}
			\caption{P$K$NN - Ripples}
		\end{subfigure}
		\begin{subfigure}{.19\textwidth}
			\centering
			\includegraphics[width=\textwidth]{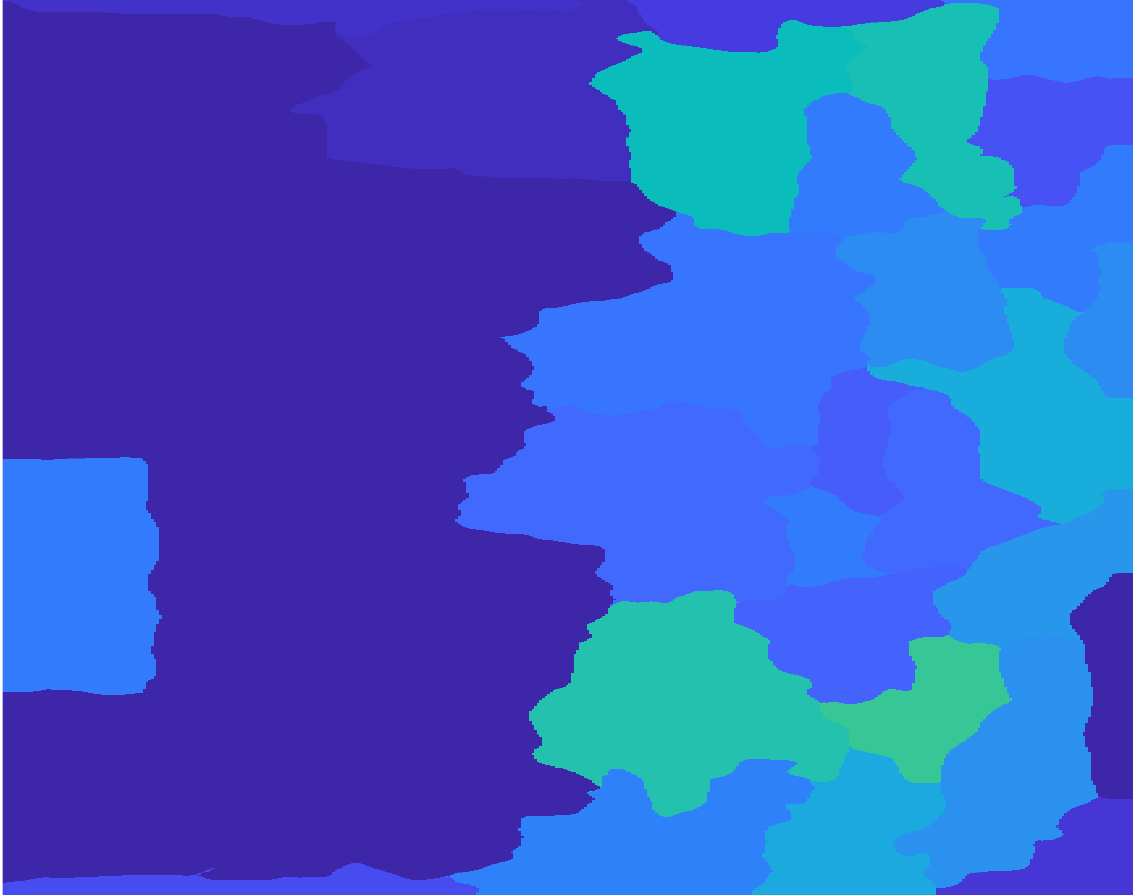}
			\caption{P$K$NN - Rocky}
		\end{subfigure}
		\caption{Example image one from fold two. The first column is actual image and superpixel segmentation. Of the remaining columns, the top row is PFLICM product maps and the bottom row is P$K$NN typicality maps.}
		\centering
		\label{fig:Img3}
	\end{figure}

\begin{figure}[htb]
		\centering
		\begin{subfigure}{.19\textwidth}
			\centering
			\includegraphics[width=\textwidth]{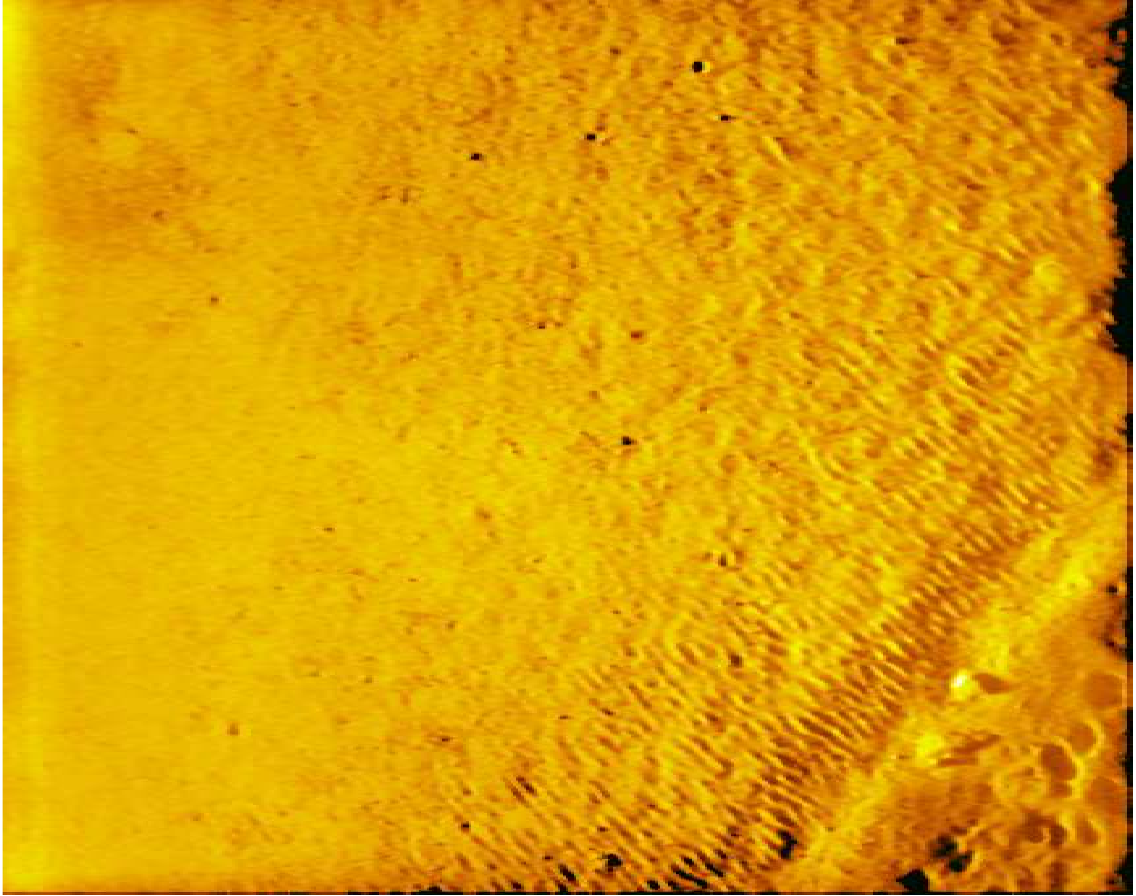}
			\caption{Original image}
		\end{subfigure}  
		\begin{subfigure}{.19\textwidth}
			\centering
			\includegraphics[width=\textwidth]{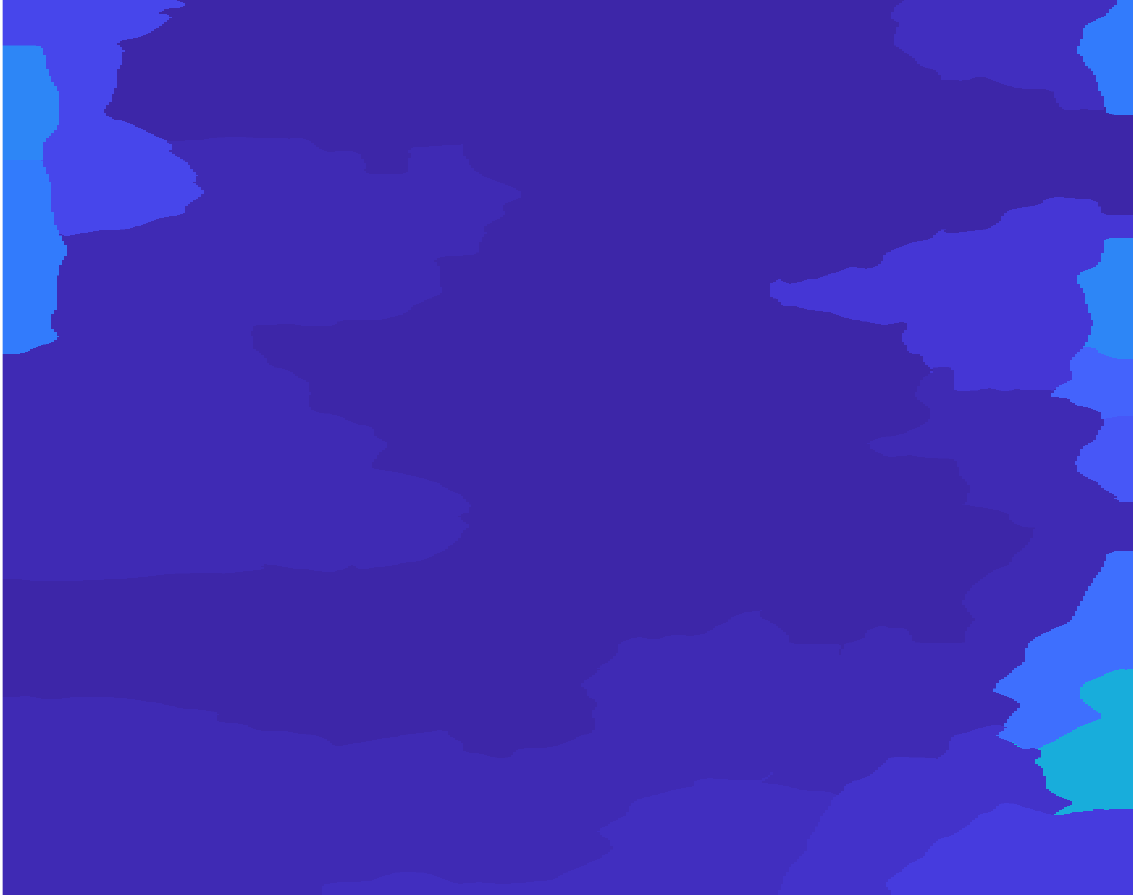}
			\caption{PFLICM - Crater}
		\end{subfigure}
		\begin{subfigure}{.19\textwidth}
			\centering
			\includegraphics[width=\textwidth]{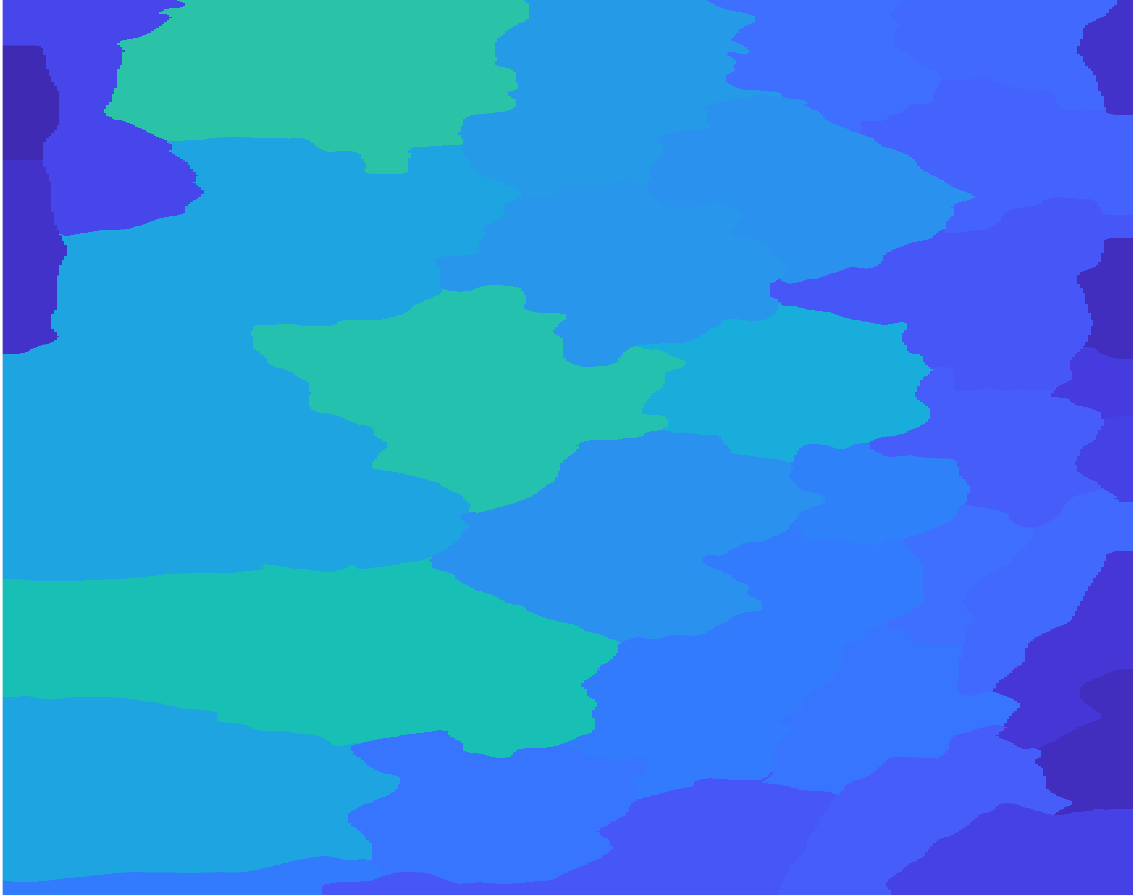}
			\caption{PFLICM - Flat}
		\end{subfigure}
		\begin{subfigure}{.19\textwidth}
			\centering
			\includegraphics[width=\textwidth]{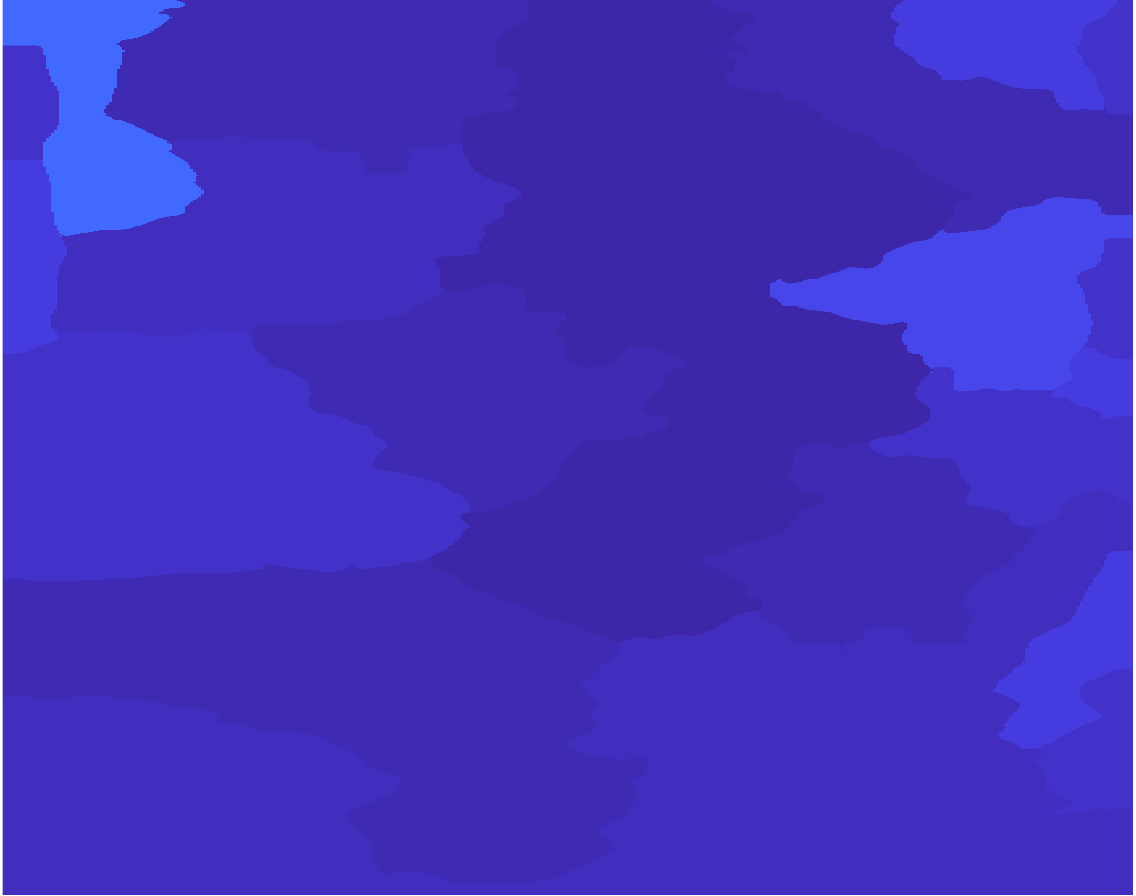}
			\caption{PFLICM - Ripples}
		\end{subfigure}
		\begin{subfigure}{.19\textwidth}
			\centering
			\includegraphics[width=\textwidth]{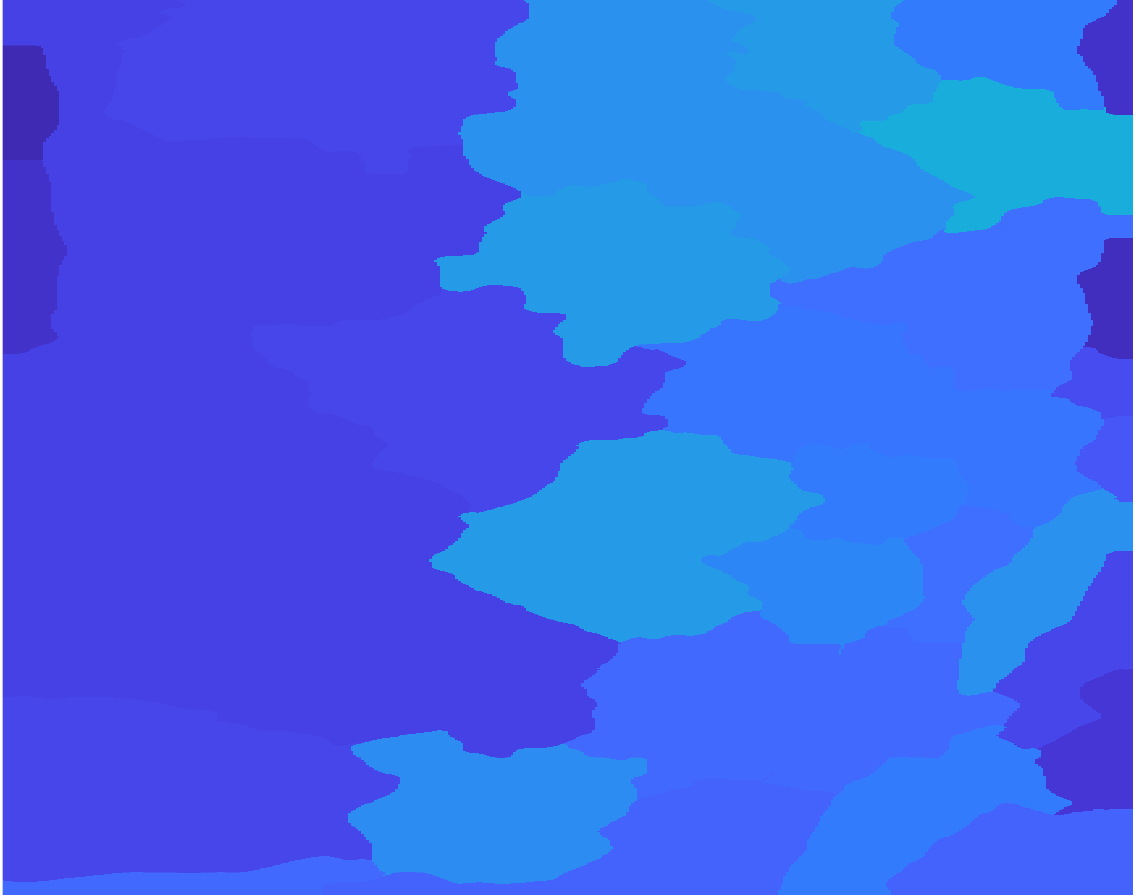}
			\caption{PFLICM - Rocky}
		\end{subfigure}
		
		\begin{subfigure}{.19\textwidth}
			\centering
			\includegraphics[width=\textwidth]{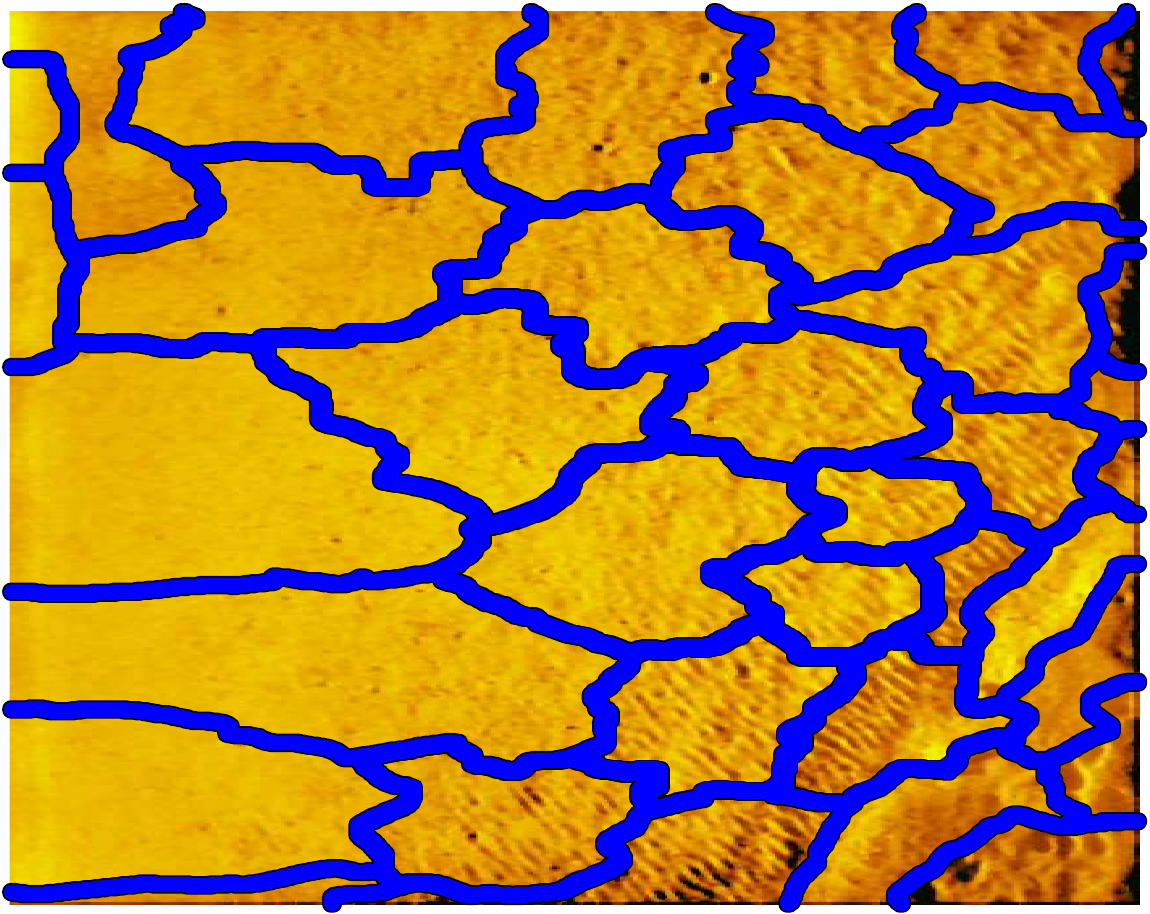}
			\caption{Superpixels}
		\end{subfigure}
		\begin{subfigure}{.19\textwidth}
			\centering
			\includegraphics[width=\textwidth]{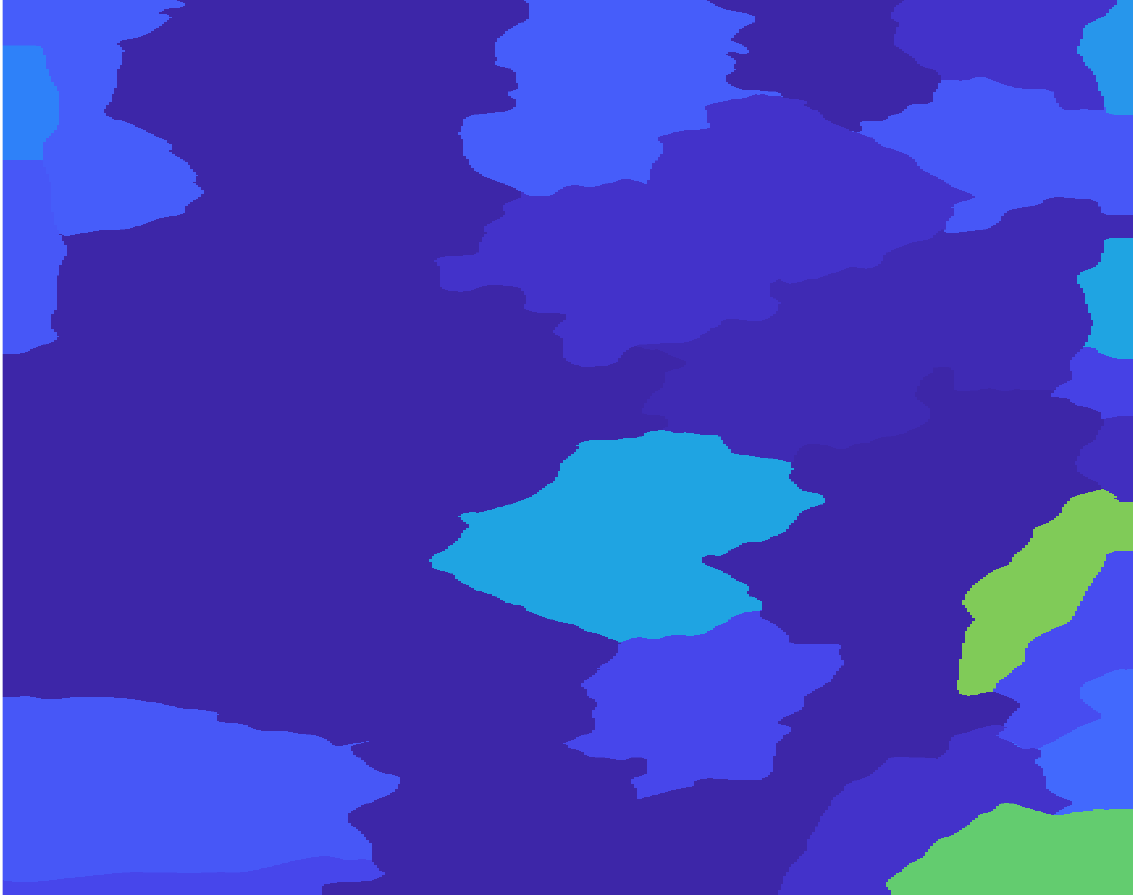}
			\caption{P$K$NN - Crater}
		\end{subfigure}
		\begin{subfigure}{.19\textwidth}
			\centering
			\includegraphics[width=\textwidth]{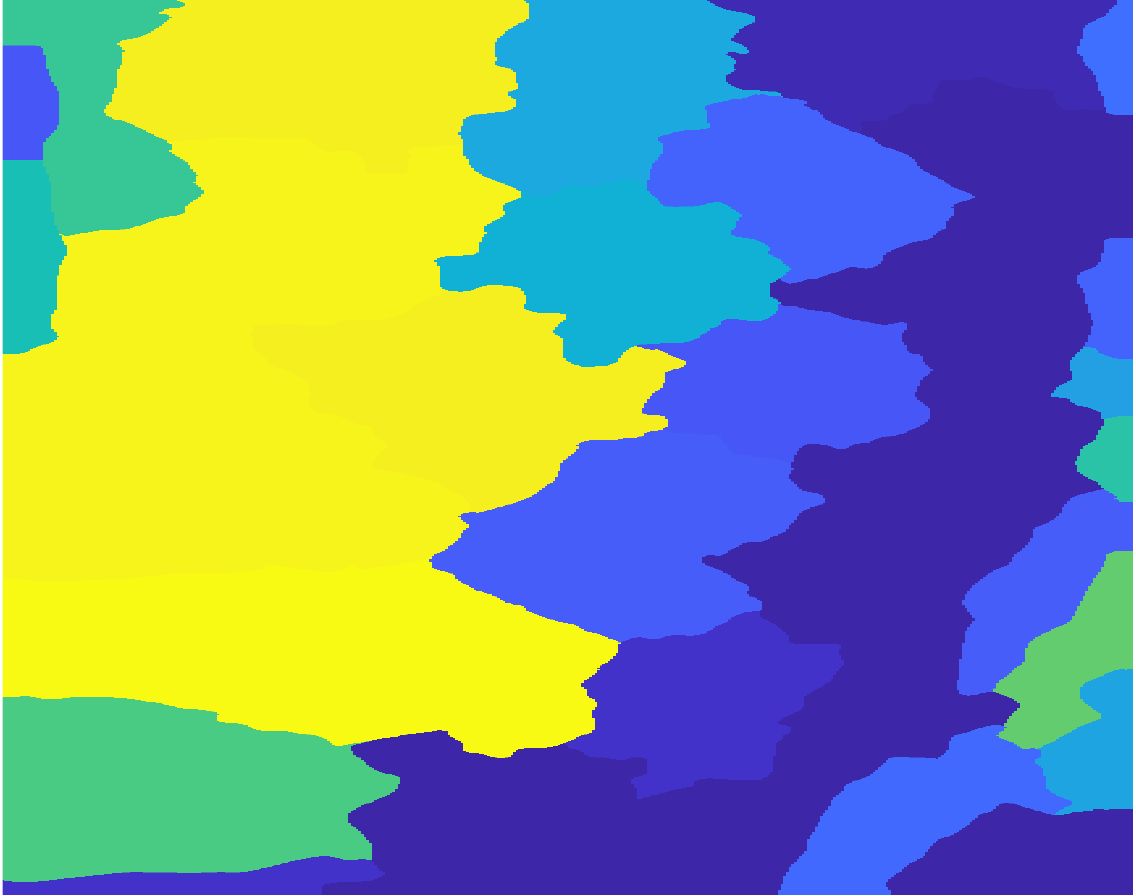}
			\caption{P$K$NN - Flat}
		\end{subfigure}
		\begin{subfigure}{.19\textwidth}
			\centering
			\includegraphics[width=\textwidth]{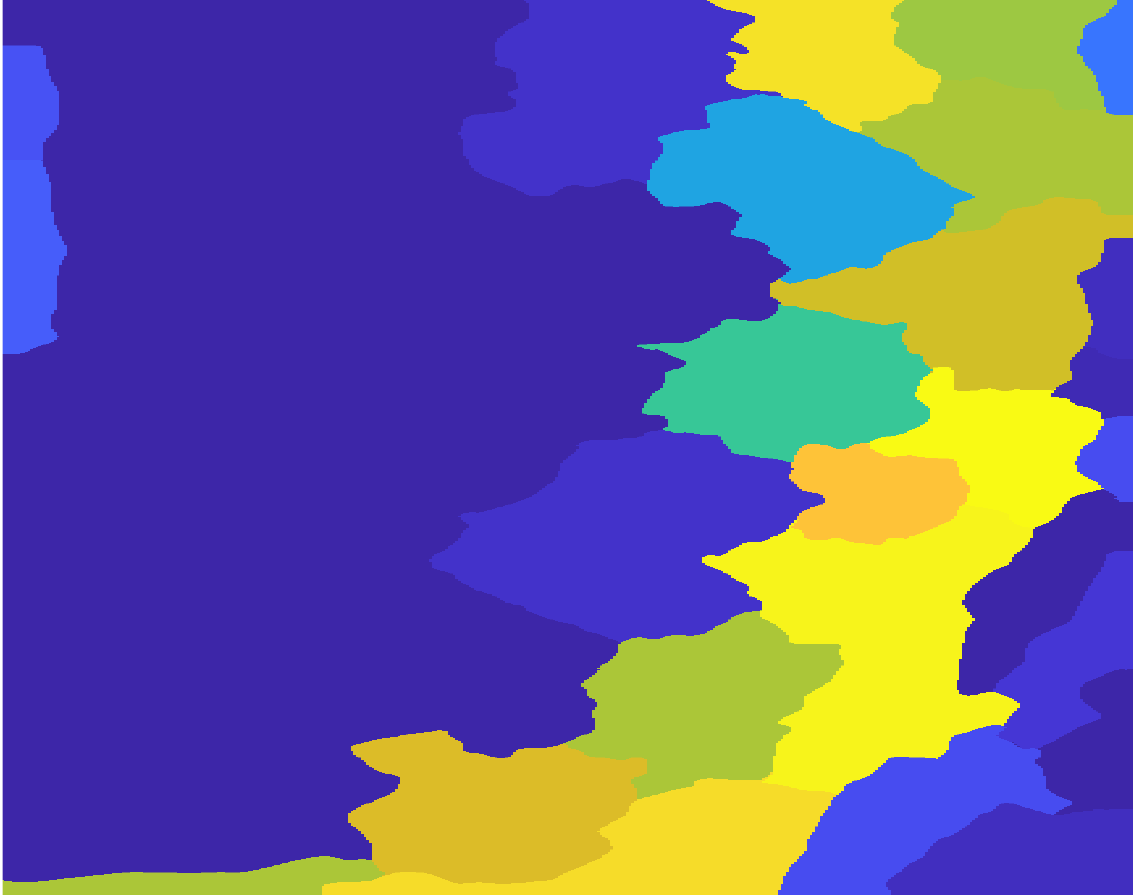}
			\caption{P$K$NN - Ripples}
		\end{subfigure}
		\begin{subfigure}{.19\textwidth}
			\centering
			\includegraphics[width=\textwidth]{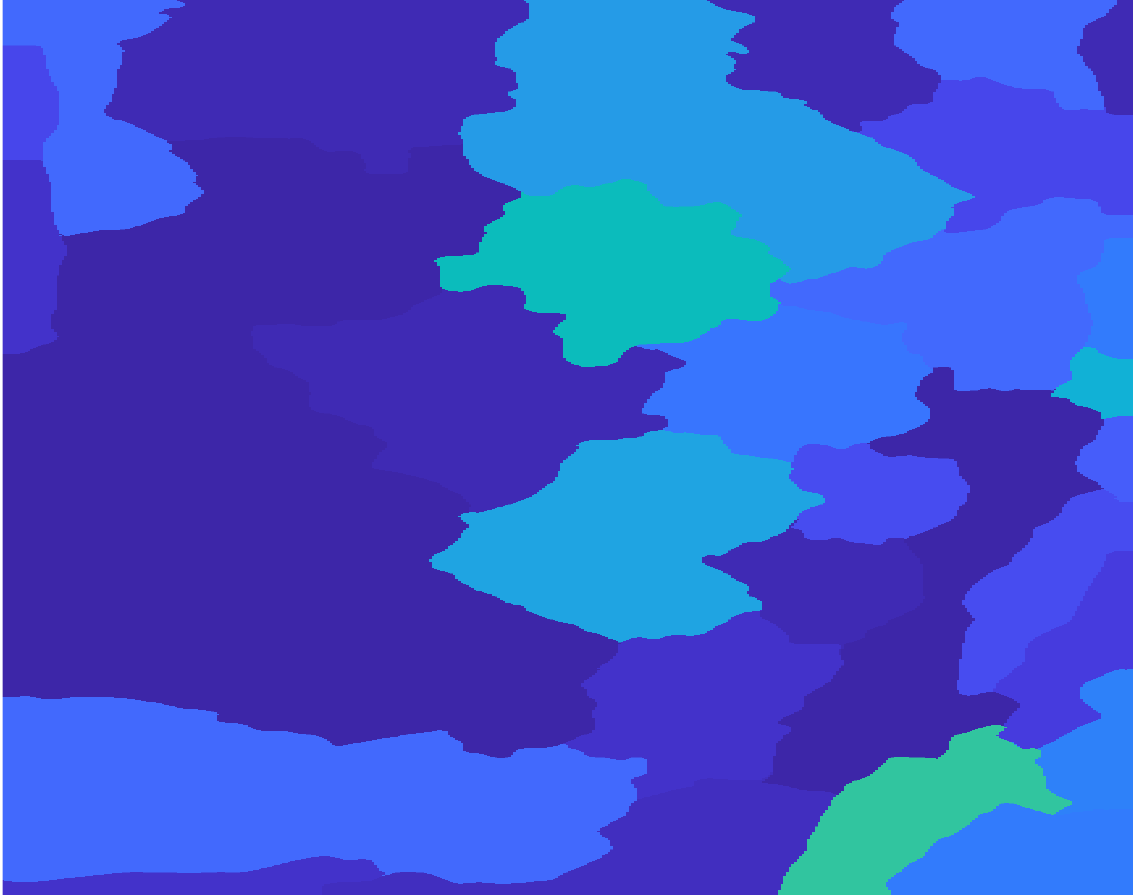}
			\caption{P$K$NN - Rocky}
		\end{subfigure}
		\caption{Example image two from fold two. The first column is actual image and superpixel segmentation. Of the remaining columns, the top row is PFLICM product maps and the bottom row is P$K$NN typicality maps.}
		\centering
		\label{fig:Img4}
	\end{figure}

\begin{figure}[htb]
		\centering
		\begin{subfigure}{.19\textwidth}
			\centering
			\includegraphics[width=\textwidth]{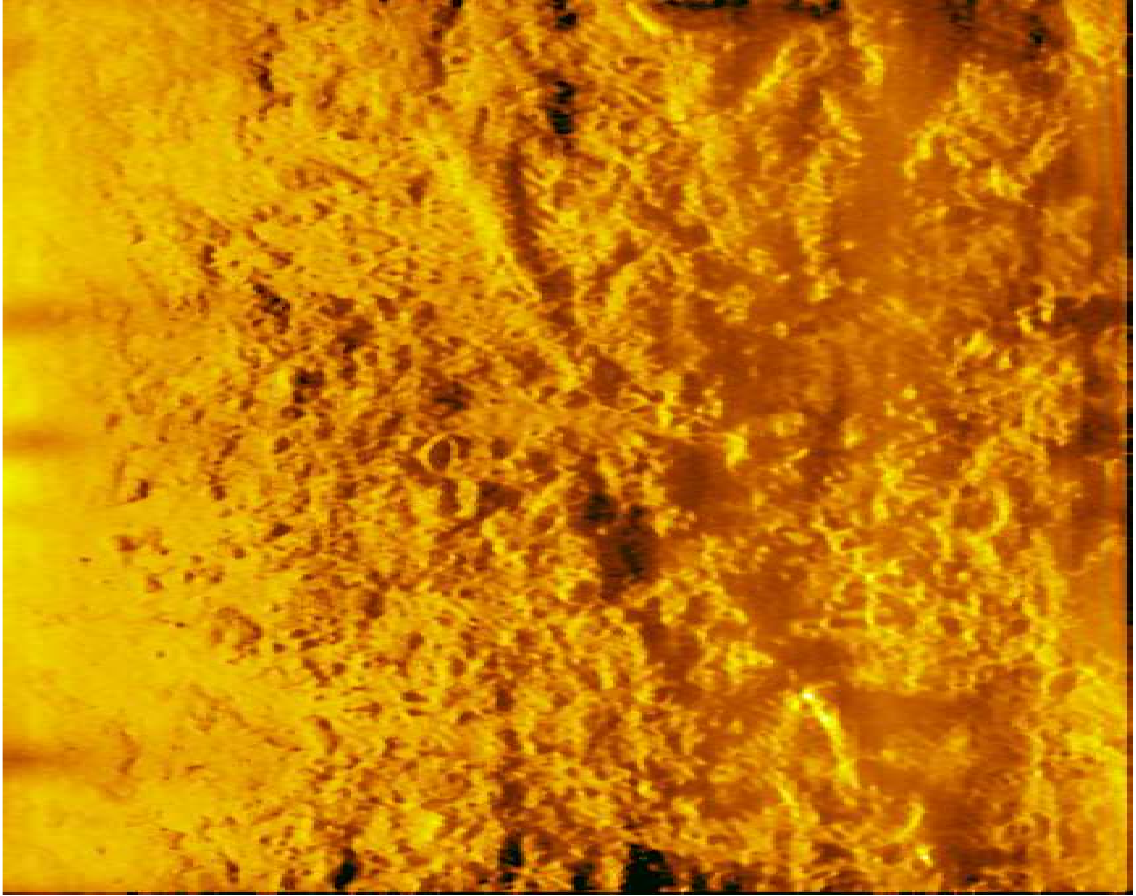}
			\caption{Original image}
		\end{subfigure}  
		\begin{subfigure}{.19\textwidth}
			\centering
			\includegraphics[width=\textwidth]{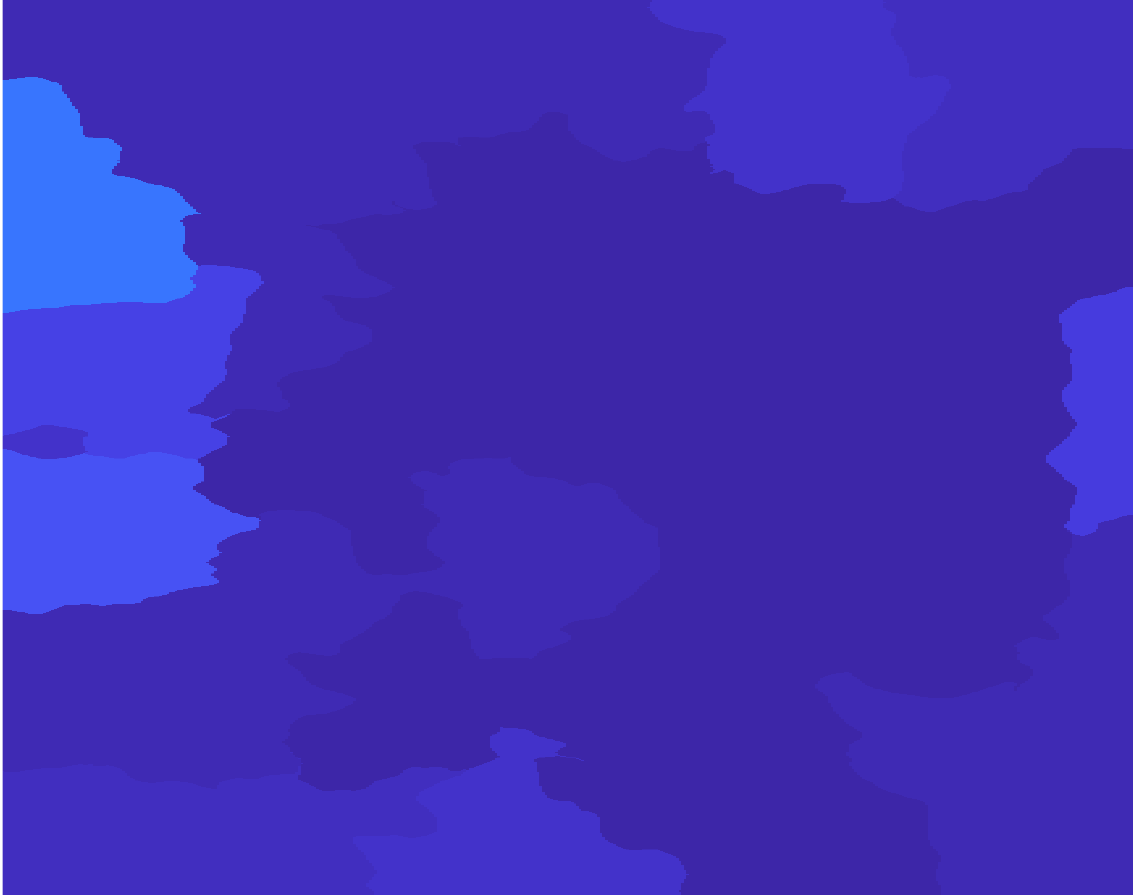}
			\caption{PFLICM - Crater}
		\end{subfigure}
		\begin{subfigure}{.19\textwidth}
			\centering
			\includegraphics[width=\textwidth]{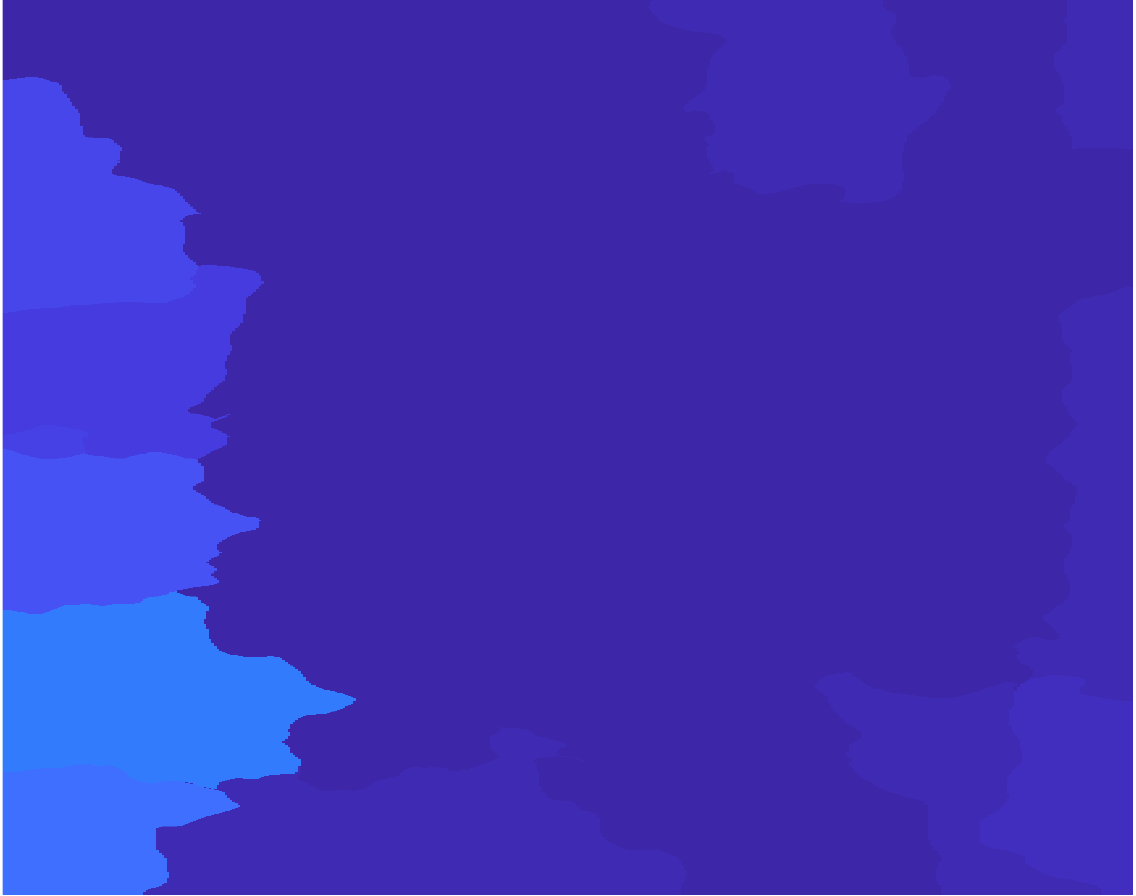}
			\caption{PFLICM - Flat}
		\end{subfigure}
		\begin{subfigure}{.19\textwidth}
			\centering
			\includegraphics[width=\textwidth]{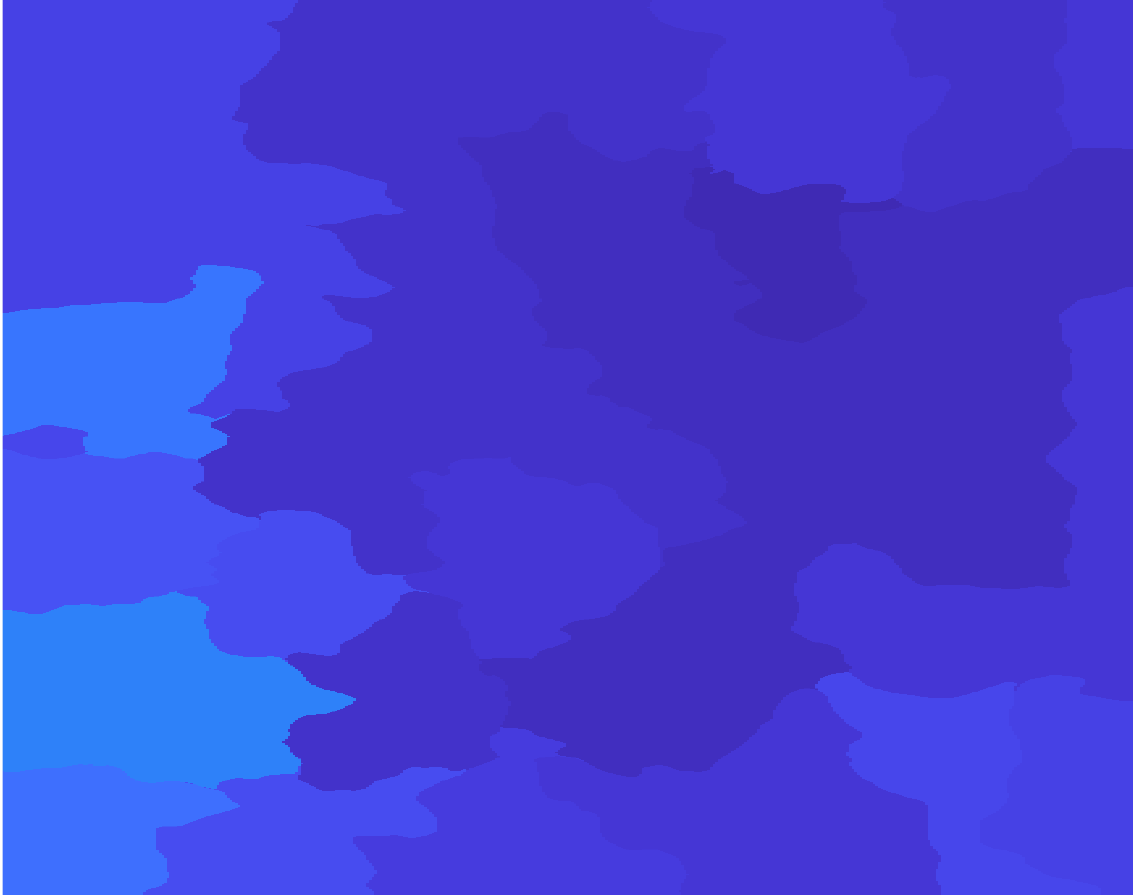}
			\caption{PFLICM - Ripples}
		\end{subfigure}
		\begin{subfigure}{.19\textwidth}
			\centering
			\includegraphics[width=\textwidth]{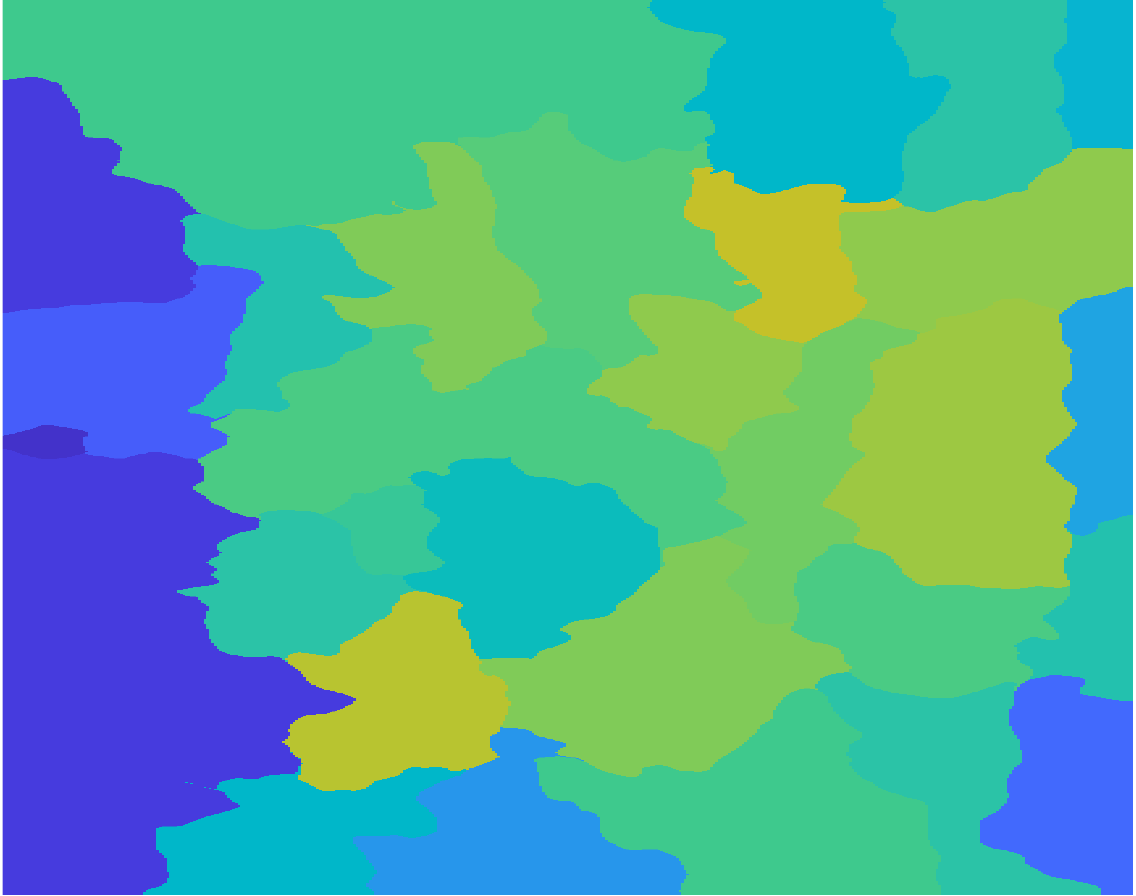}
			\caption{PFLICM - Rocky}
		\end{subfigure}
		
		\begin{subfigure}{.19\textwidth}
			\centering
			\includegraphics[width=\textwidth]{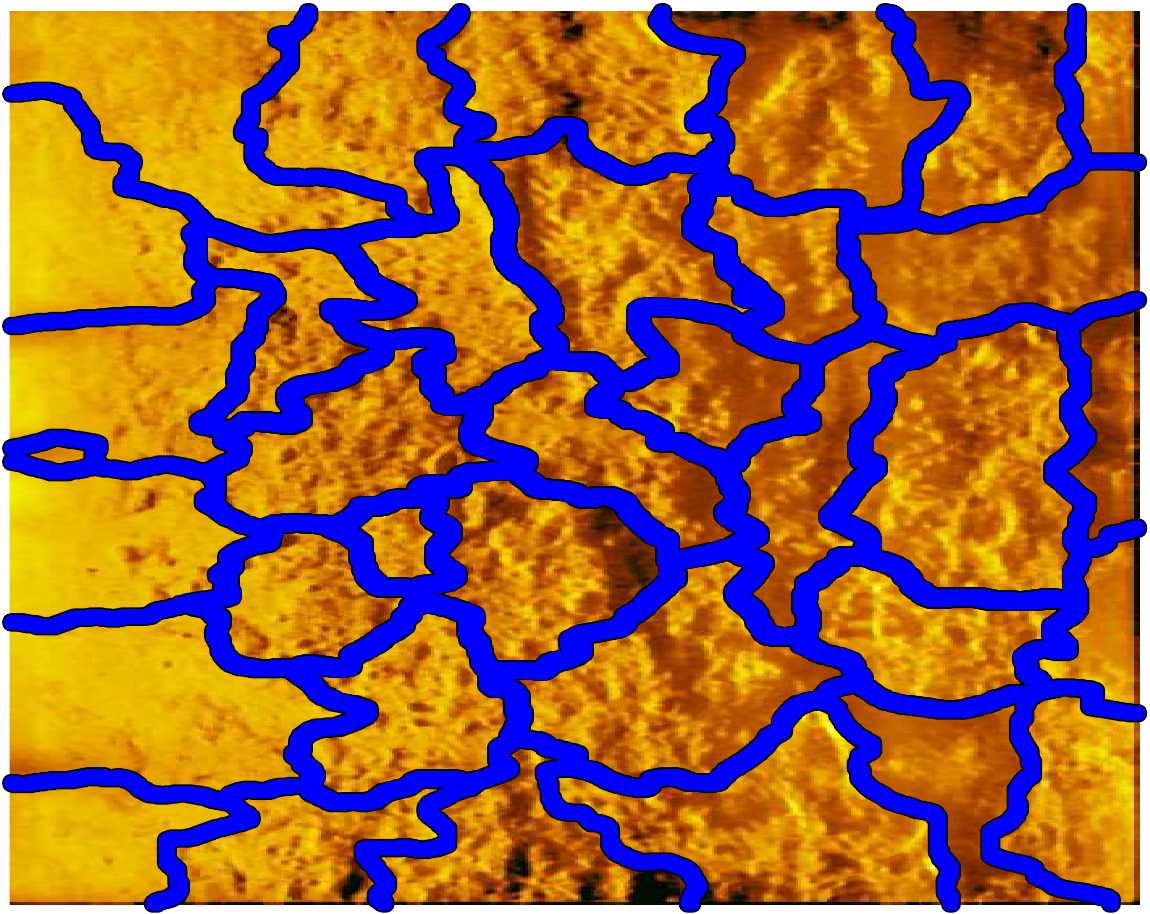}
			\caption{Superpixels}
		\end{subfigure}
		\begin{subfigure}{.19\textwidth}
			\centering
			\includegraphics[width=\textwidth]{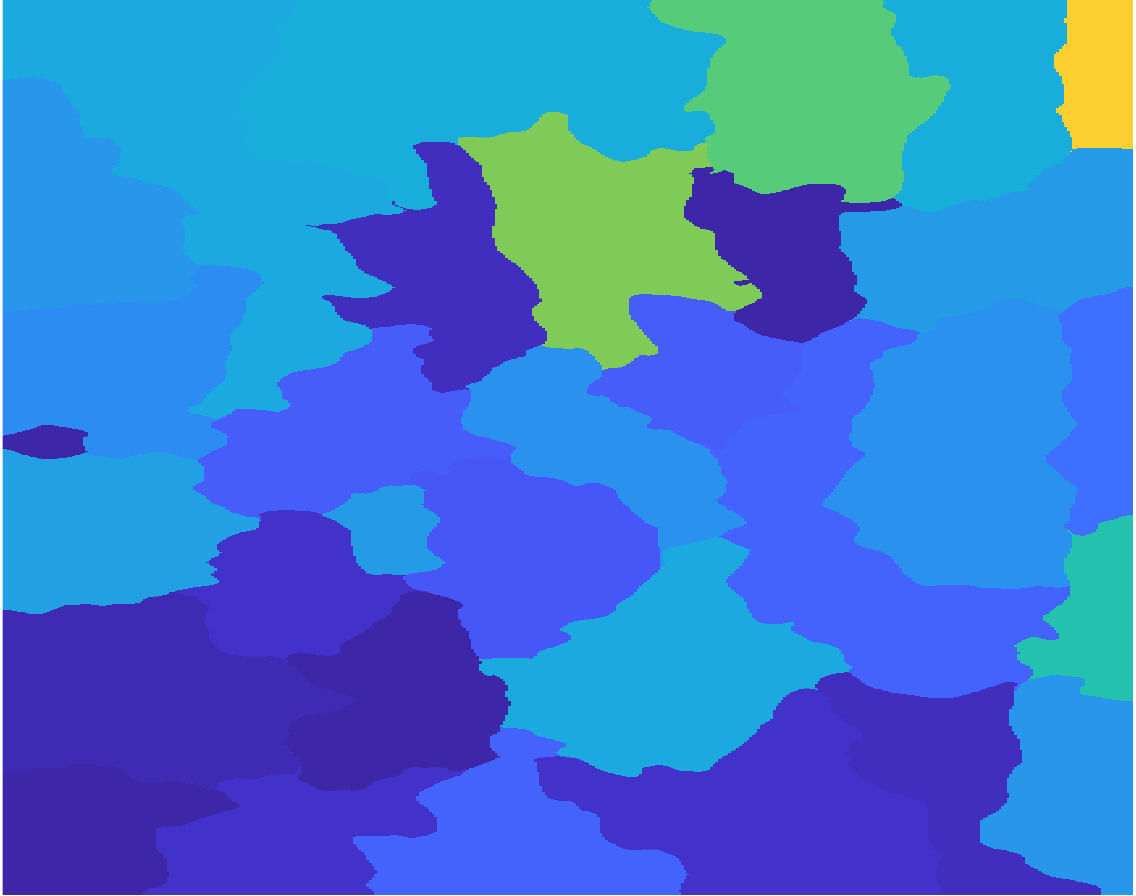}
			\caption{P$K$NN - Crater}
		\end{subfigure}
		\begin{subfigure}{.19\textwidth}
			\centering
			\includegraphics[width=\textwidth]{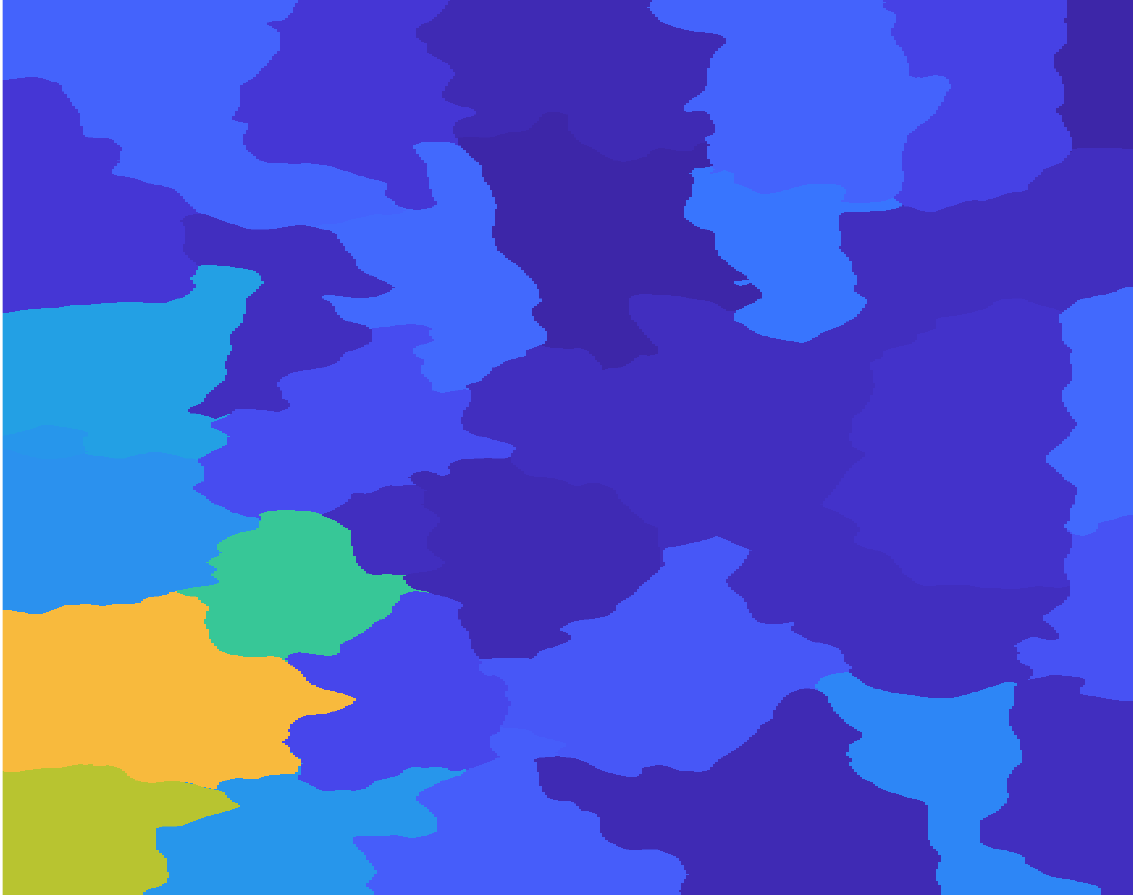}
			\caption{P$K$NN - Flat}
		\end{subfigure}
		\begin{subfigure}{.19\textwidth}
			\centering
			\includegraphics[width=\textwidth]{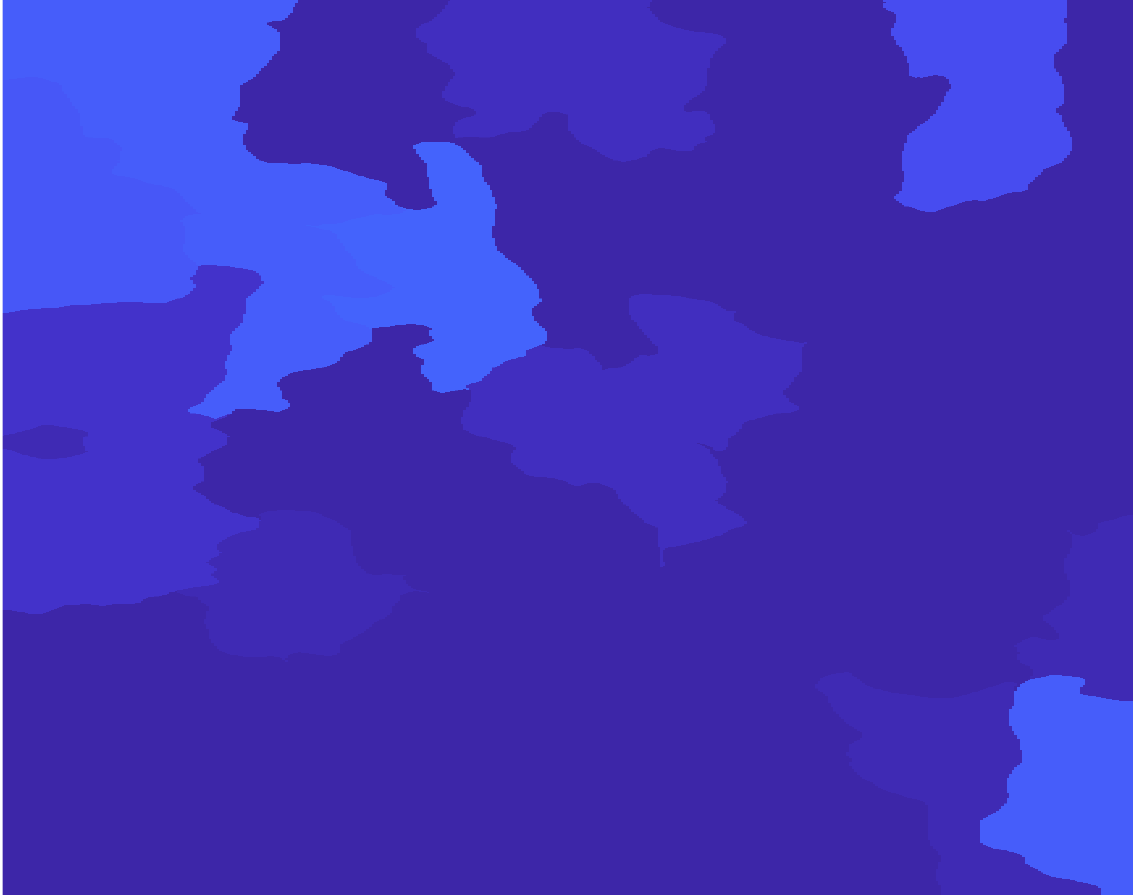}
			\caption{P$K$NN - Ripples}
		\end{subfigure}
		\begin{subfigure}{.19\textwidth}
			\centering
			\includegraphics[width=\textwidth]{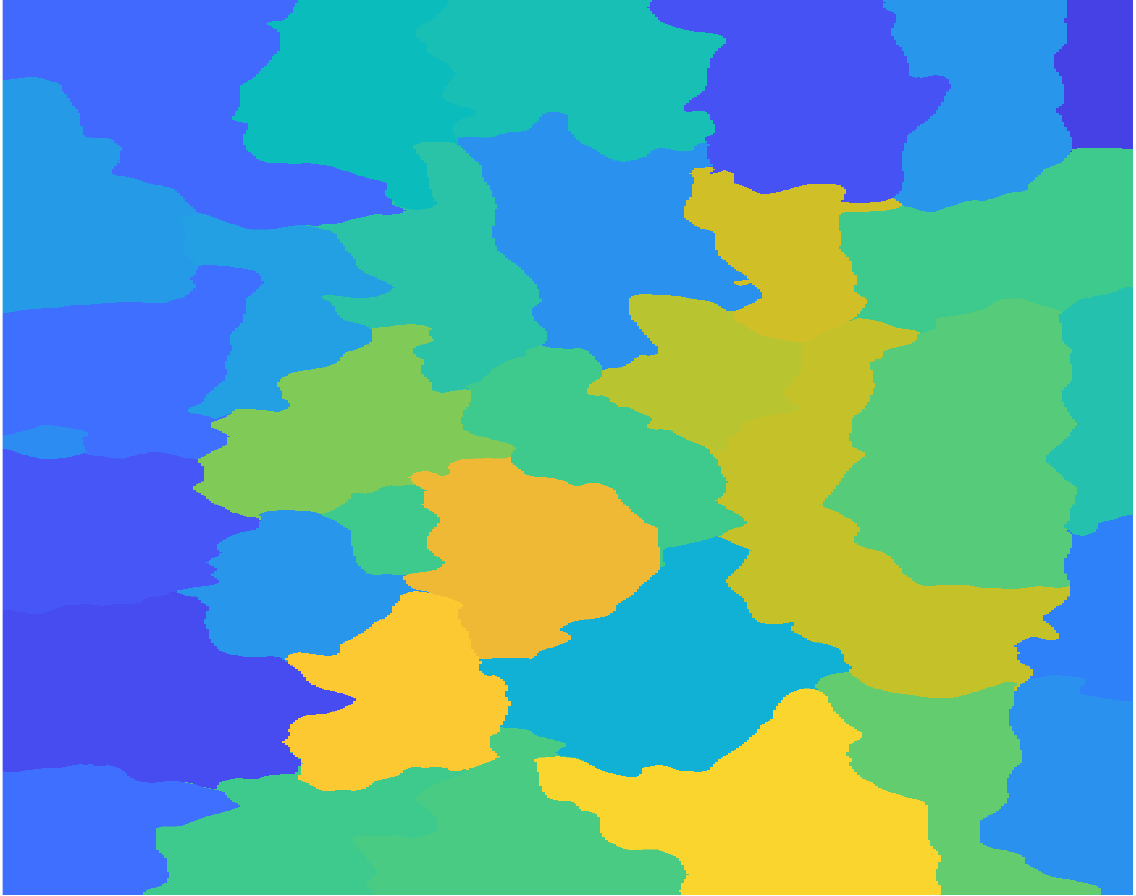}
			\caption{P$K$NN - Rocky}
		\end{subfigure}
		\caption{Example image one from fold three. The first column is actual image and superpixel segmentation. Of the remaining columns, the top row is PFLICM product maps and the bottom row is P$K$NN typicality maps.}
		\centering
		\label{fig:Img5}
	\end{figure}

\begin{figure}[htb]
		\centering
		\begin{subfigure}{.19\textwidth}
			\centering
			\includegraphics[width=\textwidth]{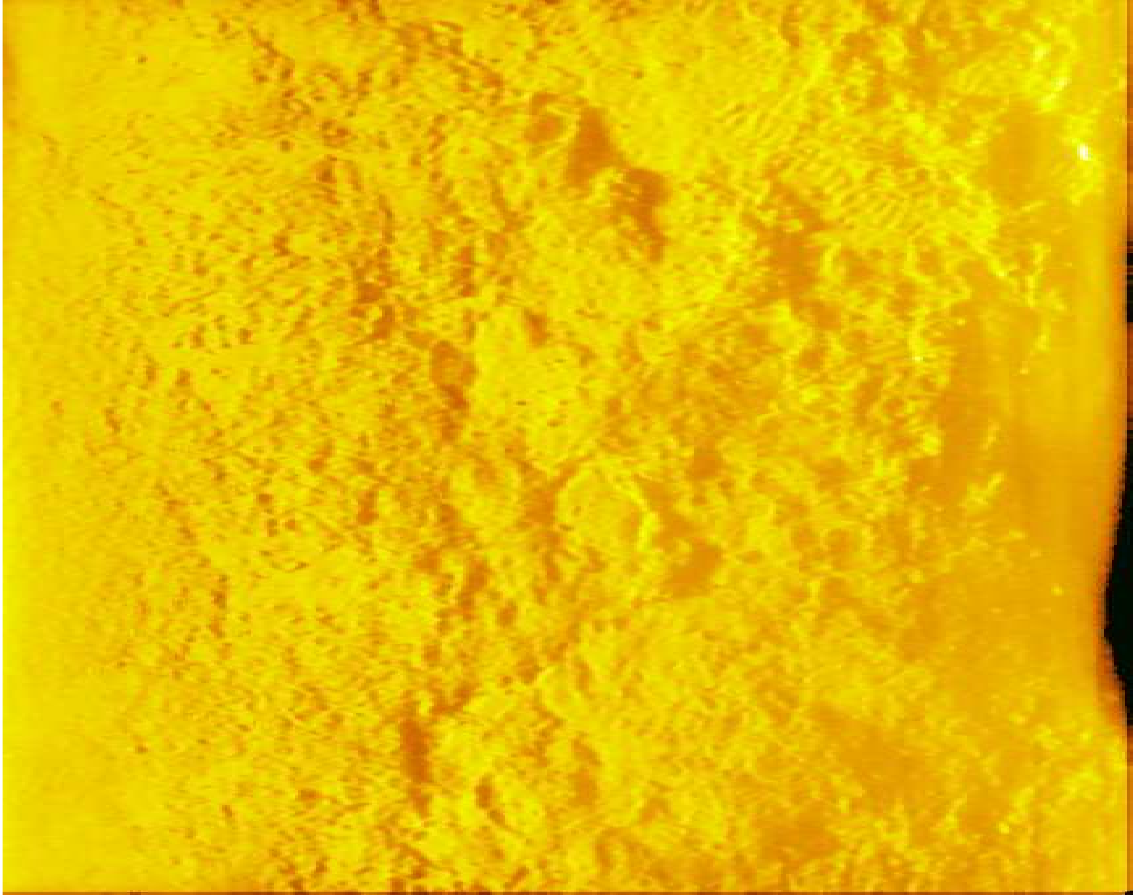}
			\caption{Original image}
		\end{subfigure}  
		\begin{subfigure}{.19\textwidth}
			\centering
			\includegraphics[width=\textwidth]{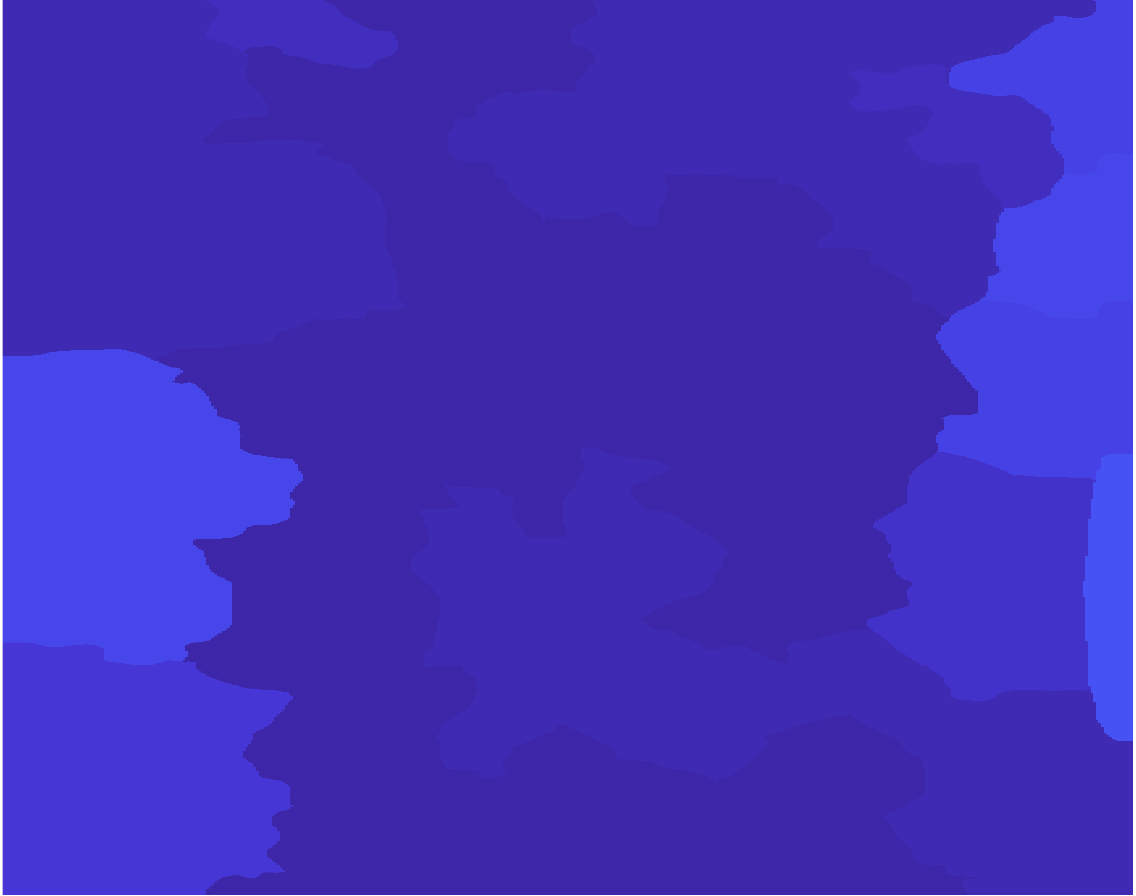}
			\caption{PFLICM - Crater}
		\end{subfigure}
		\begin{subfigure}{.19\textwidth}
			\centering
			\includegraphics[width=\textwidth]{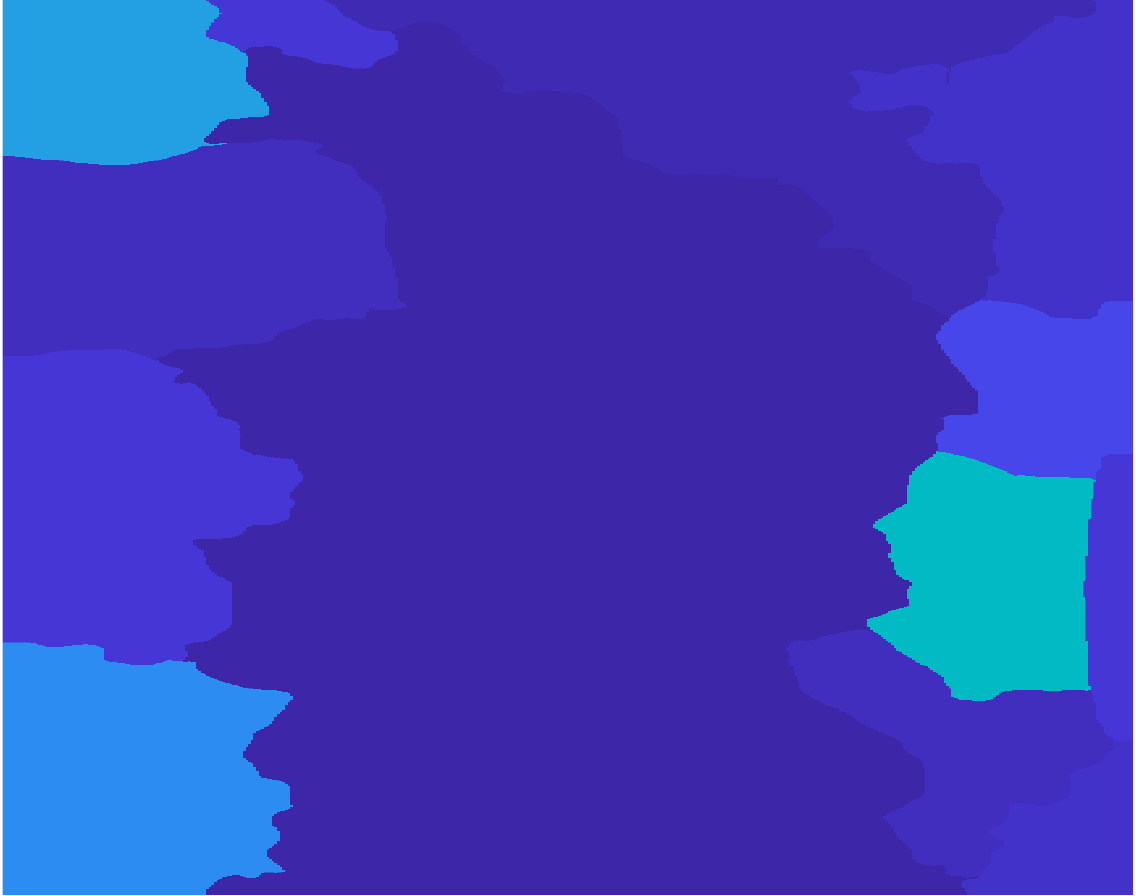}
			\caption{PFLICM - Flat}
		\end{subfigure}
		\begin{subfigure}{.19\textwidth}
			\centering
			\includegraphics[width=\textwidth]{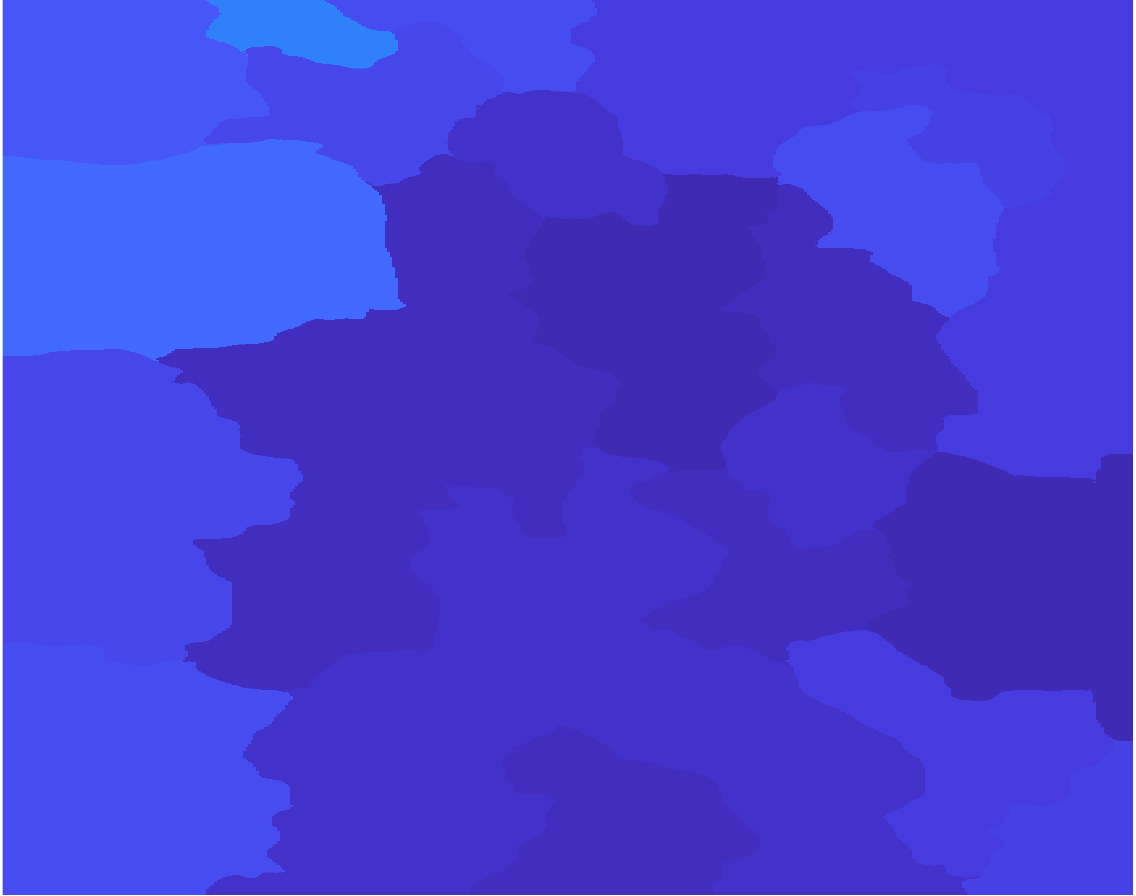}
			\caption{PFLICM - Ripples}
		\end{subfigure}
		\begin{subfigure}{.19\textwidth}
			\centering
			\includegraphics[width=\textwidth]{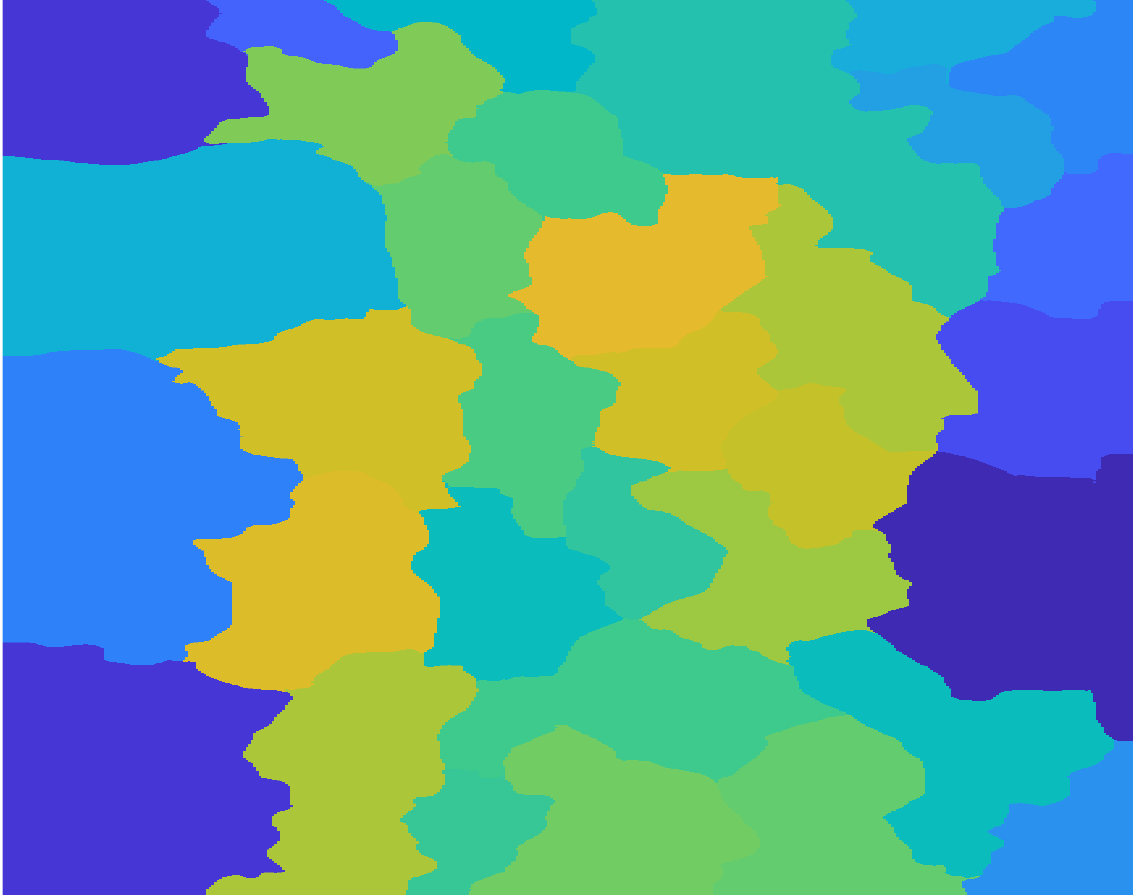}
			\caption{PFLICM - Rocky}
		\end{subfigure}
		
		\begin{subfigure}{.19\textwidth}
			\centering
			\includegraphics[width=\textwidth]{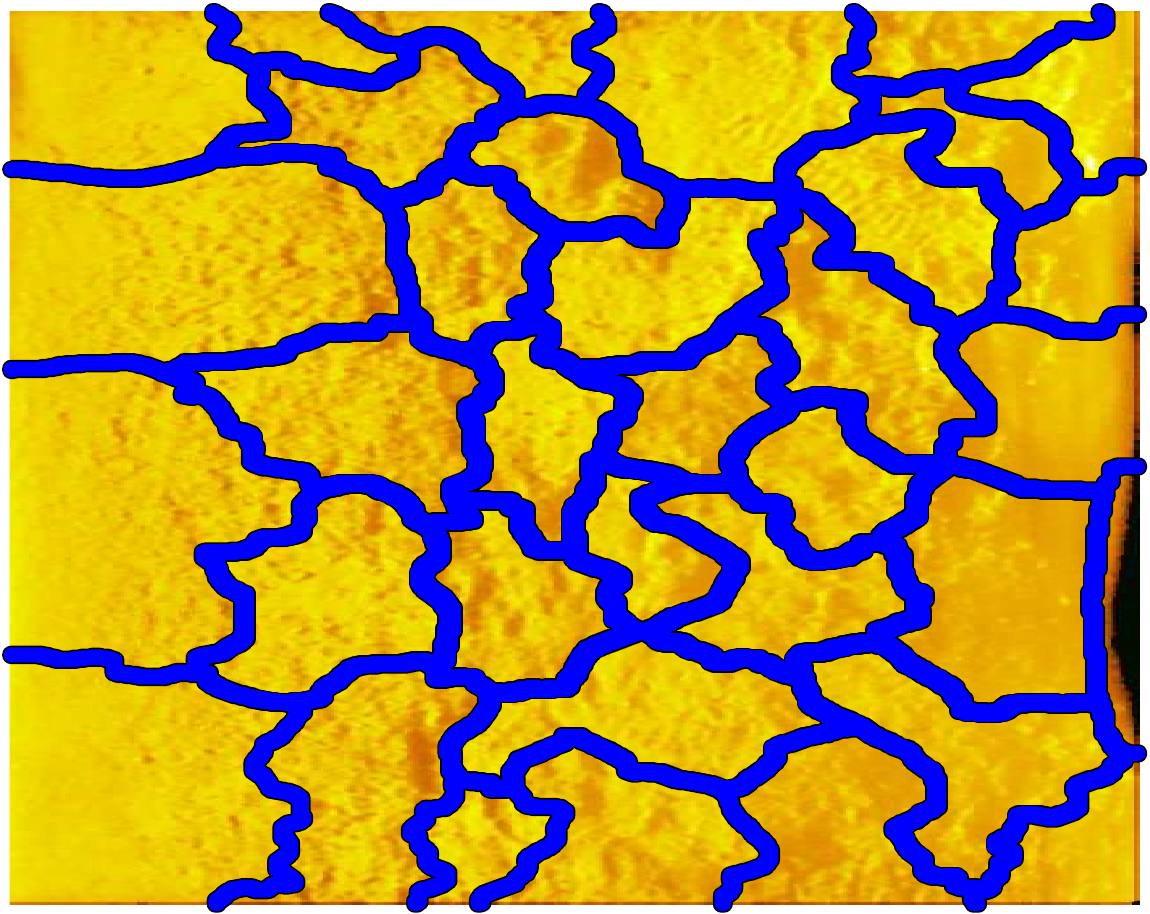}
			\caption{Superpixels}
		\end{subfigure}
		\begin{subfigure}{.19\textwidth}
			\centering
			\includegraphics[width=\textwidth]{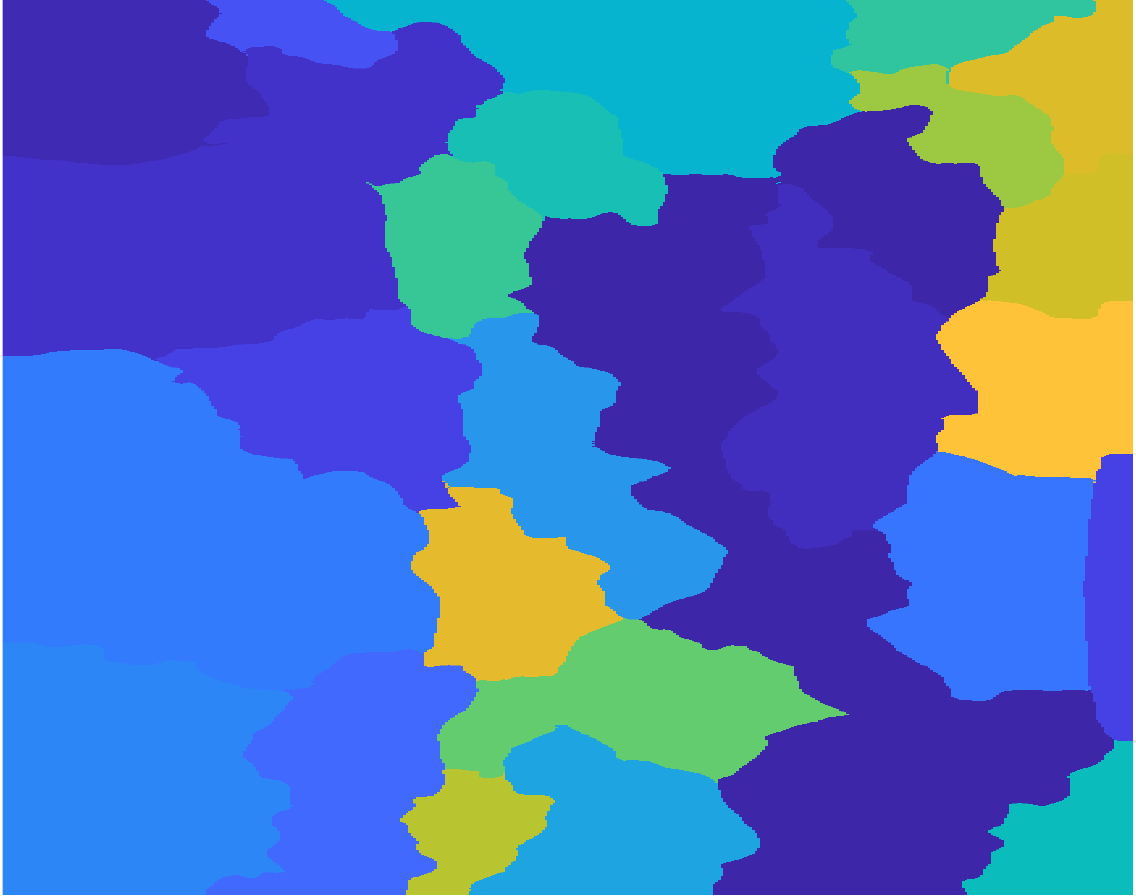}
			\caption{P$K$NN - Crater}
		\end{subfigure}
		\begin{subfigure}{.19\textwidth}
			\centering
			\includegraphics[width=\textwidth]{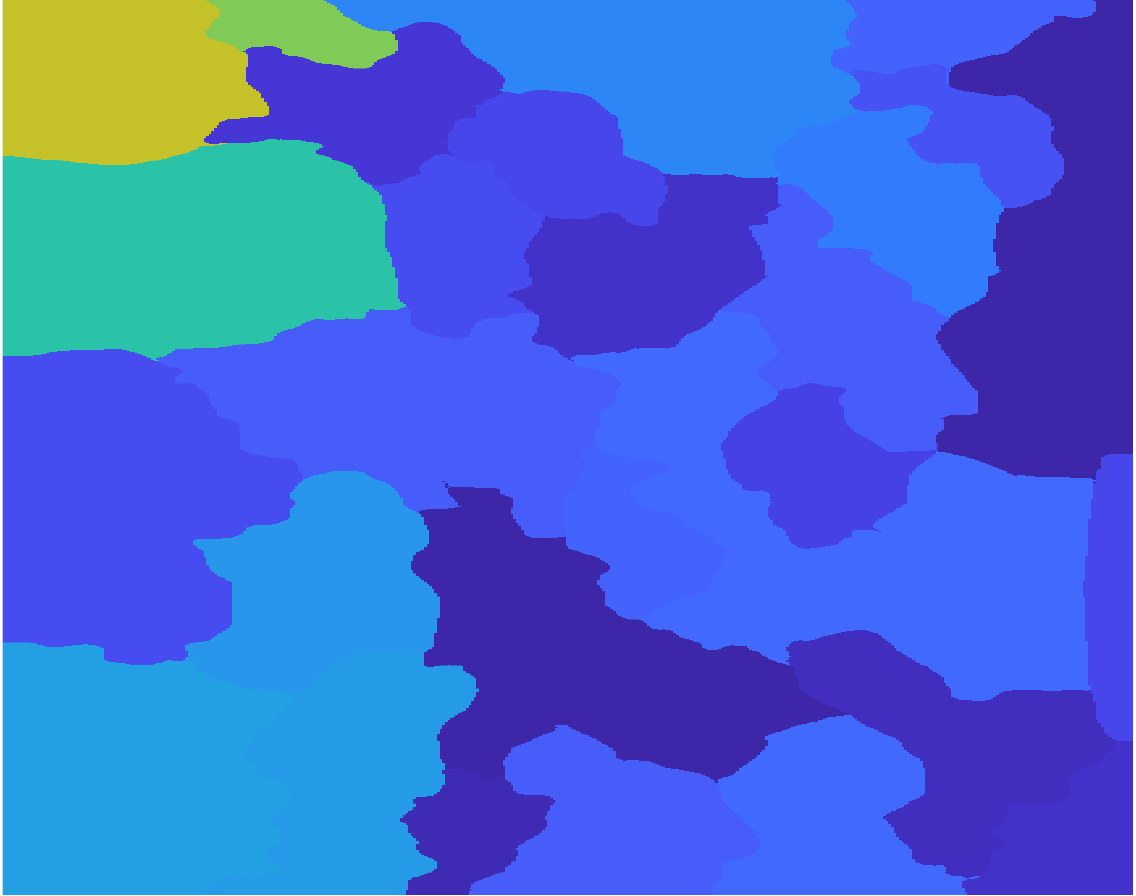}
			\caption{P$K$NN - Flat}
		\end{subfigure}
		\begin{subfigure}{.19\textwidth}
			\centering
			\includegraphics[width=\textwidth]{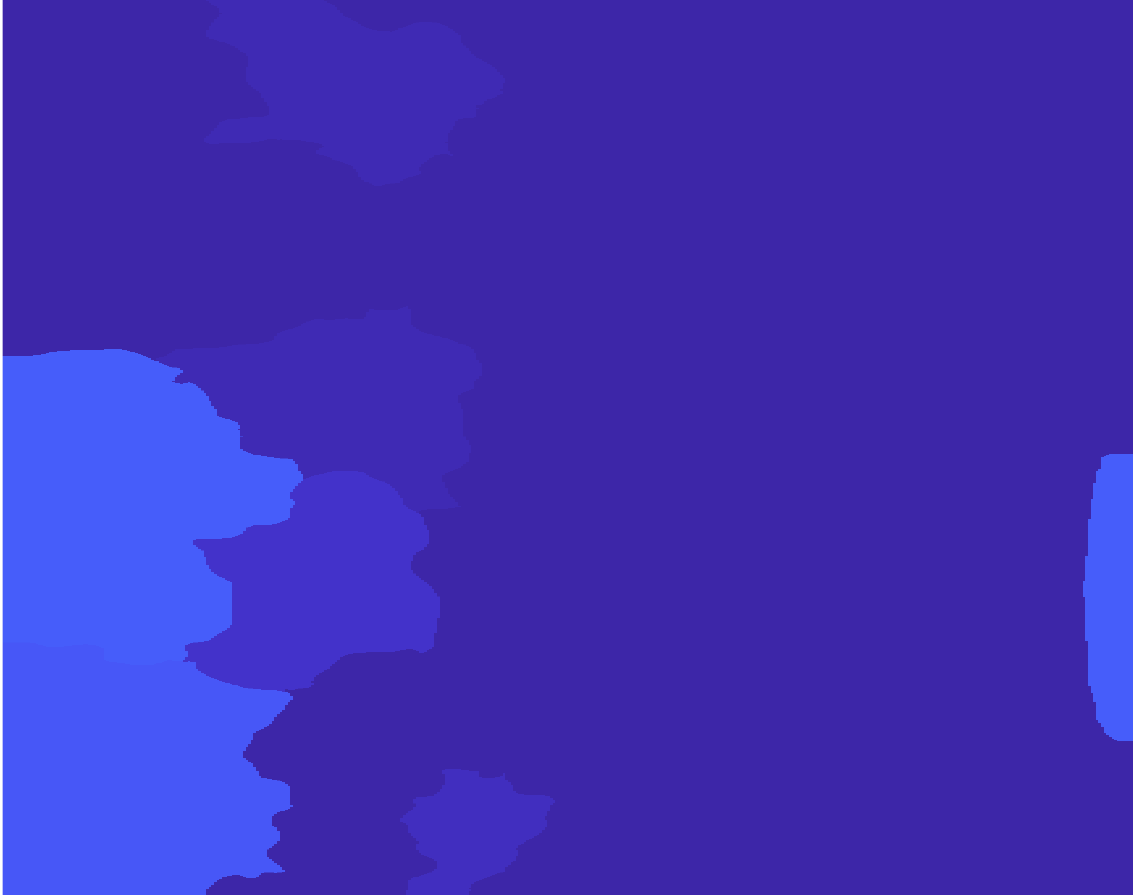}
			\caption{P$K$NN - Ripples}
		\end{subfigure}
		\begin{subfigure}{.19\textwidth}
			\centering
			\includegraphics[width=\textwidth]{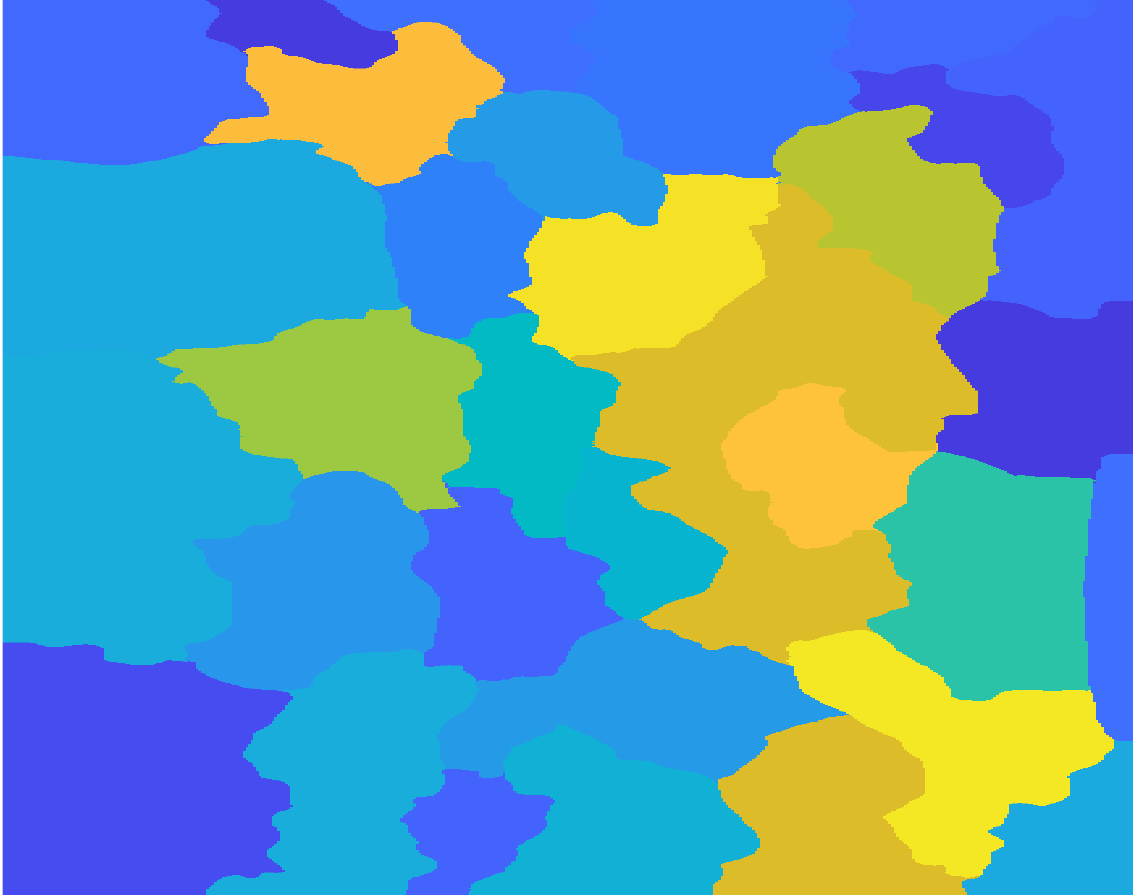}
			\caption{P$K$NN - Rocky}
		\end{subfigure}
		\caption{Example image two from fold three. The first column is actual image and superpixel segmentation. Of the remaining columns, the top row is PFLICM product maps and the bottom row is P$K$NN typicality maps.}
		\centering
		\label{fig:Img6}
	\end{figure}

	The parameters were determined manually for each algorithm. For PFLICM, the parameters were set to the following: membership weight $a$ = 14, typicality weight $b$ = 1.4, fuzzifier for fuzzy clustering term $m$ = 1.8, fuzzifier for possibilistic clustering term $q$ = 2.8, and number of clusters = 4. For P$K$NN, the parameters were to set the following: number of neighbors $k$ = 6, fuzzifier for weight term $m$= 2, and value to determine an outlier $\eta$ = 0.01. In order to perform a quantitative assessment of performance, crisp labels were assigned to each superpixel in the ground truth. The class with the maximum typicality values (for P$K$NN) and product of typicality and membership values (for PFLICM) were assigned to the superpixel as a predicted label. The confusion matrices for each fold are shown in Figure \ref{fig:ConfMats}. The segmentation results of each algorithm are shown in Figure \ref{fig:Img1} through \ref{fig:Img6}. The product maps are shown for PFLICM and the typicality maps for P$K$NN.
	
	\section{Discussion}
	Each algorithm achieves comparable performance both qualitatively and quantitatively across each fold. In the segmentation maps (Figures \ref{fig:Img1} through \ref{fig:Img6}), PFLICM and P$K$NN partitions are very similar to one another. The P$K$NN typicality maps had higher responses than the product maps from PFLICM for correctly identified textures. A possible reason for this is that P$K$NN was designed as a classifier while PFLICM is a clustering method. P$K$NN will discriminate between the textures at higher responses while PFLICM only groups areas of the image that are similar, resulting in less distinctive segments of textures.
	
	From the confusion matrices in Figure \ref{fig:ConfMats}, both PFLICM and P$K$NN perform the best on flat textures. A majority of the SAS images chosen in each fold contained a majority of superpixels in flat regions of the seafloor so the result is as expected since most of the training data is comprised of this class. Both algorithms did not perform as well on craters due to there being few superpixels containing pure crater regions. In future work, expanding the dataset to be more representative of each class could help mitigate some of the difference across each texture. Also, both algorithms perform soft classification so using ``crisp" metrics such as accuracy, precision, and recall from a confusion matrix may not be the best way to quantitatively evaluate segmentation. However, this lays a baseline for a measurable comparison of PFLICM and P$K$NN.

	Another metric used for comparison was the computational efficiency of each algorithm. For 
	training, the time complexity for PFLICM and P$K$NN are $O(mc^2i)$ and $O(mlog^2(m) + 
	mklog(m) + mlk)$ respectively ($m$ is the number of training samples, $c$ is the number of 
	clusters, $i$ is the number of iterations, $k$ is the number of neighbors, and $l$ is the 
	number of classes). For testing, the time complexity for PFLICM and P$K$NN are $O(nc^2)$ 
	and $O(nklog(m) + nlk)$ respectively ($n$ is the number of testing samples). 
	The terms $c$ and $k$ are mostly negligable since for most cases $c,k << m,n$, thus leaving
	the amount of training and testing as the dominant factors.
	The logarithmic terms in the P$K$NN complexity come from the use of a kd-tree 
	\cite{bentley1975multidimensional} for calculating the nearest neighbors.
	From these complexities it can be seen that both algorithms scale linearly with the number 
	of training and testing samples meaning that both algorithms could be effectively used
	with large amounts of testing data, with some limitations for P$K$NN. However, PFLICM does 
	scale better since the testing complexity does not rely on the number of training samples. 
	This implies that PFLICM could use a much larger training set and not impact testing time, 
	whereas the testing phase of P$K$NN scales logarithmically with the size of the training set.
	
	The labelling cost of each algorithm is another source of comparison for analysis. The P$K$NN does require that the training data is labelled at the superpixel level while PFLICM only requires labels assigned to each cluster center after training is completed. Despite increased labelling expense, P$K$NN will be more effective against poor labelling. P$K$NN computes possibilistic weights for each of the neighbors and these weights determine the level of importance each neighbor will have in the class confidences of a test sample despite the class label associated with each superpixel. For PFLICM, there is no mechanism in place to account for wrong labels on the cluster center level. As a result, P$K$NN accounts for errors made during labelling which is a desirable property for the segmentation process.
	\section{CONCLUSION}
	A comparison of PFLICM and P$K$NN are presented in this work. The performance of both algorithms is comparable to one another, both quantitatively and qualitatively. P$K$NN is significantly less computationally expensive than PFLICM in training as the training dataset increases, but PFLICM can have a much lower testing complexity if the training set is very large. P$K$NN also has an advantage over PFLICM in that P$K$NN can account for poor labelling that may occur within the training data. In future work, P$K$NN can be used to identify new textures in images through the possibilistic aspect. If the typicality values are low for each class, a new texture type can be identified. In order to add new seafloor types, a streaming/sequential clustering approach could be used\cite{ackerman2014incremental}. For an ATR system, the typicality values of P$K$NN could serve as environmental weights to assist in the classification of targets and create the environmental context for ATR applications.    
	\acknowledgments 
	
	This material is based upon work supported by the National Science Foundation Graduate Research Fellowship under Grant No. DGE-1842473 and by the Office of Naval Research grant N00014-16-1-2323.  
	
	\bibliographystyle{spiebib} 
	\bibliography{SPIE_references_2019} 

\end{document}